\documentclass[
preprintnumbers,
preprint,
a4paper,
showpacs, 
amsmath, 
amssymb, 
floatfix,
nofootinbib,
]{revtex4}

\usepackage{graphicx,color}

\begin{document}

\title{How does a collapsing star look?}

\author{Hirotaka Yoshino$^{(1)}$}
\author{Kazuma Takahashi$^{(2)}$}
\author{Ken-ichi Nakao$^{(2),(3)}$}

\affiliation{$^{(1)}$Advanced Mathematical Institute, Osaka City University, Osaka 558-8585 Japan}

\affiliation{$^{(2)}$Department of Mathematics and Physics,
Graduate School of Science, Osaka City University,
Osaka 558-8585, Japan}

\affiliation{$^{(3)}$Nambu Yoichiro Institute of Theoretical and Experimental Physics (NITEP), 
Osaka City University, Osaka 558-8585, Japan}

\preprint{OCU-PHYS-507, AP-GR-156, NITEP 27}

\date{Submitted 23 August 2019; published 28 October 2019}


%
%

\begin{abstract}
Time evolution of an optical image of a
pressureless star under
gravitational collapse is studied in the
geometric optics approximation.
The star surface is assumed to emit radiation 
obeying Lambert's cosine law but with an arbitrary spectral intensity
in the comoving frame. 
We develop a formalism 
for predicting observable quantities 
by photon counting and by radiometry, in particular,
spectral photon flux and spectral radiant flux.
Then, this method is applied to the two cases: 
One is monochromatic radiation, and 
the other is blackbody radiation. 
The two kinds of spectral flux are calculated numerically for each case.
It is reconfirmed that 
the redshift factor remains finite and  
the star becomes gradually invisible 
due to decay of the photon flux. We also 
develop an approximate method to 
present analytic formulas that describe
the late time behavior.
A possible connection of our study to observation of high-energy
neutrinos is briefly discussed. 
\end{abstract}

\maketitle

%
%

\section{Introduction}
\label{Sec:I}

We are now in the era to observe phenomena
caused by strong gravity. 
The great discoveries of gravitational waves have been made 
at advanced LIGO and Virgo detectors
(\cite{Abbott:2016} for the first one), and it is widely believed that
they were from mergers of binary black holes/neutron stars.
Also, the Event Horizon Telescope Collaboration
has succeeded in taking the observational image of
a shadow caused by the massive object at the center of the 
galaxy M87, which is supposed to be a supermassive black hole,
\footnote{See also \cite{Bambi:2019} for a discussion on the possibility
that the central object is a non-black-hole exotic object.}
through radio observations \cite{EHTCollaboration:2019}.
The primary factor to determine the shape of black hole shadows are 
circular orbits of photons 
on the photon spheres or the photon surfaces \cite{Virbhadra:1999,Claudel:2000}
or its extension, the fundamental photon orbits \cite{Cunha:2017}.
Roughly speaking, they 
are typically located at positions whose distance
from the event horizon is order of the black hole mass, $\sim M$
(in the unit where the Newtonian constant of gravitation becomes $G_{\rm N}=1$).

The black hole is defined as a complement of the causal past of the future null infinity or, in other words, a spacetime region
from which no information can escape to infinity.
This definition implies that a black hole and its any physical
influences must not exist in the view 
of any observer staying outside the black hole.
Therefore, as commented in Ref.~\cite{Cardoso:2017}, 
{\it ``observational proof for black holes is, by definition,
  impossible to obtain.''}

Black holes in our Universe will form through gravitational collapse of massive objects. As is well known, collapsing massive objects 
form black holes in the infinitely far future in the view of
any distant observer: The distant observers will see objects collapsing 
forever, even though they will form black holes. One of the very nontrivial {\it observable} predictions made by general relativity must not be the formation of black holes 
but is the process in which the spacetime geometry of the domain around a collapsing object asymptotically 
approaches the one described by the metric of the Kerr-Newman family. 
In this paper, we 
consider electromagnetic signals from
a dynamically collapsing object.
To be specific, 
consider a star under the gravitational collapse
that emits radiation from its surface. At a distant place,
an observer looks at the collapse with photons, e.g.,
by a telescope. Then, what kind of an image does the observer see,
and how does it change as time goes on?
We would like to give an answer to these questions
in the framework of general relativity. 
The asymptotic behavior of the image of the collapsing object 
gives us the information about the near horizon geometry
which will appear around the collapsing star.

It might be the usual first step to study
a simplified model rather than a realistic one.
Hence, let us assume the collapsing star to be a 
spherically symmetric one with uniform density and zero pressure, that is, 
the Oppenheimer-Snyder collapse \cite{Oppenheimer:1939}. 
Naively, one may expect that due to the combined effects
of the gravitational redshift and the Doppler shift, the frequency of
observed photons arriving at the observer
from the surface of the star would become 
smaller and smaller as time goes on, 
and eventually, it would become undetectable.

This is a classical subject: It was studied
in several literatures using the
geometric optics approximation, by approximately solving
null geodesic equations \cite{Podurets:1964,Ames:1968,Jaffe:1969,Lake:1979}
(early works are summarized in p.~847 of \cite{MTW73}, and 
see also \cite{Shapiro:1989,Shapiro:1996} for related works).
In the early works \cite{Podurets:1964,Ames:1968}, it was claimed that
the dimensionless quantity for the strength of the redshift,
$z:=\omega_{\rm e}^\prime/\omega_{\rm o}-1$, 
of the primary photons remains
finite as $z\approx 2$ 
due to approximately circular motion of
photons around the photon sphere $r=3M$, 
where $\omega_{\rm e}^\prime$ and $\omega_{\rm o}$ are the angular frequencies 
of emitted photons in the comoving frame 
attached to the surface of the star and that 
detected by a distant static observer, respectively.
Later, it was pointed out that the above analyses
had failed to take into account the contribution of photons that
are emitted into the directions almost or just tangential
to the surface of the star in its comoving frame, 
and thus, initially propagate in inward directions
in the static frame \cite{Jaffe:1969,Lake:1979}. Because 
the Doppler effect causes blueshift for such photons,
the value of $z$ of the primary component
becomes even smaller. 
Relatively recently, the same subject was studied 
but the analyses were restricted to the case that monochromatic waves
are emitted from the star surface with a very short period of
time \cite{Frolov:2005} or finite duration \cite{Frolov:2007}.

Although the subject was tackled more than fifty years ago,
as far as we know, this subject has not been fully explored.
In the above works, the calculations of spectral radiant flux   
were limited to the case that the star surface
radiates monochromatic waves, and the result is incomplete~\cite{Ames:1968}.
Although the formalism for calculating spectral radiant flux has been
given in Refs.~\cite{Frolov:2005,Frolov:2007},
in the application, approximate solutions are used 
for null geodesics and primary attention 
is paid only to the total luminosity and the
redshift factor. For this reason, 
we would like to revisit this problem to 
derive a formalism for calculating quantities
related to photon counting and radiometry,
under the condition that the radiator obeys Lambert's cosine law
with arbitrary spectral intensity. 
In particular, we present expressions for spectral photon flux and
spectral radiant flux that can be directly applied to numerical calculations. 
We apply our formalism to the 
two types of radiation: One is monochromatic radiation
and the other is blackbody one. 
This subject is studied with the method of ray tracing,
and it turns out that 
in order to see clearly the behavior of these quantities,
highly accurate numerical solutions of 
many null geodesics propagating 
from the surface of the star to the static observer are required.  
If we proceed with it with a personal computer, 
recent machines are definitely required.
Also, we present analytic 
formulas to describe the late time behavior of observable
quantities by developing a method of 
nonperturbative approximation to photon's worldlines.

One might consider that our analysis does not have a
practical application because typically, 
a collapsing star is surrounded by
optically thick plasma, and radiated photons do not directly escape
to a distant place. 
However, our analysis 
can be applied also to the case of the emission of high-energy
neutrinos from, for example, the neutrino-sphere of
a collapsing hot neutron star. Even if the plasma surrounding
the collapsing star 
is optically thick for photons, it can be transparent for neutrinos.
In order to see time evolution of the spectrum, 
an observation over a free fall time
(e.g., a few hundred $\mu$s in the case of the gravitational collapse 
of a hot neutron star with a few solar masses) is required,
and the time resolution of the observation should
be much less than the free fall time. 
The super-Kamiokande 
has the ability to determine the energy of each neutrino
with a time resolution of a few $\mathrm{ns}$ \cite{Abe:2013}.
Therefore, our study is a good starting point
for predicting the time dependence of the spectrum
of high-energy neutrinos emitted 
from a collapsing star, which will give us information
about the near horizon geometry around the collapsing star.

This paper is organized as follows.
In the next section, we
explain the setup of the problem.
In Sec.~\ref{Sec:III}, we study null geodesic equations
in the Schwarzschild spacetime and present several useful formulas.
In Sec.~\ref{Sec:IV}, we derive the formulas for
observable quantities 
by photon counting and radiometry.
In Sec.~\ref{Sec:V}, numerical methods and codes are explained.
In Sec.~\ref{Sec:VI}, we apply our formalism
to two examples, i.e., the cases where 
the star emits monochromatic radiation 
and blackbody radiation. 
In Sec.~\ref{Sec:VII}, we develop 
a nonperturbative approximate  method  to describe
the late time behavior of observable quantities.
Analytic formulas for the late time behavior are
derived to give support to the numerical results.
Section~\ref{Sec:VIII} is devoted to a summary and discussions.
In the Appendix, an approximate analysis 
of the redshift factor and some physical quantities
near the center of the image are presented.
Throughout the paper, we use
basically the units $c = G_{\rm N} = \hbar = k_{\rm B}=1$ where 
$c$, $G_{\rm N}$, $\hbar$, and $k_{\rm B}$ are the speed of light,
the Newtonian constant of gravitation, the reduced Planck constant, and the
Boltzmann constant, respectively.
But they will be recovered if necessary for clarity.

%
%
\section{Setup}
\label{Sec:II}

In this section, the setup of the system is explained.
In Sec.~\ref{Sec:IIA}, we introduce the Oppenheimer-Snyder model of
gravitational collapse of a star.
In Sec.~\ref{Sec:IIB},  
the properties of radiation from the star surface
are specified. The setup on 
the observer is explained in Sec.~\ref{Sec:IIC}.

\subsection{Oppenheimer-Snyder collapse}
\label{Sec:IIA}

As mentioned in Sec.~\ref{Sec:I},
we assume spherical symmetry of the system.
The region outside the star is assumed to be vacuum, and hence 
its geometry is described by the Schwarzschild metric, 
\begin{equation}
  ds^2=-f(r)dt^2+f(r)^{-1}dr^2+r^2(d\theta^2 + \sin^2\theta d\phi^2),
  \qquad
  f(r)=1-\frac{2M}{r}, \label{Sch-metric}
\end{equation}
where $M$ is a constant that corresponds to the total gravitational mass,
i.e., the Arnowitt-Deser-Misner mass in the present case. 
The star is supposed to be static, 
to keep its energy density uniform due to the pressure for $t< t_{\rm B}$,
and to begin collapsing
at $t=t_{\rm B}$ due to sudden disappearance of the pressure.
The radius of the static star is denoted by $R$.

By assumption, for $t> t_{\rm B}$, 
the star collapses  gravitationally. 
Such a situation is described by the Oppenheimer-Snyder model~\cite{Oppenheimer:1939}. 
The geometry inside the star is given
by the Lem\^itre-Friedmann-Robertson-Walker metric,  
\begin{equation}
ds^2=-d\tau^2+a^2\left[\frac{d\tilde{R}^2}{1-k\tilde{R}^2}+\tilde{R}^2(d\theta^2+\sin^2\theta d\phi^2)\right],
\end{equation}
with
\begin{subequations}
\begin{eqnarray}
  \tau&=&\frac{1}{2\sqrt{k}}(\xi+\sin\xi), \label{tau-xi}\\
  a&=&\frac12(1+\cos\xi), \label{a-xi}
\end{eqnarray}
\end{subequations}
for $0\le \tilde{R}\le R$.
If we set $k=2M/R^3$, the two metrics are connected
continuously and smoothly at $\tilde{R}=R$.

The world sheet of the star surface coincides with
a collection of radial timelike geodesics. Although the result is well known, 
we explicitly show the geodesic equations and their radial solution for 
later convenience. The world sheet of the star surface 
is denoted by $t=t_{\rm e}(\tau_{\rm e}^\prime)$
and $r=r_{\rm e}(\tau_{\rm e}^\prime)$,
where $\tau_{\rm e}^\prime$ is the proper time of the star surface
(the subscript ``e'' indicates
that the star surface is an emitter of radiation). 
Just at the beginning of the gravitational collapse, the star surface
is momentarily at rest,  i.e., $\dot{r}_{\rm e}=0$,
and thus the normalization condition of the timelike geodesic
tangent leads to $\dot{t}_{\rm e}=[f(R)]^{-1/2}$,  
where the dot denotes the derivative with
respect to the proper time $\tau_{\rm e}^\prime$.
The $t$-component of the geodesic equation 
leads to
\begin{equation}
  f(r_{\rm e})\dot{t}_{\rm e}=\sqrt{f(R)},
  \label{Eq:geodesic_dottre}
\end{equation}
after integrating once. Then, 
from the normalization of the timelike geodesic tangent, we have 
\begin{equation}
\dot{r}_{\rm e}=-\sqrt{f(R)-f(r_{\rm e})},
  \label{Eq:geodesic_dortre}
\end{equation}
where the sign is chosen to be consistent with the fact that the
star is collapsing. 
These two equations can be integrated by 
using the auxiliary variable $\xi$ in Eqs.~\eqref{tau-xi} and \eqref{a-xi}, 
and we have 
\begin{subequations}
\begin{eqnarray}
  \tau_{\rm e}^\prime
  &=& \frac{R}{2}\sqrt{\frac{R}{2M}}\left(\xi+\sin\xi\right),
\label{timelike-geodesic-tau}\\
  r_{\rm e}&=&\frac{R}{2}\left(1+\cos\xi\right),
  \label{timelike-geodesic-r}
\end{eqnarray}
\begin{equation}
  t_{\rm e}-t_{\rm B}=2M
  \log\left[\frac{\sqrt{\frac{R}{2M}-1}+\tan(\xi/2)}
    {\sqrt{\frac{R}{2M}-1}-\tan(\xi/2)}\right]
  +\sqrt{\frac{R}{2M}-1}\left[\frac{R}{2}\sin\xi
  +\left(2M+\frac{R}{2}\right)\xi\right].
\label{timelike-geodesic-t}
\end{equation}
\end{subequations}
Here, the integral constant is chosen so that
$r_{\rm e}=R$ and $t_{\rm e}=t_{\rm B}$ 
at $\tau_{\rm e}^\prime=0$.
The surface of the star crosses the horizon at
\begin{equation}
\xi=\xi_{\rm h}:=2\arctan\sqrt{\frac{R}{2M}-1}
\label{horizon_crossing_xih}
\end{equation}
and $t_{\rm e}$ diverges there 
reflecting the fact that the Schwarzschild time $t$
becomes degenerate on the horizon.

\subsection{Radiation from the star surface}
\label{Sec:IIB}

In this paper, we assume that radiation is emitted from the star surface
in accordance with Lambert's cosine law.
If $dN$ number of photons are emitted
from a surface in the interval $d\tau_{\rm e}^\prime$
of its proper time, photon flux is defined as
\begin{equation}
  \mathcal{F}_{\rm e}^{\rm (N)} = \frac{dN}{d\tau_{\rm e}^\prime},
  \label{Def:photon_flux_emission}
\end{equation}
where the superscript ``(N)'' of $\mathcal{F}_{\rm e}^{\rm (N)}$ is added to
stress that we are considering the photon number flux
(not the radiation energy flux). 
The photon flux within an infinitesimal interval of angular frequency
$d\omega_{\rm e}$ emitted from an 
areal element $dS_{\rm e}^\prime$ on the star surface
within an infinitesimal solid angle $d\Omega_{\rm e}^\prime$ 
with the direction at an angle $\vartheta_{\rm e}^\prime$ to the normal of the area element is given by
\begin{equation}
  d^3\mathcal{F}_{\rm e}^{\rm (N)}
  =I_{\rm e}^{\rm (N)}
  \cos\vartheta_{\rm e}^\prime d\Omega_{\rm e}^\prime dS_{\rm e}^\prime
  d\omega_{\rm e}^\prime,
  \label{Lambert1}
\end{equation} 
Here, $I_{\rm e}^{\rm (N)}$ is called spectral photon radiance.\footnote{
  In terms of radiated energy $E_{\rm e}$,
  the spectral radiance $I_{\rm e}^{\rm (E)}$ is introduced by
  $d^3\left({dE_{\rm e}}/{d\tau_{\rm e}^\prime}\right)
  =I_{\rm e}^{\rm (E)} \cos\vartheta_{\rm e}^\prime d\Omega_{\rm e}^\prime dS_{\rm e}^\prime
  d\omega_{\rm e}^\prime.$
  The two kinds of radiance are related as $I_{\rm e}^{\rm (E)}=\omega_{\rm e}^\prime I_{\rm e}^{\rm (N)}$ in the unit $\hbar=1$.
}
In general, $I_{\rm e}^{\rm (N)}$ is a function of
time $\tau_{\rm e}^\prime$, angular frequency $\omega_{\rm e}^\prime$,
and the angle $\vartheta_{\rm e}^\prime$. 
If $I_{\rm e}^{\rm (N)}$ does not depend on the angle $\vartheta_{\rm e}^\prime$,
the radiation is said to follow Lambert's cosine law. 
Here, all quantities with prime are measured in
the comoving frame of the star surface, or more precisely,
with respect to the tetrad frame,
\begin{subequations}
\begin{align}
  (e_0^\prime)_{\mu}&=\left(-f(r_{\rm e})\dot{t}_{\rm e},f^{-1}(r_{\rm e}) \dot{r}_{\rm e}, 0,0\right), \label{tetrad-comoving-0} \\
  (e_1^\prime)_{\mu}&=\left(-\dot{r}_{\rm e}, \dot{t}_{\rm e}, 0,0\right), \\
  (e_2^\prime)_{\mu}&=\left(0,0,r_{\rm e},0\right), \\
  (e_3^\prime)_{\mu}&=\left(0,0,0,r_{\rm e}\sin\theta\right). \label{tetrad-comoving-3}
\end{align}
\end{subequations}
Note that by virtue of Eqs.~\eqref{Eq:geodesic_dottre}
and \eqref{Eq:geodesic_dortre}, each component of $(e_0^\prime)_{\mu}$
and $(e_1^\prime)_{\mu}$ becomes a function of
$r_{\rm e}$. Introducing the 
null vector tangent to the worldline 
of an emitted photon from the star surface as 
\begin{equation}
k^\mu :=\frac{dx^\mu}{d\lambda}= (t_{,\lambda},\ r_{,\lambda},\ 0,\ \phi_{,\lambda}),
\label{def:null-tangent-vector}
\end{equation}
where the worldline is a geodesic
parametrized by the affine parameter $\lambda$, 
the angle $\vartheta_{\rm e}^\prime$ is 
given by
\begin{equation}
  \tan\vartheta_{\rm e}^\prime
=\frac{k^\mu (e_3^\prime)_{\mu}}{k^\nu (e_1^\prime)_{\nu}},
  \label{definition-thetaeprime}
\end{equation}
in terms of the tetrad vectors,
Eqs.~\eqref{tetrad-comoving-0}--\eqref{tetrad-comoving-3}. 
Note that due to the spherical symmetry, 
it is sufficient to consider geodesics 
in the equatorial plane $\theta=\pi/2$.

The value of $I_{\rm e}^{\rm (N)}$ may change 
as the collapse proceeds.  
Since the radius of the star surface $r_{\rm e}$ is
a monotonically decreasing function of time, $I_{\rm e}^{\rm (N)}$ 
is regarded as a function of not only
$\omega_{\rm e}^\prime$ but also $r_{\rm e}$,
i.e., $I_{\rm e}^{\rm (N)}(r_{\rm e},\omega_{\rm e}^\prime)$. 
We also introduce the following quantity called photon radiance,
\begin{equation}
  J_{\rm e}^{\rm (N)}(r_{\rm e}):=\int_0^\infty
  I_{\rm e}^{\rm (N)}(r_{\rm e},\omega_{\rm e}^\prime) d\omega_{\rm e}^\prime.
\end{equation}
The photon radiance and the photon flux
are related as 
\begin{equation}
  d^2\mathcal{F}_{\rm e}^{\rm (N)}
  =J_{\rm e}^{\rm (N)}(r_{\rm e}) \cos\vartheta_{\rm e}^\prime d\Omega_{\rm e}^\prime dS_{\rm e}^\prime.
  \label{Lambert2}
\end{equation}

In this paper, we develop a formalism
to calculate observable quantities,
in particular, spectral photon flux and spectral radiant
flux for arbitrary 
$I_{\rm e}^{\rm (N)}(r_{\rm e},\omega_{\rm e}^\prime)$.
We apply our formalism to radiation with the following two kinds of spectra.

\subsubsection{Monochromatic radiation}
\label{Sec:IIB1}

The first case is that the emitted 
radiation is monochromatic with the angular frequency
$\bar{\omega}_{\rm e}^\prime$. In this case, $I_{\rm e}^{\rm (N)}$ takes the form
\begin{equation}
  I_{\rm e}^{\rm (N)}(r_{\rm e},\omega_{e}^\prime) = J_{\rm e}^{\rm (N)}(r_{\rm e})
  \delta(\omega_{\rm e}^\prime -\bar{\omega}_{\rm e}^\prime).
  \label{intensity-monochromatic-wave-emission}
\end{equation}
In addition, we assume that $J_{\rm e}^{\rm (N)}(r_{\rm e})$ be constant
throughout the collapse for simplicity:
\begin{equation}
  J_{\rm e}^{\rm (N)}(r_{\rm e})\equiv J.
  \label{constancy-of-intensity}
\end{equation}
The extension to the case that $J(r_{\rm e})$ varies in time
is straightforward.

\subsubsection{Blackbody radiation}
\label{Sec:IIB2}

The second case is blackbody radiation. 
In this case, we have 
\begin{equation}
  I_{\rm e}^{\rm (N)}(\omega_{\rm e}^\prime, r_{\rm e})
  =\frac{\omega_{\rm e}^{\prime 2}}{4\pi^3\left[\exp(\omega_{\rm e}^\prime/T_{\rm e}^\prime)-1\right]}.
  \label{intensity-Planck-radiation}
\end{equation}
Note that we are adopting the unit $\hbar=k_{\rm B}=1$, and
this formula is not for the energy flux but for the photon flux.
For this formula, $J_{\rm e}^{\rm (N)}(r_{\rm e})$ is calculated as
\begin{equation}
  J_{\rm e}^{\rm (N)}(r_{\rm e})=\frac{\zeta(3)}{2\pi^3}{T_{\rm e}^{\prime}}^3.
  \label{J-Planck-radiation}
\end{equation}
In general, $I_{\rm e}^{\rm (N)}(\omega_{\rm e}^\prime, r_{\rm e})$
and $J_{\rm e}^{\rm (N)}(r_{\rm e})$ are dependent
on $r_{\rm e}$ through the change in the temperature $T_{\rm e}^\prime$.
In this paper, for simplicity, we assume
\begin{equation}
  T_{\rm e}^\prime=\mathrm{constant},
  \label{constancy-of-temperature}
\end{equation}
and therefore, the value of $I_{\rm e}^{\rm (N)}(\omega_{\rm e}^\prime)$
does not depend on the radius of the star surface.

\subsection{The observer}
\label{Sec:IIC}

%
\begin{figure}[tb]
\centering
\includegraphics[width=0.7\textwidth,bb=0 0 394 197]{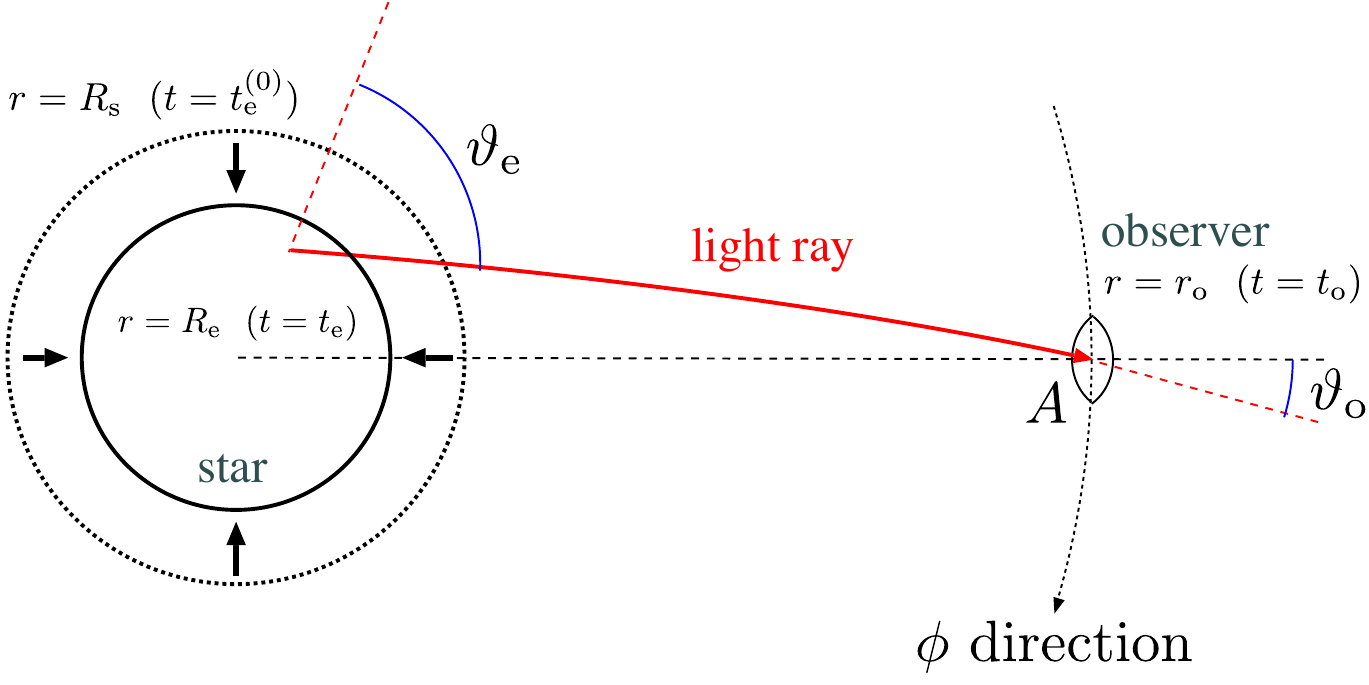}
 \caption{
Setup of the system and an example of a light ray.
}
 \label{Fig:setup}
\end{figure}
%

Figure~\ref{Fig:setup} is a schematic picture of relative positions of 
the collapsing star and the observer. 
The observer is at rest with respect to
the Schwarzschild frame at the radius $r_{\rm o}$.
More specifically, the observer's worldline in the four-dimensional spacetime
is parametrically given by
$t=t_{\rm o}(\tau_{\rm o})$, 
$r=r_{\rm o}(\tau_{\rm o})$, $\theta=\theta_{\rm o}(\tau_{\rm o})$, and $\phi=\phi_{\rm o}(\tau_{\rm o})$, 
where $\tau_{\rm o}$ is the proper time of the observer  
(the subscript ``o'' indicates the observer). 
By assumption, $r_{\rm o}$, $\theta_{\rm o}$ and $\phi_{\rm o}$ are constant.
Because the spacetime is spherically symmetric, 
we set $\theta_{\rm o}=\pi/2$, $\phi_{\rm o}=0$
without loss of generality.
The proper time  
$\tau_{\rm o}$ and the coordinate time $t_{\rm o}$ are related as
$d\tau_{\rm o}/dt_{\rm o}=\sqrt{f(r_{\rm o})}$. 
Furthermore, we assume that $r_{\rm o}$ is much larger than the radius of the
initial star surface $R$. 
The observer detects photons from the star surface using a telescope,
and determines the photon/radiant flux with 
respect to the proper time $\tau_{\rm o}$ of the observer.
The telescope is directed toward the center of the star,
and its lens has the area $A$.
In the following sections, we calculate the flux of photon number and
the flux of photon energy 
that come into the telescope through its lens.

We determine the origin of time coordinate $t$ so that the observer finds that the star begins to collapse at $t_{\rm o}=0$. 
A radially propagating photon emitted from $r=R$
takes a period of time $\mathcal{T}$ given by 
\begin{equation}
  \mathcal{T}= r_{\rm o}-R+2M\log\left(\frac{r_{\rm o}-2M}{R-2M}\right),
  \label{te0}
\end{equation}
in order to arrive at the observer staying at $r=r_{\rm o}$. 
Hence, the beginning time of the gravitational collapse,
$t_{\rm B}$, introduced 
in the paragraph including Eq.~\eqref{Sch-metric} is equal to $-\mathcal{T}$. 
Note that the observer finds that the star begins to collapse
through the time variation in the redshift at the center of the image.

%
%
\section{Null geodesics in the Schwarzschild spacetime}
\label{Sec:III}

In the geometric optics approximation,
all the information we need about detected photons
(e.g., the angular frequency and the photon flux)
are included in the null geodesic congruence 
that connects the emitter with the observer. 
In this section, 
we explain basic properties of null geodesics in this spacetime.
Useful analytic
formulas are also presented.

An example of a light ray is depicted in Fig.~\ref{Fig:setup}.
A photon is emitted from the star surface 
at $(t,r)=(t_{\rm e}, r_{\rm e})$ and arrives
at the observer at $(t,r)=(t_{\rm o}, r_{\rm o})$. 
At the emission and observation events, 
we introduce the angles $\vartheta_{\rm e}$ and $\vartheta_{\rm o}$,
respectively, between the 
spatial direction of photon's motion and the outward radial direction 
measured in the static frame.
To be more precise, introducing
the tetrad basis associated with the static frame,
\begin{subequations}
\begin{align}
  (e_0)_{\mu} &= \left(-f^{1/2},0,0,0\right), \label{tetrad-static-0}\\
  (e_1)_{\mu} &= \left(0,f^{-1/2},0,0\right), \label{tetrad-static-1}\\
  (e_2)_{\mu} &= \left(0,0,r,0\right), \\
  (e_3)_{\mu} &=\left(0,0,0, r\sin\theta\right), \label{tetrad-static-3}
\end{align}
\end{subequations}
and the tangent vector of the null geodesic
given by Eq.~\eqref{def:null-tangent-vector}, 
$\vartheta_{\rm e}$ and $\vartheta_{\rm o}$
are defined as
\begin{equation}
  \tan\vartheta_{\rm e}=
\left. \frac{k^\mu (e_3)_{\mu}}{k^\nu (e_1)_{\nu}}\right|_{(t,r)=(t_{\rm e},r_{\rm e})}, \qquad
    \tan\vartheta_{\rm o}=
 \left.\frac{k^\mu (e_3)_{\mu}}
     {k^\nu (e_1)_{\nu}}\right|_{(t,r)=(t_{\rm o},r_{\rm o})}.
     \label{definition-thetao-thetae}
\end{equation}
Here, note that $\vartheta_{\rm e}$ is different from 
$\vartheta_{\rm e}^\prime$ introduced 
in Eq.~\eqref{Lambert1}, and these two quantities must be distinguished.

Here, we mention the basic idea of the ray tracing method
whose numerical details will be presented in Sec.~\ref{Sec:V}. 
Let us consider a photon that comes into the telescope
at $t=t_{\rm o}$ with the angle $\vartheta_{\rm o}$.
By numerically solving the null geodesic equation 
\begin{equation}
  \frac{dk^\mu}{d\lambda}+\Gamma^{\mu}_{~\alpha\beta}k^{\alpha}k^{\beta}=0,
  \label{null-geodesic-equations}
\end{equation}
backward in time, we obtain 
the worldline of the photon in the form, 
$t=t(\lambda)$, $r=r(\lambda)$, $\theta=\pi/2$ and $\phi=\phi(\lambda)$.
If the value of $\vartheta_{\rm o}$ is not too large,
the worldline of the photon intersects the world sheet
of the star surface, 
which is given by Eqs.~\eqref{timelike-geodesic-r}
and \eqref{timelike-geodesic-t}. 
The intersection corresponds to the event of photon emission 
from the star surface.

For a photon traveling in the equatorial plane $\theta=\pi/2$, 
the geodesic equations \eqref{null-geodesic-equations} lead to
the conservation laws for energy and angular momentum of a photon,
\begin{subequations}
  \begin{align}
    ft_{,\lambda} &= E_{\rm ph}, \label{null-geodesic-energy}\\
    r^2\phi_{,\lambda} &= bE_{\rm ph}, \label{null-geodesic-angular-momentum}
  \end{align}
by integrating $t$ and $\phi$ components once, respectively,
where $E_{\rm ph}$ and $b$ are integral constants
corresponding to the energy and the impact parameter
of the photon, respectively.
The null condition $k_\mu k^\mu = 0$ becomes    
  \begin{equation}    
    ft_{,\lambda}^2 = f^{-1}r_{,\lambda}^2+r^2\phi_{,\lambda}^2. 
    \label{null-geodesic-null-condition}
  \end{equation}
\end{subequations}
From Eqs.~\eqref{null-geodesic-energy}--\eqref{null-geodesic-null-condition}, 
we have
\begin{subequations}
  \begin{eqnarray}
    \frac{dr}{dt} &=& \pm f\sqrt{1-\frac{b^2f}{r^2}},
\label{geodesic-drdt}
    \\
    \frac{d\phi}{dt} &=& \frac{bf}{r^2}.
    \label{geodesic-dphidt}
  \end{eqnarray}
\end{subequations}
Here, we would like to make a remark on a previous work. 
When the sign of Eq.~\eqref{geodesic-drdt}
is plus (respectively, minus), the photon is moving outward (respectively, inward).
As we 
will see in Fig.~\ref{Fig:Photon_trajectory_example} later,
there are photons that initially travel inward but turn outward and 
arrive at the observer. 
In this case, the right-hand side of Eq.~\eqref{geodesic-drdt} changes
its sign during the propagation.  Since 
it is difficult to handle such an equation numerically,
we solve the second-order differential equation for
$r$ instead of Eq.~\eqref{geodesic-drdt} (see Sec.~\ref{Sec:V}). 
By contrast, the analysis of Ref.~\cite{Ames:1968}
used Eq.~\eqref{geodesic-drdt} only with the plus sign, and thus,
it has missed the contribution from
photons that change the sign of Eq.~\eqref{geodesic-drdt}.

Below, we list up the useful relations
that will be used later.
These relations can be used also 
to check the accuracy of numerical calculations.

\subsection{Radial photon orbit}
\label{Sec:IIIA}

In the case of $b=0$, photons propagate in the radial direction. 
Equations~\eqref{geodesic-drdt} and \eqref{geodesic-dphidt} 
are integrated as $\phi={\rm const}$, and 
\begin{equation}
  t-t_{\rm o} = r-r_{\rm o}+2M\log\left(\frac{r-2M}{r_{\rm o}-2M}\right),
  \label{radial-solution}
\end{equation}
where we have chosen the plus sign of Eq.~\eqref{geodesic-drdt}. 
Equation~\eqref{te0} is based on this formula:
The value of $\mathcal{T}$ is obtained by substituting 
$(t,r)=(-\mathcal{T}, R)$ and $t_{\rm o}=0$ into Eq.~\eqref{radial-solution}.

\subsection{Photon orbit with critical impact parameter}
\label{Sec:IIIB}

Another special case where  
worldlines of photons are analytically obtained is 
that the impact parameter $b$ has the following value:
\begin{equation}
b_{\rm crit}:=3\sqrt{3}M, \label{bcrit}
\end{equation}
where $b_{\rm crit}$ is called the critical impact parameter.
Setting $b=b_{\rm crit}$, Eq.~\eqref{geodesic-drdt} becomes
\begin{equation}
\frac{dt}{dr}=\pm\frac{r^{5/2}}{(r-2M)(r-3M)\sqrt{r+6M}}.
\end{equation}
It is seen that $r=3M$ for arbitrary $t$ is a solution to this
equation, which represents the worldline of 
a circularly orbiting photon. Below, we solve for
worldlines of photons whose radial positions change in time.
This equation can be integrated as
\begin{equation}
  t=\pm F(r)+\mathrm{const}, 
  \label{orbit-critical-impact-parameter-1}
\end{equation}
where the function $F(r)$ is given by
\begin{equation}
  F(r)=\bar{F}(r) -3\sqrt{3}M
  \log\left|\frac{\sqrt{3r}+\sqrt{r+6M}}{\sqrt{3r}-\sqrt{r+6M}}\right|
  \label{orbit-critical-impact-parameter-2}
\end{equation}
with
\begin{equation}
  \bar{F}(r):=
  \sqrt{r(r+6M)}
  +4M\log\left(\sqrt{\frac{r}{M}}+\sqrt{\frac{r}{M}+6}\right)
  +2M\log\left(\frac{2\sqrt{r}+\sqrt{r+6M}}{2\sqrt{r}-\sqrt{r+6M}}\right).
  \label{orbit-critical-impact-parameter-3}
\end{equation}
Because $F(r)$ diverges at $r=3M$, these worldlines never cross $r=3M$.
For this reason, $t=F(r)$ implicitly includes two kinds of solutions:
$t=F^{\rm (out)}(r)$ with
\begin{equation}
  F^{\rm (out)}(r) := \left.F(r)\right|_{3M<r<\infty},
  \label{bcrit_geodesic_outer_region}
\end{equation}
whose worldline is confined in the outside region $3M<r<\infty$,
and $t=F^{\rm (in)}(r)$ with
\begin{equation}
  F^{\rm (in)}(r) = \left. F(r)\right|_{2M<r<3M}.
  \label{bcrit_geodesic_inner_region}
\end{equation}
whose worldline is confined in the inside region $2M<r<3M$. 
As a result, we have four kinds of solutions, 
$t=\pm F^{\rm (out)}(r)+{\rm const}$ and $t=\pm F^{\rm (in)}(r)+{\rm const}$, in total.

Among these four solutions, the solution of special importance is
\begin{equation}
  t - t_{\rm o} = F^{\rm (out)}(r)-F^{\rm (out)}(r).
  \label{orbit_bcrit_observation_point}
\end{equation}
This solution represents 
a worldline of a photon that originally
orbits in the domain very close to a circle on $r=3M$ in the past,
but gradually leaves that domain outward 
and arrives at the observation point $(t,r)=(t_{\rm o}, r_{\rm o})$.
For $t\ll -M$, this orbit is approximated as
\begin{equation}
  \frac{r}{M}-3\approx \exp\left[\frac{t-t_{\rm o}-C+F(r_{\rm o})}{3\sqrt{3}M}\right]
  \label{asymptotic_behavior_geodesic_bcrit}
\end{equation}
with 
\begin{equation}
  \frac{C}{M}=3\sqrt{3}+4\log(\sqrt{3}+3)
  +2\log\left(\frac{2\sqrt{3}+3}{2\sqrt{3}-3}\right)-3\sqrt{3}\log 18.
  \label{definition-C}
\end{equation}
As will be shown in Secs.~\ref{Sec:VI} and \ref{Sec:VII},
this orbit plays a crucial role for the late time behavior of observables.

In Sec.~\ref{Sec:VII}, two other kinds of solutions,
$t=-F^{\rm (out)}(r)$ and $t=-F^{\rm (in)}(r)$, will be used.
These solutions represent worldlines of photons
that asymptotically approach the photon sphere for large $t$
from the outside and inside regions, respectively.

\subsection{Relation between $r_{\rm e}$ ($r_{\rm o}$) and $\vartheta_{\rm e}$ ($\vartheta_{\rm o}$)}
\label{Sec:IIIC}

Due to the conservation laws \eqref{null-geodesic-energy} and \eqref{null-geodesic-angular-momentum} of a photon 
and the null condition~\eqref{null-geodesic-null-condition},
the tangent vectors $k^\mu$ 
of a geodesic of a photon, Eq.~\eqref{def:null-tangent-vector}, 
is written in terms of $E_{\rm ph}$, $b$, and $r$.
At the points of the emission and observation events,
the tangent vectors can be rewritten as
\begin{subequations}
\begin{eqnarray}
  \left.k^\mu\right|_{r=r_{\rm e}}
  &=&
  E_{\rm ph}(f(r_{\rm e})^{-1},\ \cos\vartheta_{\rm e},\ 0,\
  r_{\rm e}^{-1}f(r_{\rm e})^{-1/2}\sin\vartheta_{\rm e}),
  \label{null-tangent-vector-emission}
  \\
  \left.k^\mu\right|_{r=r_{\rm o}}
  &=&
  E_{\rm ph}(f(r_{\rm o})^{-1},\ \cos\vartheta_{\rm o},\ 0,\
  r_{\rm o}^{-1}f(r_{\rm o})^{-1/2}\sin\vartheta_{\rm o}),
  \label{null-tangent-vector-observation}
\end{eqnarray}
\end{subequations}
by eliminating $b$
using the definitions for $\vartheta_{\rm e}$ and $\vartheta_{\rm o}$
in Eq.~\eqref{definition-thetao-thetae}, respectively.
Then, the conservation of the angular momentum,
Eq.~\eqref{null-geodesic-angular-momentum}, implies 
\begin{equation}
  b=\frac{r_{\rm e}}{\sqrt{f(r_{\rm e})}}\sin\vartheta_{\rm e}
  =\frac{r_{\rm o}}{\sqrt{f(r_{\rm o})}}\sin\vartheta_{\rm o}.
  \label{b-thetae-thetao}
\end{equation}
From this relation, we can read off the following two implications.
First, for a given $r_{\rm o}$, Eq.~\eqref{b-thetae-thetao} implies that 
there is correspondence between $b$ and $\vartheta_{\rm o}$, 
which is one-to-one for $|\vartheta_{\rm o}|\le \pi/2$.
Therefore, one can specify the angle $\vartheta_{\rm o}$
of arriving photons using the value of $b$ instead.
Second, for a given $b$, the value of $\sin\vartheta_{\rm e}$
can be determined once the value of $r_{\rm e}$ is specified.
These facts will be used in our numerical calculation.

\subsection{Relation between $\vartheta_{\rm e}$ and $\vartheta_{\rm e}^\prime$}
\label{Sec:IIID}

$\vartheta_{\rm e}$ and $\vartheta_{\rm e}^\prime$ have been defined
as the angles between the spatial direction of photon's motion
and the outward radial direction measured 
in the static and comoving frames, respectively. 
In order to find the relation between these two angles,
let us introduce (the absolute value of)
the velocity $\beta$ of the star surface
measured in the static frame.
Since $(e_{0}^\prime)^{\mu}$ is equivalent to
the four-velocity of the star surface, we have 
\begin{equation}
  \beta:=\left|
  \frac{(e_{0}^\prime)^{\mu} (e_1)_{\mu}}{(e_{0}^\prime)^{\nu} (e_0)_{\nu}}
  \right|
  =-f^{-1}(r_{\rm e})\frac{\dot{r}_{\rm e}}{\dot{t}_{\rm e}}
  =\sqrt{1-\frac{f(r_{\rm e})}{f(R)}}. 
  \label{definition-beta}
\end{equation}
It can be checked that the comoving frame is related
to the static frame through the boost transformation,
\begin{subequations}
\begin{eqnarray}
  (e_{0}^\prime)^{\mu}&=&\gamma\left[{(e_{0})^\mu-\beta (e_{1})^\mu}\right],
  \label{Lorentz_0}\\
  (e_{1}^\prime)^{\mu}&=&\gamma\left[{-\beta (e_{0})^\mu+(e_{1})^\mu}\right],
  \label{Lorentz_1}\\
  (e_{2}^\prime)^{\mu}&=&(e_{2})^\mu,
  \label{Lorentz_2}\\
(e_{3}^\prime)^{\mu}&=&(e_{3})^\mu,
\label{Lorentz_3}
\end{eqnarray}
\end{subequations}
where $\gamma = 1/\sqrt{1-\beta^2}$. 
From Eqs.~\eqref{definition-thetaeprime} and \eqref{definition-thetao-thetae},
we have 
\begin{equation}
  \cos\vartheta_{\rm e}=\frac{\cos\vartheta_{\rm e}^\prime-\beta}{1-\beta\cos\vartheta_{\rm e}^\prime}.
  \label{relation_thetae_thetaeprime}
\end{equation}
One of the implications of this relation is as follows. 
Let us consider photons making the limb of the observed image.
Such photons are 
emitted in the directions tangential
to the star surface in the comoving frame, i.e.,
$\vartheta_{\rm e}^\prime=\pi/2$. 
Equation \eqref{relation_thetae_thetaeprime} leads to 
\begin{equation}
\cos\vartheta_{\rm e}=-\sqrt{1-\frac{f(r_{\rm e})}{f(R)}} \label{cos}
\end{equation}
for these photons, and thus 
the value of $\vartheta_{\rm e}$ is greater
than $\pi/2$. 
This fact implies that these photons are emitted in the inward direction
in the static frame.

%
%
%
\section{Observable quantities}
\label{Sec:IV}

In this section, we present a formalism to calculate 
observable quantities. The derivations of 
the redshift, the quantities related to photon counting,
and the quantities related to radiometry are 
shown here, one by one.

\subsection{Redshift}
\label{Sec:IVA}

In describing the strength of the redshift
of photons, the quantity 
$z:=\omega_{\rm e}^\prime/\omega_{\rm o}-1$ is commonly used,
where $\omega_{\rm o}$ and $\omega_{\rm e}^\prime$
are angular frequency of observed photons and of emitted photons,
respectively.
But in this paper, instead of $z$, we use  
\begin{equation}
  \alpha:=\frac{1}{1+z} 
  =\frac{\omega_{\rm o}}{\omega_{\rm e}^\prime}. 
  \label{redshift_definition}
\end{equation}
We call $\alpha$ the redshift factor. 
The angular frequency of a photon at the emission event is
\begin{equation}
  \omega_{\rm e}^\prime :=-\left.k^\mu (e_0^\prime)_{\mu}\right|_{r=r_{\rm e}}
  \label{omegaeprime-kmu}
\end{equation}
where, as mentioned,
$(e_{0}^\prime)^{\mu}$ is equivalent to the four-velocity of
the star surface.   
The angular frequency of an observed photon is
\begin{equation}
  \omega_{\rm o} :=-\left.k^\mu (e_0)_{\mu}\right|_{r=r_{\rm o}}
  \label{omegao-kmu}
\end{equation}
where $(e_{0})^{\mu}$ has been introduced in Eq.~\eqref{tetrad-static-0}
which is equivalent to the
four-velocity of the observer. 
For a fixed observation point $(t_{\rm o}, r_{\rm o})$,
all arriving photons with a fixed angle $\vartheta_{\rm o}$ have
the same value of the redshift factor $\alpha$.
Since there is a one-to-one relation between $b$ and $\vartheta_{\rm o}$
by virtue of Eq.~\eqref{b-thetae-thetao},  
$\alpha$ is regarded 
a function of the impact parameter $b$   
and observer's coordinate time $t_{\rm o}$.

In order to derive an expression for $\alpha$,
it is convenient to consider
the gravitational redshift from the star surface to the observer, 
\begin{equation}
  \frac{\left.k^\mu (e_0)_{\mu}\right|_{r=r_{\rm o}}}
       {\left. k^\nu (e_0)_{\nu}\right|_{r=r_{\rm e}}}
       = \sqrt{\frac{f(r_{\rm e})}{f(r_{\rm o})}},
  \label{gravitational_redshift}
\end{equation}
which is derived from Eqs.~\eqref{tetrad-static-0}, 
\eqref{null-tangent-vector-emission},
and \eqref{null-tangent-vector-observation},
and the Doppler effect coming from the relative motion of the star surface to the static frame,
\begin{align}
\left.\frac{k^\mu (e_0)_{\mu}}{k^\nu (e_0^{\prime})_{\nu}}\right|_{r=r_{\rm e}}
=\frac{\gamma^{-1}}{1+\beta\cos\vartheta_{\rm e}},
\label{Doppler_shift}
\end{align}
which is derived from Eqs.~\eqref{tetrad-static-0}, \eqref{tetrad-static-1},
\eqref{null-tangent-vector-emission}, 
and \eqref{Lorentz_0}.
Multiplication of Eqs.~\eqref{gravitational_redshift}
and \eqref{Doppler_shift} corresponds to
the redshift factor 
Eq.~\eqref{redshift_definition}, and hence, we have
\begin{align}
  \alpha 
  = \frac{f(r_{\rm e})}{\sqrt{f(R)f(r_{\rm o})}}
  \left(1+\cos\vartheta_{\rm e}\sqrt{1-\frac{f(r_{\rm e})}{f(R)}}\right)^{-1}.
  \label{redshift-in-terms-of-thetae}
\end{align}
For a given impact parameter
$b$, the value of $\sin\vartheta_{\rm e}$ is given in terms of $r_{\rm e}$
from Eq.~\eqref{b-thetae-thetao}. Then,  
$\cos\vartheta_{\rm e}$ has two possible values, 
$\cos\vartheta_{\rm e}=\pm \sqrt{1-\sin^2\vartheta_{\rm e}}$,
and plus (respectively, minus) sign corresponds to an outwardly
(respectively, inwardly) emitted photon.  
Therefore, the value of $\alpha$ can be determined
if we specify the radius $r_{\rm e}$ of the star surface
and the direction of propagation of a photon 
at the emission event.

By using Eqs.~\eqref{definition-beta} and \eqref{relation_thetae_thetaeprime},
Eq.~\eqref{redshift-in-terms-of-thetae} is rewritten in the form
\begin{equation}
  \alpha = \sqrt{\frac{f(R)}{f(r_{\rm o})}}
  \left(1-\cos\vartheta_{\rm e}^\prime\sqrt{1-\frac{f(r_{\rm e})}{f(R)}}\right).
  \label{redshift-in-terms-of-thetaeprime}
\end{equation}
In the case of $\vartheta_{\rm e}^\prime=\pi/2$, 
as mentioned, photons are emitted in the tangential direction
to the star surface in the comoving frame and make the limb
of the observed image.
The redshift factor of such photons is equal to 
\begin{equation}
\alpha_{\rm limb}:=\left. \alpha\right|_{\vartheta_{\rm e}^\prime=\pi/2}=\sqrt{\frac{f(R)}{f(r_{\rm o})}}. \label{alpha_limb}
\end{equation} 
Therefore, the redshift factor on the limb of the image   
is unchanged throughout the collapse.
This happens because such 
photons initially propagate in the inward direction
in the static frame,
and thus, the Doppler effect causes
the blueshift, which cancels out the increase in the gravitational redshift.

%
\begin{figure}[tb]
 \centering
 \includegraphics[width=0.6\textwidth,bb=0 0 308 222]{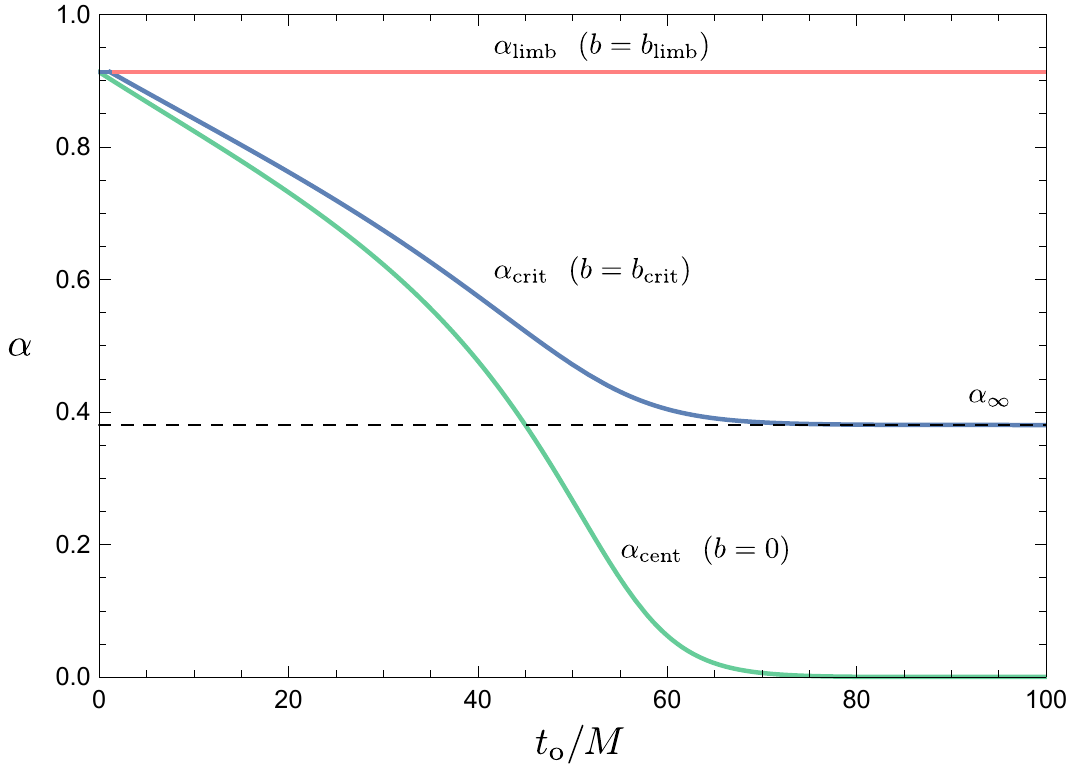}
 \caption{
   The values of the redshift factor $\alpha_{\rm cent}$, $\alpha_{\rm crit}$,
   and $\alpha_{\rm limb}$ for $b=0$, $b_{\rm crit}$ 
   and $b_{\rm limb}$, respectively, as functions of
   observer's time $t_{\rm o}$. The value of $\alpha_{\rm crit}$
   asymptotes to $\alpha_{\infty}$ given by Eq.~\eqref{alpha_infty}.
}
 \label{Fig:redshift_factor_b0_bcrit_blimb}
\end{figure}
%

As discussed in Secs.~\ref{Sec:IIIA} and \ref{Sec:IIIB}, 
worldlines of null geodesics for $b=0$ and $b=b_{\rm crit}$
are given in the analytic forms, Eq.~\eqref{radial-solution} and 
Eq.~\eqref{orbit_bcrit_observation_point}
with Eqs.~\eqref{orbit-critical-impact-parameter-2}--\eqref{bcrit_geodesic_outer_region}, respectively. 
Therefore, the redshift factor can be calculated analytically 
for photons with these values of $b$ as functions
of observer's time $t_{\rm o}$. The result is 
shown in Fig.~\ref{Fig:redshift_factor_b0_bcrit_blimb}.
Hereafter, 
the values of the redshift factor $\alpha$ for $b=0$ and $b=b_{\rm crit}$
are denoted by 
\begin{eqnarray}
  \alpha_{\rm cent}&:=&\left. \alpha\right|_{b=0},
  \label{alpha_cent}
  \\
  \alpha_{\rm crit}&:=&\left. \alpha\right|_{b=b_{\rm crit}},
  \label{alpha_crit}
\end{eqnarray}
respectively, where the subscript ``cent'' indicates 
the center of the star image, $\vartheta_{\rm o}=0$. 
The value of $\alpha_{\rm cent}$ decays to zero, and 
its late time behavior is approximately
proportional to $\exp(-t_{\rm o}/4M)$ (see the Appendix).
The redshift factor $\alpha_{\rm crit}$ for $b=b_{\rm crit}$
asymptotes to a constant value $\alpha_{\infty}$.
The value of $\alpha_{\infty}$ 
is obtained by setting $r_{\rm e}=3M$ and $\vartheta_{\rm e}=\pi/2$ 
in Eq.~\eqref{redshift-in-terms-of-thetae},
because in the limit of $t_{\rm o}\to \infty$, the analytic expression, 
Eq.~\eqref{orbit_bcrit_observation_point}
with Eqs.~\eqref{orbit-critical-impact-parameter-2}--\eqref{bcrit_geodesic_outer_region},  
for the worldline of a photon with $b=b_{\rm crit}$
suggests that the photon is emitted from the star surface
precisely on the photon sphere $r=3M$ in the tangential direction to it.
This leads to 
\begin{equation}
\alpha_{\infty}=\frac{f(3M)}{\sqrt{f(R)f(r_{\rm o})}}. \label{alpha_infty}
\end{equation}
By virtue of Eq.~\eqref{asymptotic_behavior_geodesic_bcrit}, 
the value of $\alpha_{\rm crit}$ asymptotes to
$\alpha_{\infty}$ in the manner that the value of    
$\alpha_{\rm crit}-\alpha_{\infty}$ is approximately proportional to 
$\exp(-t_{\rm o}/3\sqrt{3}M)$.
The value of $\alpha_{\rm limb}$, i.e.,
the redshift factor of photons on the limb 
whose impact parameter is denoted by $b_{\rm limb}$, 
is also shown in Fig.~\ref{Fig:redshift_factor_b0_bcrit_blimb}. 
Note that although $\alpha_{\rm limb}$ is a constant throughout the collapse, 
$b_{\rm limb}$ itself is time dependent, and its behavior must be
determined numerically.
In order to obtain the redshift factors of photons
with other impact parameters, we have to perform numerical calculations.

\subsection{Photon counting}
\label{Sec:IVB}

In photon counting, individual photons are counted
using some single-photon detector. If $dN$ photons
are counted in a time interval $d\tau_{\rm o}$,
photon flux is determined by
\begin{equation}
\mathcal{F}^{\rm (N)}:=\frac{dN}{d\tau_{\rm o}}.
\end{equation}
Below, we derive the formulas to calculate 
the photon intensity, the photon flux, and spectral photon intensity.
Before starting, 
we briefly discuss the angular range of $\vartheta_{\rm o}$
of the image of the star and 
the range of the impact parameter $b$
of arriving photons for a later convenience.

The image of the star has the shape of a disk.
The center and the limb of the image are composed of photons
with $b=0$ and $b=b_{\rm limb}$, respectively, 
whereas the other part of the image is composed of photons with
$0<b<b_{\rm limb}$. The angular diameter of the image is equal to 
$2\vartheta_{\rm o}^{\rm (limb)}$, where $\vartheta_{\rm o}^{\rm (limb)}$
is related to $b_{\rm limb}$ through Eq.~\eqref{b-thetae-thetao} as
\begin{equation}
b_{\rm limb}=\frac{r_{\rm o}}{\sqrt{f(r_{\rm o})}}\sin\vartheta_{\rm o}^{\rm (limb)}.
\end{equation}
Since $b_{\rm limb}$ depends on the observer's time as mentioned before,
we denote it as $b_{\rm limb}(t_{\rm o})$.   
The initial value of $b_{\rm limb}$, i.e. $b_{\rm limb}(0)$,  
is obtained by setting $\vartheta_{\rm e}=\pi/2$ and
$r_{\rm e}=R$ in Eq.~\eqref{b-thetae-thetao} as
\begin{equation}
  b_{\rm limb}(0)=\frac{R}{\sqrt{f(R)}}. \label{blim0}
\end{equation}
See Sec.~\ref{Sec:VIA} for more information
about the behavior of $b_{\rm limb}(t_{\rm o})$.

\subsubsection{Photon intensity}
\label{Sec:IVB1}

Photon intensity is a 
quantity that is useful in observing 
an object whose image has a finite size. 
It is defined as photon flux per unit
solid angle observed by a telescope,
${d\mathcal{F}^{\rm (N)}}/{d\Omega_{\rm o}}$.
Photon intensity is a directional quantity.

%
\begin{figure}[tb]
 \centering
   \includegraphics[width=0.35\textwidth,bb=0 0 321 155]{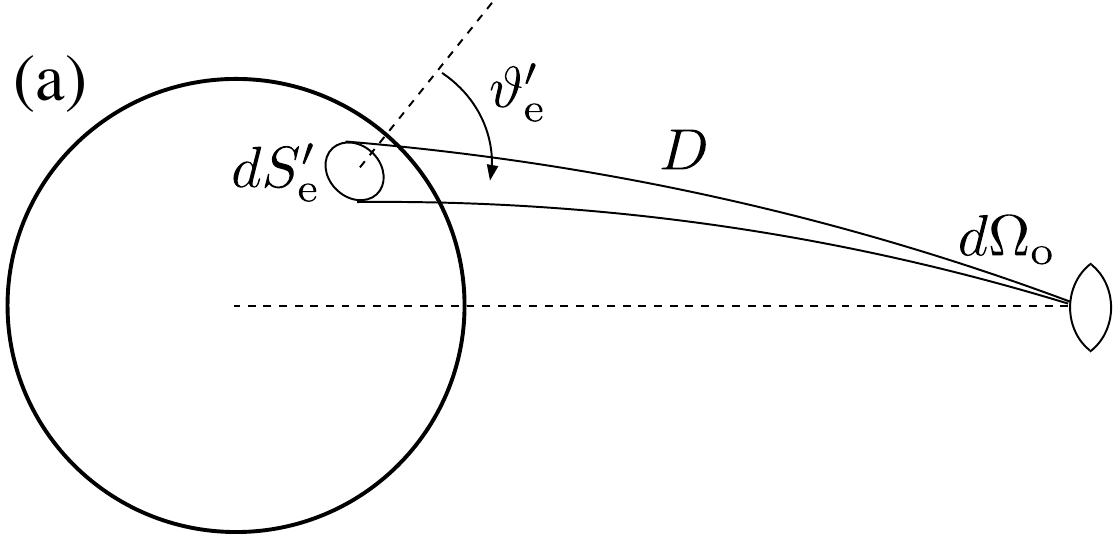}
   \hspace{1cm}
 \includegraphics[width=0.35\textwidth,bb=0 0 338 140]{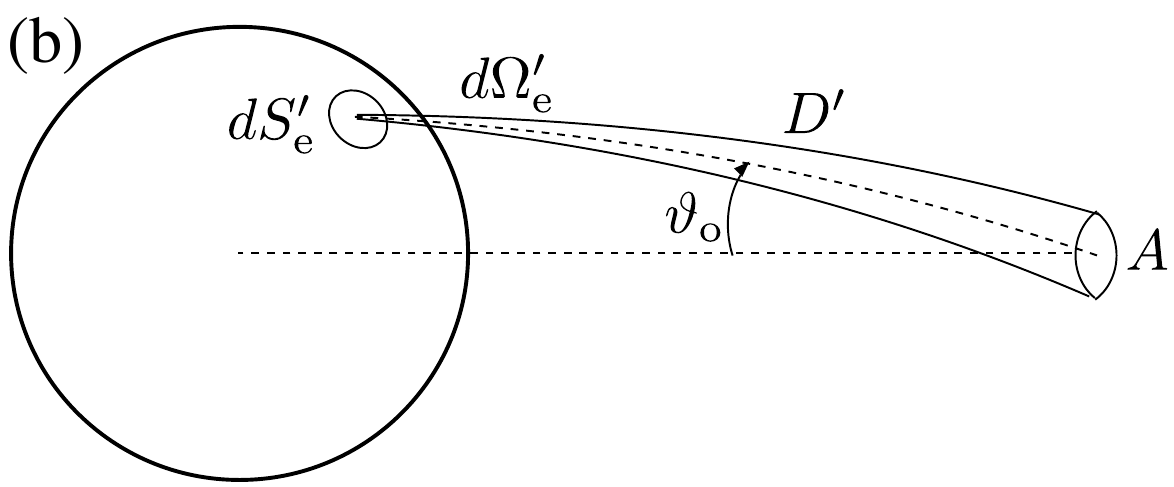}
 \caption{
   Definitions of (a) the angular diameter distance $D$
   and (b) the corrected luminosity distance $D^\prime$.
   See text for details.
}
 \label{Fig:ADD_CLD}
\end{figure}
%

Let us derive the formula to calculate this quantity.
As shown in the left picture of Fig.~\ref{Fig:ADD_CLD}, 
we consider a ray bundle that has a vertex with 
the infinitesimal solid angle $d\Omega_{\rm o}$ at the observation point. 
Evolving the ray bundle backward in time,
it intersects the star surface
at the area element $dS_{\rm e}^\prime$
at the angle $\vartheta_{\rm e}^\prime$ to
the normal to the area element (in the comoving frame).
These two quantities are related to each other through the
angular diameter distance $D$ as
\begin{equation}
  dS_{\rm e}^\prime\cos\vartheta_{\rm e}^\prime
  =D^2d\Omega_{\rm o}.
  \label{angular-diameter-distance}
\end{equation}
Alternatively, we can consider a ray bundle
with a vertex at the center of the area element $dS_{\rm e}^\prime$
on the star surface. We suppose that
at the vertex, the ray bundle have an infinitesimal solid angle $d\Omega_{\rm e}^\prime$
such that it exactly intersects the lens of the telescope with the area $A$
after evolving forward in time (the right-hand side of Fig.~\ref{Fig:ADD_CLD}).
The relation between these two quantities is expressed as
\begin{equation}
  A\cos\vartheta_{\rm o}
  ={D^\prime}^2d\Omega_{\rm e}^\prime.
  \label{corrected-luminosity-distance}
\end{equation}
Here, the quantity $D^\prime$ is called
the corrected luminosity distance \cite{Schneider:1992}.
These two relations \eqref{angular-diameter-distance}
and \eqref{corrected-luminosity-distance} 
with Eqs.~\eqref{Def:photon_flux_emission} and \eqref{Lambert2} give 
\begin{equation}
  d\left(\frac{dN}{d\tau_{\rm e}^\prime}\right) =
  J_{\rm e}^{\rm (N)}(r_{\rm e}) A\left(\frac{D}{D^\prime}\right)^2
  \cos\vartheta_{\rm o}d\Omega_{\rm o}.
\end{equation}
Here, we use the {\it reciprocity relation},
\begin{equation}
{D}/{D^\prime} = \omega_{\rm o}/\omega_{\rm e}^\prime=:\alpha,
\label{reciprocity-relation}
\end{equation}
which is a well-known relation in the context of gravitational lensing \cite{Etherington:1933} 
(see also \cite{Schneider:1992}).
Note that this relation can be applied to the cases that
both the star and the observer are moving relatively: The effects of
the relativistic beaming and the time delay caused by relativistic motion
and strong gravity are all included in this formula. 
The two time intervals $d\tau_{\rm e}^\prime$
and $d\tau_{\rm o}$,
during which the same $dN$ photons are emitted and observed,
are related as
$\omega_{\rm e}^\prime d\tau_{\rm e}^\prime=\omega_{\rm o}d\tau_{\rm o}$.
Then, we have the formula for the photon intensity,
\begin{equation}
  \frac{d\mathcal{F}^{\rm (N)}}{d\Omega_{\rm o}}
  =J_{\rm e}^{\rm (N)}(r_{\rm e})A\ \alpha^3\cos\vartheta_{\rm o}.
  \label{photon-intensity-unnormalized}
\end{equation}

We normalize this quantity by the initial value
of the photon flux, $\mathcal{F}^{\rm (N)}(0)$.
The initial photon flux can be calculated
by setting $r_{\rm e}=R$ and $\alpha = \alpha_{\rm limb}$
with Eq.~\eqref{alpha_limb}, 
and by integrating Eq.~\eqref{photon-intensity-unnormalized} 
with respect to the solid angle in the range
$0\le \vartheta_{\rm o}\le \vartheta_{\rm o}^{\rm (limb)}(0)$.
The result is
\begin{equation}
  \mathcal{F}^{\rm (N)}(0) =
  \pi J_{\rm e}^{\rm (N)}(R)A\sqrt{\frac{f(R)}{f(r_{\rm o})}}\left(\frac{R}{r_{\rm o}}\right)^2.
\end{equation}
Therefore, we have
\begin{equation}
  \frac{\left[d\mathcal{F}^{\rm (N)}/d\Omega_{\rm o}\right](t_{\rm o})}
       {\mathcal{F}^{\rm (N)}(0)}
  =\pi^{-1}
  \left(\frac{r_{\rm o}}{R}\right)^2
  \sqrt{\frac{f(r_{\rm o})}{f(R)}}
  \frac{J_{\rm e}^{\rm (N)}(r_{\rm e})}{J_{\rm e}^{\rm (N)}(R)}\ 
  \alpha^3
  \cos\vartheta_{\rm o}.
  \label{photon-intensity-normalized}
\end{equation}
Here, $r_{\rm e}$ is the radius of the star
at the moment when photons arriving at the telescope at
the observer's coordinate time $t_{\rm o}$ 
with the angle $\vartheta_{\rm o}$ were emitted.
Namely, similarly to the redshift factor $\alpha$,
the quantity $r_{\rm e}$ is regarded as a function
of $(\vartheta_{\rm o}, t_{\rm o})$.

\subsubsection{Photon flux}
\label{Sec:IVB2}

The photon flux can be obtained
by integrating Eq.~\eqref{photon-intensity-normalized}
with respect to the solid angle
$d\Omega_{\rm o}=2\pi\sin\vartheta_{\rm o}d\vartheta_{\rm o}$. 
By using the integration by substitution with
the relation between $b$ and $\vartheta_{\rm o}$
in Eq.~\eqref{b-thetae-thetao}, 
we have
\begin{equation}
  \frac{\mathcal{F}^{\rm (N)}(t_{\rm o})}
       {\mathcal{F}^{\rm (N)}(0)}
  =
  \frac{2}{R^2}
  \sqrt{\frac{f(r_{\rm o})^3}{f(R)}}
  \int_0^{b_{\rm limb}}
  \frac{J_{\rm e}^{\rm (N)}(r_{\rm e})}{J_{\rm e}^{\rm (N)}(R)}\ 
  \alpha^3 b\ db. 
  \label{photon-flux}
\end{equation}

\subsubsection{Spectral photon flux}
\label{Sec:IVB3}

Spectral photon flux is defined as
the photon flux per unit angular frequency interval, 
$d\mathcal{F}^{\rm (N)}/d\omega_{\rm o}$. 
In order to derive the formula for this quantity,
first we derive the spectral photon flux 
within an infinitesimal solid angle $d\Omega_{\rm o}$ and then,
integrate it with respect to the solid angle.

The derivation is similar to that of the photon intensity.
Using the relations~\eqref{angular-diameter-distance}
and \eqref{corrected-luminosity-distance}
with Eqs.~\eqref{Def:photon_flux_emission} and \eqref{Lambert1},
we have
\begin{equation}
  d^2\left(\frac{dN}{d\tau_{\rm e}^\prime}\right) =
  I_{\rm e}^{\rm (N)}(r_{\rm e},\omega_{\rm e}^\prime) A\left(\frac{D}{D^\prime}\right)^2
  \cos\vartheta_{\rm o}d\Omega_{\rm o}d\omega_{\rm e}^\prime.
\end{equation}
Because photons emitted with an angular frequency $\omega_{\rm e}^\prime$
arrive at the observation point as photons with
angular frequency $\omega_{\rm o}=\alpha\omega_{\rm e}^\prime$,
we have\footnote{Defining spectral photon radiance received
  by the telescope as
  $I_{\rm o}^{\rm (N)}:=d^2\mathcal{F}^{\rm (N)}/A\cos\vartheta_{\rm o}d\Omega_{\rm o}d\omega_{\rm o}$, we have the relation $I_{\rm o}^{\rm (N)}/\omega_{\rm o}^2=I_{\rm e}^{\rm (N)}/{\omega_{\rm e}^\prime}^2$ from Eq.~\eqref{d2Fn}. In terms
  of spectral radiance,
  $I_{\rm o}^{\rm (E)}:=\omega_{\rm o}I_{\rm o}^{\rm (N)}$
  and $I_{\rm e}^{\rm (E)}:=\omega_{\rm e}^\prime I_{\rm e}^{\rm (N)}$, we recover the
  conserved relation $I_{\rm o}^{\rm (E)}/\omega_{\rm o}^3=I_{\rm e}^{\rm (E)}/{\omega_{\rm e}^\prime}^3$ \cite{Podurets:1964,Ames:1968}.
} 
\begin{equation}
  d^2\mathcal{F}^{\rm (N)}
  =I_{\rm e}^{\rm (N)}(r_{\rm e},\omega_{\rm o}/\alpha)A\ \alpha^2\cos\vartheta_{\rm o}d\Omega_{\rm o}d\omega_{\rm o}.
  \label{d2Fn}
\end{equation}
Here, we used the reciprocity relation~\eqref{reciprocity-relation}
and the relation between the two time intervals,
$\omega_{\rm e}^\prime d\tau_{\rm e}^\prime=\omega_{\rm o}d\tau_{\rm o}$.
If we integrate this formula with respect to $\omega_{\rm o}$,
the formula for the photon intensity \eqref{photon-intensity-unnormalized}
is recovered. 
Integrating with respect to the solid angle
and normalizing with $\mathcal{F}^{\rm (N)}(0)$,
we have
\begin{equation}
  \frac{\left[d\mathcal{F}^{\rm (N)}/d\omega_{\rm o}\right](t_{\rm o})}{\mathcal{F}^{\rm (N)}(0)}
  =\frac{2}{R^2}
  \sqrt{\frac{f(r_{\rm o})^3}{f(R)}}
  \int_0^{b_{\rm limb}}
  \frac{I_{\rm e}^{\rm (N)}(r_{\rm e},\omega_{\rm o}/\alpha)}{J_{\rm e}^{\rm (N)}(R)}
  \alpha^2b\ db.
  \label{spectrum-photon-flux}
\end{equation}

\subsection{Radiometry}
\label{Sec:IVC}

Now, we derive formulas for quantities
related to radiometry. In radiometry, the energy of received 
electromagnetic radiation is measured. If radiant energy $dE_{\rm o}$
passes through the lens and is received by the telescope
in the interval $d\tau_{\rm o}$ of its proper time,
the radiant flux, or equivalently, the energy flux of photons  
is defined by
\begin{equation}
  \mathcal{F}^{\rm (E)}:=\frac{dE_{\rm o}}{d\tau_{\rm o}}.
\end{equation}
Similarly to the case of photon counting, radiant intensity
is defined as the radiant flux 
per unit solid angle, 
$d\mathcal{F}^{\rm (E)}/d\Omega_{\rm o}$, and 
spectral radiant flux is defined as
the radiant flux per unit angular frequency interval, 
$d\mathcal{F}^{\rm (E)}/d\omega_{\rm o}$. 

Since one photon has the energy
$\hbar\omega_{\rm o}$ at the observation event, 
the radiant flux for an infinitesimal solid angle $d\Omega_{\rm o}$ and
an infinitesimal angular frequency interval $d\omega_{\rm o}$
is given by
\begin{equation}
  d^2\mathcal{F}^{\rm (E)} = \omega_{\rm o}d^2\mathcal{F}^{\rm (N)}
  \label{d2Fe}
\end{equation}
(in the unit $\hbar=1$), where $d^2\mathcal{F}^{\rm (N)}$
is given in Eq.~\eqref{d2Fn}. 
Once Eq.~\eqref{d2Fe} is given, 
we obtain the radiant intensity
$d\mathcal{F}^{\rm (E)}/d\Omega_{\rm o}$ and 
the spectral radiant flux $d\mathcal{F}^{\rm (E)}/d\omega_{\rm o}$
by integrating Eq.~\eqref{d2Fe} with the angular frequency
and the solid angle, respectively.

We choose the initial radiant flux  
$\mathcal{F}^{\rm (E)}(0)$ as the normalization factor. 
Integrating $d^2\mathcal{F}^{\rm (E)}$ with both the solid angle
and the angular frequency for the initial value 
$r_{\rm e}=R$ and $\alpha=\alpha_{\rm limb}$ with Eq.~\eqref{alpha_limb}, we have
\begin{eqnarray}
  \mathcal{F}^{\rm (E)}(0)
  &=& \left.\langle \omega_{\rm e}^\prime\rangle\right|_{R}
  \sqrt{\frac{f(R)}{f(r_{\rm o})}}\ 
  \mathcal{F}^{\rm (N)}(0).
\end{eqnarray}
Here, we have introduced the mean value of the angular frequency
of emitted photons in the comoving frame,
\begin{equation}
  \left.\langle \omega_{\rm e}^\prime\rangle\right|_{r_{\rm e}}
  :=\frac{\int_0^\infty\omega_{\rm e}^\prime I_{\rm e}^{\rm (N)}(r_{\rm e},\omega_{\rm e}^\prime)d\omega_{\rm e}^\prime}{J_{\rm e}^{\rm (N)}(r_{\rm e})}.
\end{equation}
Then, we have
\begin{equation}
  \frac{d^2\mathcal{F}^{\rm (E)}}{\mathcal{F}^{\rm (E)}(0)}
  =
  \pi^{-1}\left(\frac{r_{\rm o}}{R}\right)^2
  \frac{f(r_{\rm o})}{f(R)}
  \alpha^2
  \cos\vartheta_{\rm o}d\Omega_{\rm o}
  \frac{I_{\rm e}^{\rm (N)}(r_{\rm e}, \alpha^{-1}\omega_{\rm o})}{J_{\rm e}^{\rm (N)}(R)}
  \frac{\omega_{\rm o}}{\left.\langle\omega_{\rm e}^\prime\rangle\right|_{R}}
  d\omega_{\rm o}.
  \label{d2fe-normalized}
\end{equation}

\subsubsection{Radiant intensity}
\label{Sec:IVC1}

Integrating Eq.~\eqref{d2fe-normalized} with respect to
the angular frequency $\omega_{\rm o}$, we obtain the formula
for the radiant intensity, 
\begin{equation}
  \frac{\left[d\mathcal{F}^{\rm (E)}/d\Omega_{\rm o}\right](t_{\rm o})}{\mathcal{F}^{\rm (E)}(0)}
  =
  \pi^{-1}\left(\frac{r_{\rm o}}{R}\right)^2
  \frac{f(r_{\rm o})}{f(R)}
  \frac{J_{\rm e}^{\rm (N)}(r_{\rm e})}{J_{\rm e}^{\rm (N)}(R)}
  \frac{\left.\langle\omega_{\rm e}^\prime\rangle\right|_{r_{\rm e}}}{\left.\langle\omega_{\rm e}^\prime\rangle\right|_{R}}
  \alpha^4
  \cos\vartheta_{\rm o}.
  \label{radiant-intensity}
\end{equation}
The radiant intensity is also called the surface brightness
in astronomy.

\subsubsection{Radiant flux}
\label{Sec:IVC2}

Integrating Eq.~\eqref{radiant-intensity} with respect
to the solid angle and rewriting $\vartheta_{\rm o}$ in the integral with $b$
using Eq.~\eqref{b-thetae-thetao}, 
we obtain the formula for the radiant flux, 
\begin{equation}
  \frac{\mathcal{F}^{\rm (E)}(t_{\rm o})}{\mathcal{F}^{\rm (E)}(0)}
  =
  \frac{2}{R^2}
  \frac{f(r_{\rm o})^2}{f(R)}
  \int_0^{b_{\rm limb}}
  \frac{J_{\rm e}^{\rm (N)}(r_{\rm e})}{J_{\rm e}^{\rm (N)}(R)}
  \frac{\left.\langle\omega_{\rm e}^\prime\rangle\right|_{r_{\rm e}}}{\left.\langle\omega_{\rm e}^\prime\rangle\right|_{R}}
  \alpha^4
  b\ db.
  \label{radiant-flux}
\end{equation}

\subsubsection{Spectral radiant flux}
\label{Sec:IVC3}

Integrating  Eq.~\eqref{d2fe-normalized} with respect to
the solid angle 
and rewriting $\vartheta_{\rm o}$ in the integral with $b$
using Eq.~\eqref{b-thetae-thetao}, we obtain the formula for
the spectral radiant flux, 
\begin{equation}
  \frac{\left[d\mathcal{F}^{\rm (E)}/d\omega_{\rm o}\right](t_{\rm o})}{\mathcal{F}^{\rm (E)}(0)}
  =\frac{2}{R^2}
  \frac{f(r_{\rm o})^2}{f(R)}
  \int_{0}^{b_{\rm limb}}
  \frac{I_{\rm e}^{\rm (N)}(r_{\rm e}, \alpha^{-1}\omega_{\rm o})}{J_{\rm e}^{\rm (N)}(R)}
  \frac{\omega_{\rm o}}{\left.\langle\omega_{\rm e}^\prime\rangle\right|_{R}}
  \alpha^2
  b\ db.
  \label{spectral-radiant-flux}
\end{equation}

%
%
\section{Numerical method}
\label{Sec:V}

As explained previously, in the ray tracing method,
worldlines of photons are numerically generated
from the observer to the past direction, and 
the intersection points of the worldlines of photons and the world sheet of
the surface of the star are determined. Then, the observable quantities
are calculated from the positions of the intersections.
We have made two codes to carry out this procedure,
HY's code and KT's code,
which are based on fairly different methods.
For both codes, we have checked that numerical 
data reproduce the analytic worldlines of photons
for $b=0$ and $b=b_{\rm crit}$ discussed 
in Secs.~\ref{Sec:IIIA} and \ref{Sec:IIIB}. 
Also, consistency between the two codes
has been checked in situations 
where the both codes give reliable results.

\subsection{HY's code}
\label{Sec:VA}

As stated before, all of the observable quantities are expressed in terms of
the redshift factor $\alpha$, 
and the redshift factor $\alpha$ is calculated
once the radius $r_{\rm e}$ and the direction of photon's propagation
at the emission event are given.
Using this fact, HY's code solves only the radial
positions of photons. Differentiating Eq.~\eqref{geodesic-drdt},
we can eliminate $b$ from that equation as 
\begin{equation}
  \frac{d^2r}{dt^2}
  +\frac{f^{-1}}{r}\left(1-\frac{5M}{r}\right)\left(\frac{dr}{dt}\right)^2
  =\frac{f}{r}\left(1-\frac{3M}{r}\right).
  \label{nullgeodesic-equation-r-and-t}
\end{equation}
This equation can be applied for both plus and minus signs of Eq.~\eqref{geodesic-drdt}.
In order to handle emission events in the region very close to
the horizon with sufficient precision
(that is, to calculate extremely small $\alpha$ at
the late stage of the collapse), we adopt the tortoise coordinate $r_*$
and solve for $r_*(t)$. The tortoise coordinate is 
defined by $dr/dr_*=f(r)$, or after integration, 
\begin{equation}
  r_*=r+2M\log\left(\frac{r}{M}-2\right).
  \label{Definition_tortoise}
\end{equation}
In this coordinate, the horizon is located at $r_*=-\infty$,
and thus, a sufficient resolution is obtained in the neighborhood of $r=2M$. 
Rewriting Eq.~\eqref{nullgeodesic-equation-r-and-t},
the equation for $r_*(t)$ is 
\begin{equation}
  \frac{d^2r_*}{dt^2}
  = 
  \frac{1}{r}\left(1-\frac{3M}{r}\right)
  \left[1-\left(\frac{dr_*}{dt}\right)^2\right].
  \label{nullgeodesic-equation-tortoise-and-t}
\end{equation}
It is easily seen that $r_*=t+\mathrm{const}$ is a solution to this
equation. This solution corresponds to the outgoing radial null geodesic
given in Eq.~\eqref{radial-solution}.

The mass of the star is set to be $M=1$
in numerical calculations.
In solving Eq.~\eqref{nullgeodesic-equation-tortoise-and-t}, 
the value of $r$ is required for a given $r_*$. 
Because the inverse relation of Eq.~\eqref{Definition_tortoise}
cannot be given analytically,
the relation between $r_*$ and $x:=r-2$ is calculated in advance
for discretized values of $r_*=i\times \Delta r_*$ with $\Delta r_*=0.01$
for integers $i$,  
by solving Eq.~\eqref{Definition_tortoise}
with the Newton-Raphson method. If $x$ is smaller than $10^{-15}$,
the approximate formula $x=\exp(r_*/2-1)$ is used. 
Here, $x$ is chosen rather than $r$ in order to avoid  
cancellation of significant digits in subtracting two nearly equal numbers.
Then, the value of $x$ are generated by 
the seventh-order Lagrange interpolation for a given $r_*$.

The equation for $r_*(t)$ is solved using the Runge-Kutta method
with the time step $\Delta t=0.01$ for a given impact parameter
$b$ from $t=0$ to negative time direction,
$t=-n\Delta t$ ($n=1$, $2$, ...).
The information of the impact parameter $b$ is included
in the ``initial condition'', 
$r=r_{\rm o}$ and $dr_*/dt=\cos\vartheta_{\rm o}$ at $t=0$, where
$\vartheta_{\rm o}$ is evaluated by Eq.~\eqref{b-thetae-thetao}.
In this way, we obtain a numerical solution 
$r_*=p(t)$ for discrete values of $t$.
Because of the time-translational symmetry of the
Schwarzschild spacetime, any null geodesic
with the same initial conditions except for the arrival
time $t_{\rm o}$ can be obtained just by shifting 
this solution to
the future direction, $r_*=p(t-t_{\rm o})$.

Once geodesics are generated numerically, 
the emission point is calculated for each of the
given parameters $t_{\rm o}$ and $b$. 
For this purpose, we generate the data of the world sheet
of the star surface in the form $r_*=s(t)$
for discrete moments $t=n^\prime\Delta t$ ($n^\prime=0$, $\pm 1$, $\pm2$, ...),
using Eqs.~\eqref{timelike-geodesic-r} 
and \eqref{timelike-geodesic-t} for $t\ge t_{\rm B}$
(before the collapse starts, i.e., for $t< t_{\rm B}$, 
the world sheet is chosen to be that of the static star).
Using these data, we calculate the distance in the tortoise coordinate between
the photon and the star surface, $\mathcal{D} :=p(t-t_{\rm o})-s(t)$,
for each moment of $t=t_{\rm o}-n\Delta t$ ($n=0,1,2,...$).
For an interval between two moments in which $\mathcal{D}$ changes its sign,
we make an interpolant function of $\mathcal{D}$, and 
the equation $\mathcal{D} =0$ is solved to obtain the emission time $t_{\rm e}$
and the corresponding tortoise coordinate $r_*=p(t_{\rm e})$.
Translating from the tortoise coordinate to the
circumferential radial coordinate,
we obtain the radius $r_{\rm e}$ of the star surface 
at the emission event.
Then, the redshift factor $\alpha$ are calculated using
the formulas \eqref{redshift-in-terms-of-thetae}
and \eqref{b-thetae-thetao}. Here, the sign of $\cos\vartheta_{\rm e}$
must be chosen appropriately as remarked after
Eq.~\eqref{redshift-in-terms-of-thetae}.

%
\begin{figure}[tb]
 \centering
 \includegraphics[width=0.60\textwidth,bb=0 0 300 280]{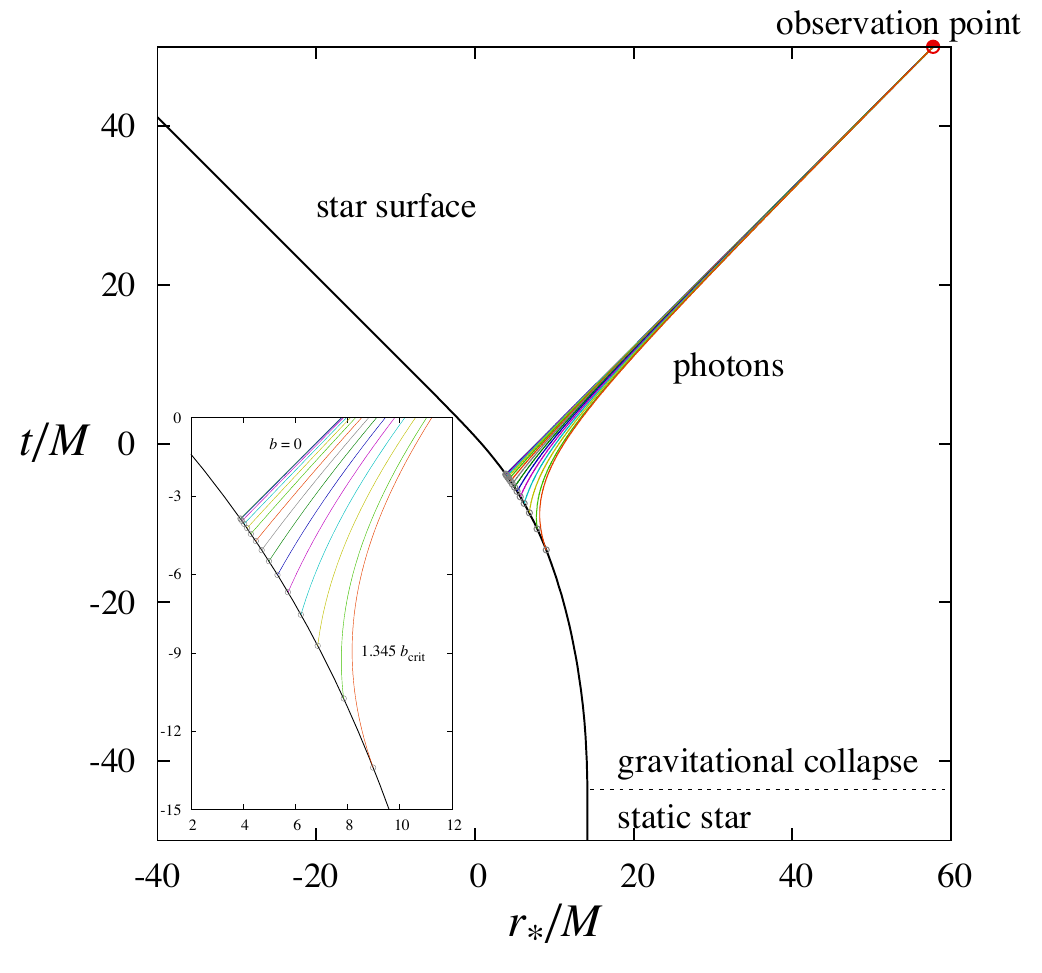}
 \caption{
   The world sheet of the star surface and worldlines of
   photons that arrive at the observer at
   $(r_{\rm o}, t_{\rm o})=(50M, 50M)$ in the $(r_*,t)$-plane.
   The impact parameters of photons are $b/b_{\rm crit}=0$, $0.1$, ..., $1.3$
   and $1.345$, where $b_{\rm crit}=3\sqrt{3}M$ is the critical impact parameter.
   The inset enlarges the emission region.
}
 \label{Fig:Photon_trajectory_example}
\end{figure}
%

Figure~\ref{Fig:Photon_trajectory_example}
shows the world sheet of the star surface 
and photons' worldlines with $b/b_{\rm crit}=0$, $0.1$,
..., $1.3$ and $1.345$ in the $(r_*,t)$-plane
that arrive at the point
$(r_{\rm o}, t_{\rm o})=(50M, 50M)$ of the observation event.
The intersections of geodesics and the world sheet of the
star surface are marked by circles ($\circ$). 
The photon with
$b=1.345\times b_{\rm crit}$ 
is emitted in an approximately parallel direction to the star
surface in the comoving frame during the collapse, and therefore, it initially
propagates in the inward direction.

In order to calculate the observable quantities, 
we need the dependence of the redshift factor $\alpha$
on the impact parameter $b$.
For this purpose, how to put grid points 
in the $b$ space is very important,
especially at the late stage of the gravitational collapse,
because the value of $d\alpha/db$ grows unboundedly large 
around $b=b_{\rm crit}$. 
In order to keep sufficient resolution,
we apply the nested grid method in the $b$ space.
Namely, we put nine computation domains 
in the $b$ space, each of which we call the $N$-th layer
with $N=1$, ..., $9$.
The first layer covers the whole range of $b$
and grid points are put as $b=j\times \Delta b^{(1)}$
$(0\le j\le 2151)$
with the grid size $\Delta b^{(1)}/b_{\rm crit}=10^{-3}$.
In each of the $N$-th layers with $N\ge 2$,
five hundred grid points are put as $b=b_{\rm crit}+j\times \Delta b^{(N)}$
$(-100\le j\le 400)$ with $\Delta b^{(N)}/b_{\rm crit} = 10^{-2-N}$
to give better resolution in the range 
$1-1\times 10^{-N}\le b/b_{\rm crit}\le 1+4\times 10^{-N}$. 
In order to proceed with the numerical calculation effectively,
not all of the layers are always used: 
If the $N^\prime$-th layer is the highest layer
in which $b_{\rm limb}$ is included, we use up to the $N^\prime$-th layer.
At an observer's time $t_{\rm o}=150M$, up to the ninth layer
is necessary. In addition to these layers,
the ``outer layer'' is put 
for $2.1400\le b/b_{\rm crit}\le 2.1516$ with the resolution
$\Delta b^{\rm (out)}/b_{\rm crit} = 10^{-4}$,
because a better resolution is required in this domain
to correctly describe the behavior of $\alpha(b)$
when the limb of the star image begins shrinking.

There is an upper limit of $t_{\rm o}$
until which our numerical results are reliable
because of the unstable nature
of the circular orbit of a photon. 
Most of the photons observed in the late time
have the impact parameter $b\approx b_{\rm crit}$ and 
arrive at the observer after orbiting  
approximately on the photon sphere $r=3M$.
The duration of this orbiting phase becomes longer
as $t_{\rm o}$ is increased.
Let us consider a geodesic with the impact parameter precisely 
equal to $b_{\rm crit}$. 
As shown in Eqs.~\eqref{orbit-critical-impact-parameter-1}--\eqref{bcrit_geodesic_outer_region},
such a photon asymptotically orbits around the photon sphere $r=3M$
in the past direction. 
However, a numerical solution of the same geodesic with $b=b_{\rm crit}$
finally leaves the photon sphere due to
the growth of numerical errors. 
This gives the upper bound on the duration of the
orbiting phase describable by numerical calculations. 
As a result, our numerical results are reliable
up to $t_{\rm o}\approx 150M$ 
(for the choice $R=10M$ and $r_{\rm o}=50M$).
Beyond this time, a different numerical techniques 
or an analytic approximate method for asymptotic behavior
must be developed 
(see Sec.~\ref{Sec:VII}).

\subsection{KT's code}
\label{Sec:VB}

In KT's code, 
the geodesic equations \eqref{null-geodesic-equations}
for $(t(\lambda), r(\lambda), \phi(\lambda))$
are solved on the equatorial plane backward in time 
using the fourth-order Runge-Kutta method. 
Here, we would like to note that, unlike HY's code,
not the tortoise coordinate $r_*$ but the circumferential radius $r$ is
used as the radial coordinate. 
The unit of the length is adopted to be $M$
also in this code (i.e., $M=1$), and 
the step of the affine parameter is chosen as $\delta\lambda=10^{-2}$.
In KT's code, geodesics are specified by the value of $\vartheta_{\rm o}$
(rather than $b$) and they are solved by putting uniform grid points 
in the $\vartheta_{\rm o}$ space as $\vartheta_{\rm o}=j\times (\delta\vartheta_{\rm o})$
with $\delta\vartheta_{\rm o}=6.0\times 10^{-5}$ 
for the integers $0\le j\le 5000$.
Among them, the geodesics with $0\le j\le 3681$ cross the world sheet
of the star surface, whereas the others do not.
The ``initial'' conditions are chosen as $x^\mu=(0,\ r_{\rm o},\ \pi/2,\ 0)$
and
$k^\mu=\left(f(r_{\rm o})^{-1/2},\ f(r_{\rm o})^{1/2}\cos\vartheta_{\rm o},\ 0,\ r_{\rm o}^{-1}\sin\vartheta_{\rm o}\right)$
to generate a null geodesic which arrives at the observer
at  $t_{\rm o}=0$ with the angle $\vartheta_{\rm o}$.
During the numerical calculation, 
the null condition and the conserved quantities (i.e.,
the energy and the angular momentum) were monitored to check the accuracy.

The point $(t_{\rm e}, r_{\rm e})$ of an emission event is determined 
in a different way from HY's code. Solving the null geodesic equations,
photon's worldline is obtained for discretized values
of the affine parameter as $(t(\lambda_n), r(\lambda_n))$,
where $n$ is a non-negative integer.
Expressing the world sheet of the star surface
\eqref{timelike-geodesic-r} and \eqref{timelike-geodesic-t} 
in the form $t=t_{\rm e}(r)$, the distance in the time coordinate 
between the photon and the star surface,
$\mathcal{D}^\prime:=t(\lambda_n)-t_{\rm e}(r(\lambda_n))$,
is calculated for each $n$. 
If $\mathcal{D}^\prime$ changes its sign at $(t(\lambda_n), r(\lambda_n))$, 
the geodesic equation is solved again forward in time
starting from $(t(\lambda_n), r(\lambda_n))$
but with a smaller step of the affine parameter
as $\Delta \lambda^{(1)}=\Delta\lambda/4$.
Then, within the four steps, $\mathcal{D}^\prime$ changes its sign
from negative to positive. Then, again, adopting
a further smaller step of the affine parameter
as $\Delta\lambda^{(2)}=\Delta\lambda^{(1)}/4$,
the geodesic equation is solved backward in time.
Iterating these steps, we can obtain fairly accurate value
of the intersection point $(t_{\rm e}, r_{\rm e})$. 
The redshift factor $\alpha$ is evaluated based on 
the definitions of the angular frequency, 
Eqs.~\eqref{redshift_definition}--\eqref{omegao-kmu}.

Because KT's code solves $\phi(\lambda)$ together,
it is possible to obtain the orbit in the equatorial 
plane and the null tangent vector at every point,
which are the information that HY's code has discarded. 
By contrast, due to the use of a uniform grid 
in the $\vartheta_{\rm o}$ space, the resolution becomes
insufficient for $t_{\rm o}\gtrsim 80M$.
Also, because of the use of the circumferential radius $r$,
the coordinate position of
the emission event cannot be
determined with sufficient accuracy in the neighborhood of the horizon 
due to cancellation of significant digits. 
For this reason, basically
the data taken by HY's code are shown in the next section.
For a figure generated by KT's code, 
we specify it in caption.

%
%
\section{Numerical Results} 
\label{Sec:VI}

In this section, 
we apply our formalism developed
in the previous sections to specific examples
and present the numerical results
for observable quantities. 
In Secs.~\ref{Sec:VIA}--\ref{Sec:VIC}, we discuss
the quantities independent of the spectral property
of the radiator, i.e., the redshift factor (Sec.~\ref{Sec:VIA}),
the photon intensity and 
the radiant intensity (Sec.~\ref{Sec:VIB}), and the photon flux and
the radiant flux
(Sec.~\ref{Sec:VIC}).
Then, in Sec.~\ref{Sec:VID}, we present the results for 
the spectral photon flux and the spectral radiant flux
for the cases of monochromatic and blackbody radiation, one by one.
Throughout this section, the initial radius of the star
is chosen as $R=10 M$ and the location of the observer
as $r_{\rm o}=50 M$.

\subsection{Redshift factor}
\label{Sec:VIA}

%
\begin{figure}[tb]
 \centering
 \includegraphics[width=0.45\textwidth,bb=0 0 360 262]{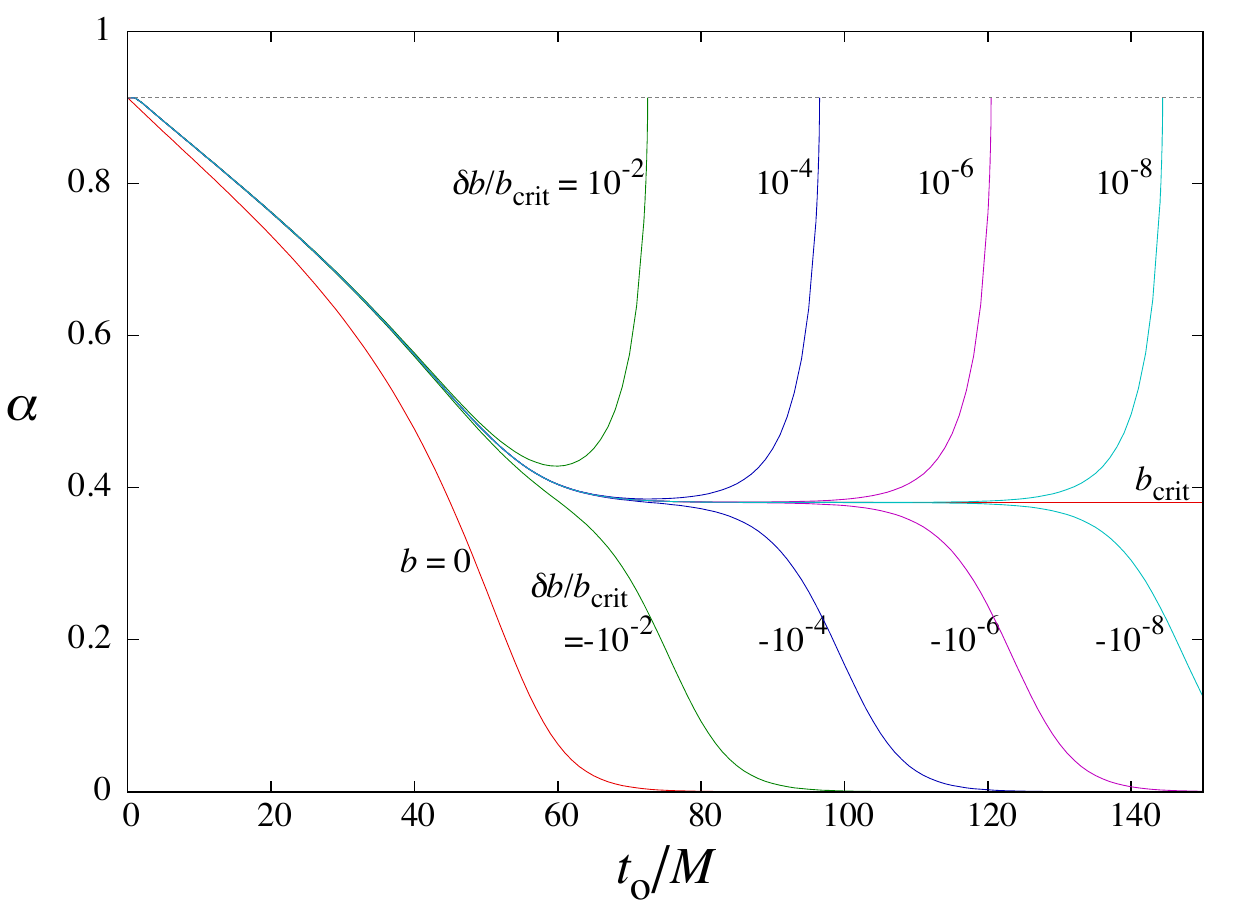}
 \includegraphics[width=0.45\textwidth,bb=0 0 365 262]{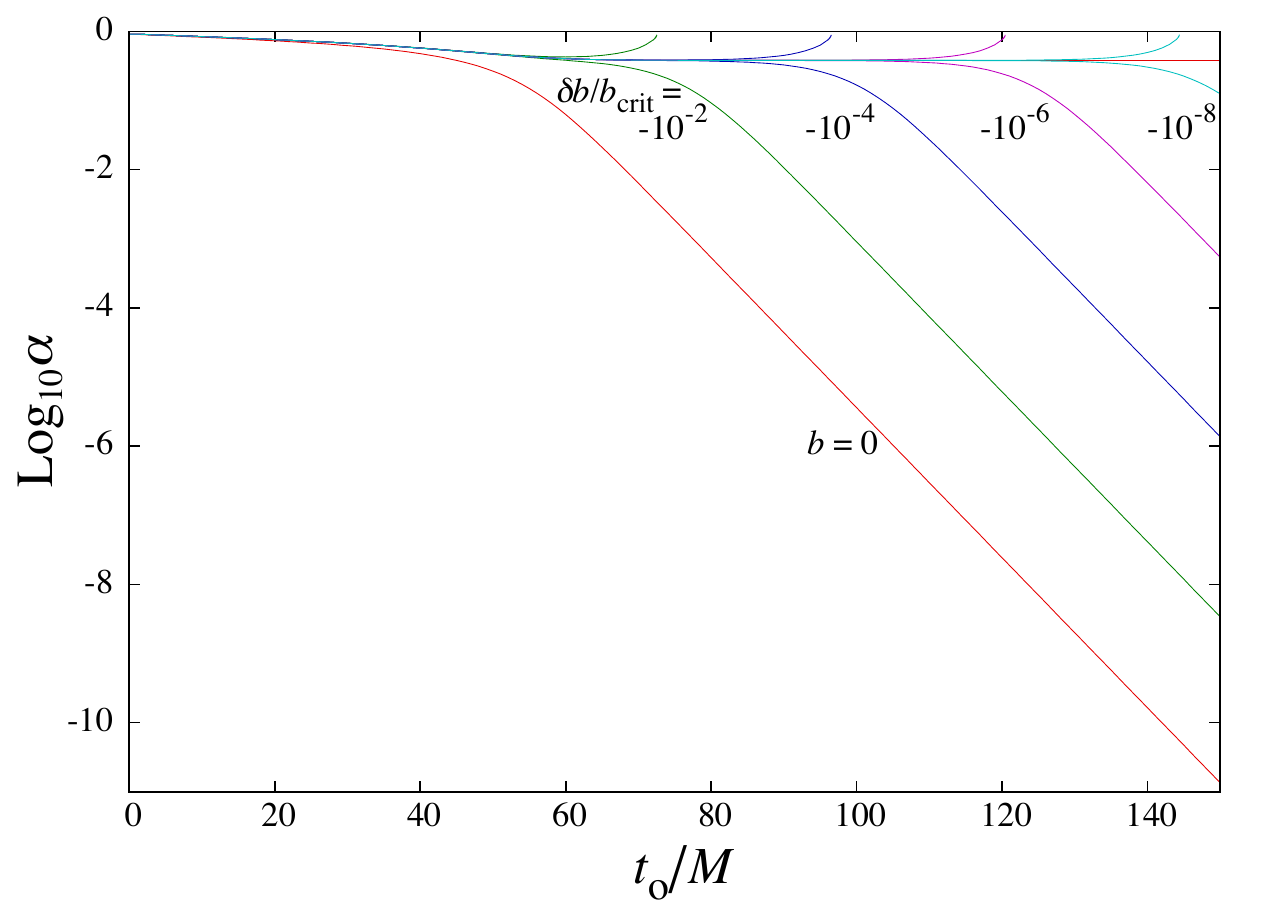}
 \caption{
   Left panel: The time evolution of $\alpha$ for fixed impact parameters $b$
   in the unit $M=1$.
   The cases $b=0$ and $b_{\rm crit}$,
   and $\delta b/b_{\rm crit}= \pm 10^{-2}$, $\pm 10^{-4}$,
   $\pm 10^{-6}$, $\pm 10^{-8}$ are shown, where $\delta b:=b-b_{\rm crit}$.
   Right panel: Same as left panel but the vertical axis is shown
   in the logarithmic scale.
}
 \label{Fig:alpha-t-numerical}
\end{figure}
%

The left panel of Fig.~\ref{Fig:alpha-t-numerical} shows the time evolution
of the redshift factor $\alpha$ for fixed values of the
  impact parameter $b$, 
or equivalently, for fixed directions $\vartheta_{\rm o}$. 
The cases $b=0$ and $b_{\rm crit}$ are already shown 
in Fig.~\ref{Fig:redshift_factor_b0_bcrit_blimb}, and the same behavior
is reproduced by numerical calculation. 
Additionally, we show the cases $\delta b/b_{\rm crit}= \pm 10^{-2}$,
$\pm 10^{-4}$,
$\pm 10^{-6}$, and $\pm 10^{-8}$, where 
\begin{equation}
\delta b:=b-b_{\rm crit}. \label{Delta-b}
\end{equation}
For $b>b_{\rm crit}$, the value of $\alpha$ once decreases,
but after some time it begins increasing
until it becomes $\alpha=\alpha_{\rm limb}$.
After this moment, $\alpha$ loses its value
because the null geodesic does not intersect the world sheet of the
star surface. On the other hand, for $b<b_{\rm crit}$, the value of
$\alpha$ continues to decrease to zero, because 
the intersection of the null geodesic and the world sheet
of the star surface, i.e., the emission event, 
becomes close to the horizon and the effect of the
gravitational redshift becomes unlimitedly strong. 
The right panel of Fig.~\ref{Fig:alpha-t-numerical}
shows the same figure but the scale of the vertical axis is labeled by
$\log_{10}\alpha$. Each curve decays
exponentially in time for large $t_{\rm o}$. 
The asymptotic behavior is 
analytically evaluated as 
$\propto \exp(-t_{\rm o}/4M)$ (see the Appendix),
and our numerical solutions correspond to this behavior
with the error less than $10^{-6}$.

%
\begin{figure}[tb]
 \centering
 \includegraphics[width=0.45\textwidth,bb=0 0 360 252]{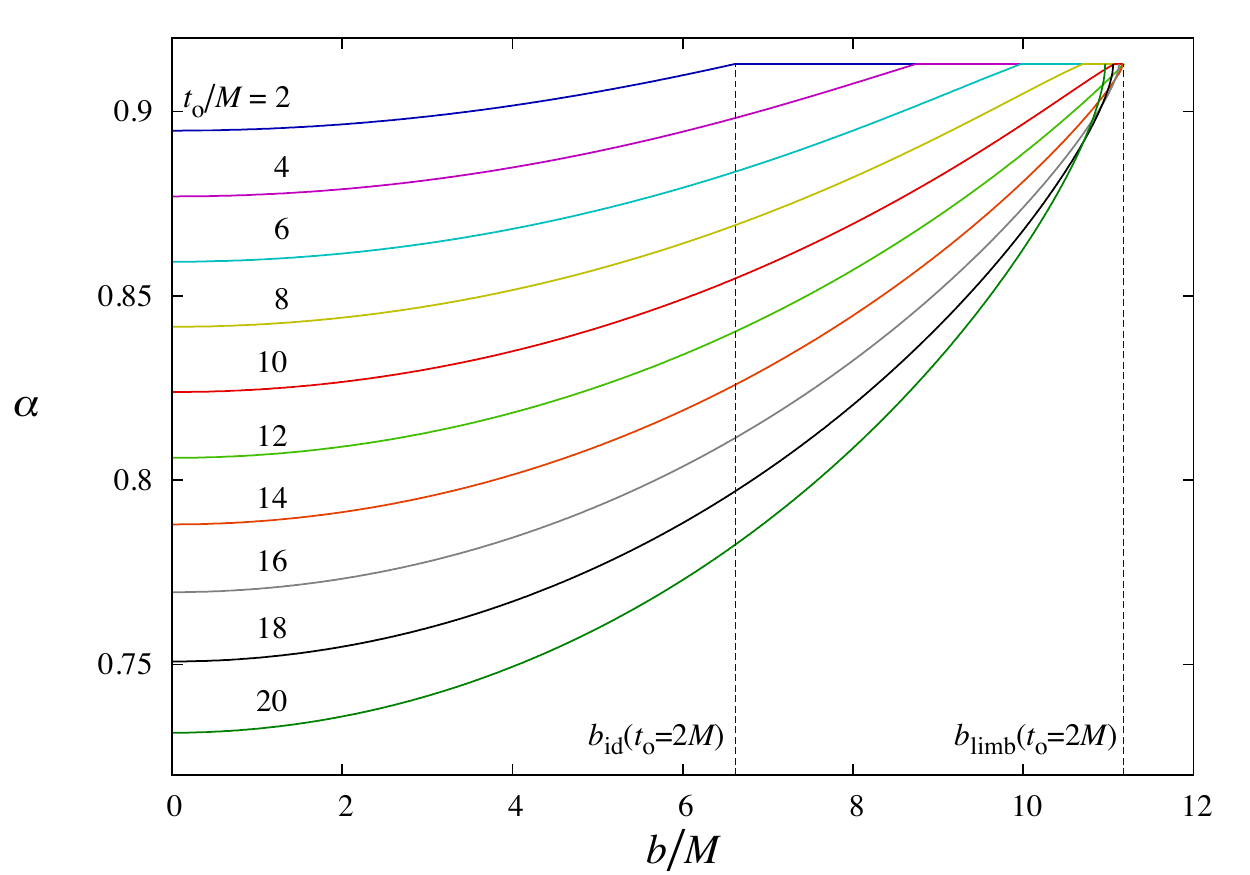}
 \includegraphics[width=0.45\textwidth,bb=0 0 360 252]{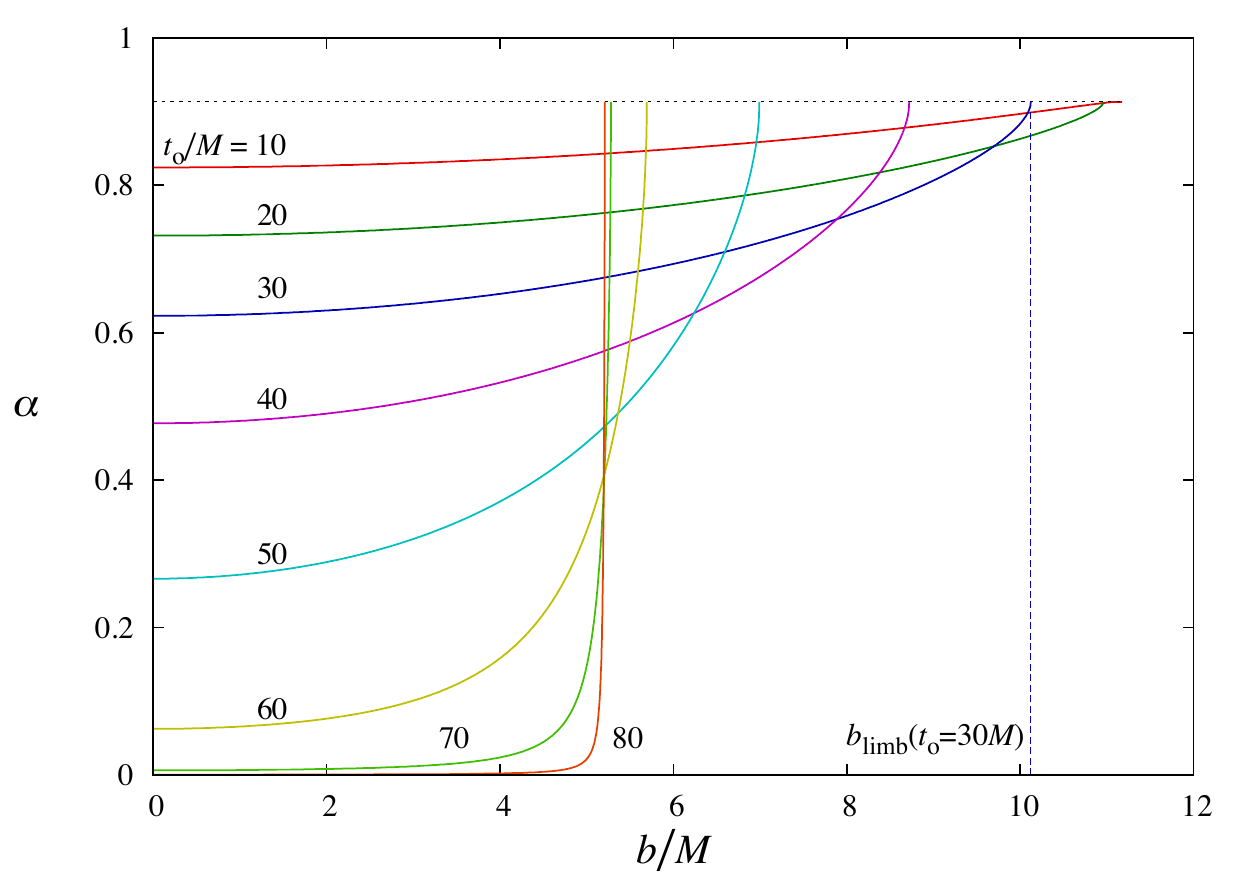}
 \includegraphics[width=0.45\textwidth,bb=0 0 360 252]{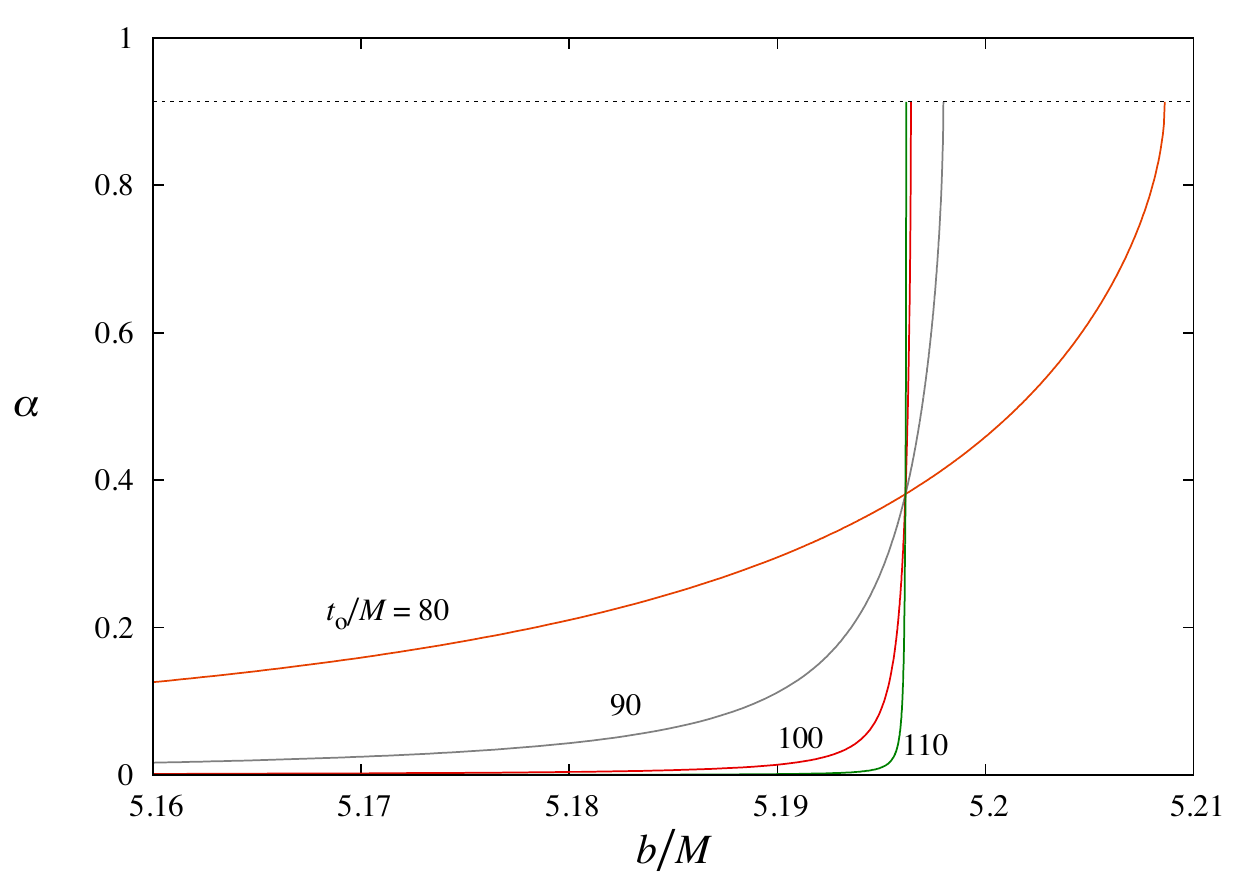}
 \includegraphics[width=0.45\textwidth,bb=0 0 360 252]{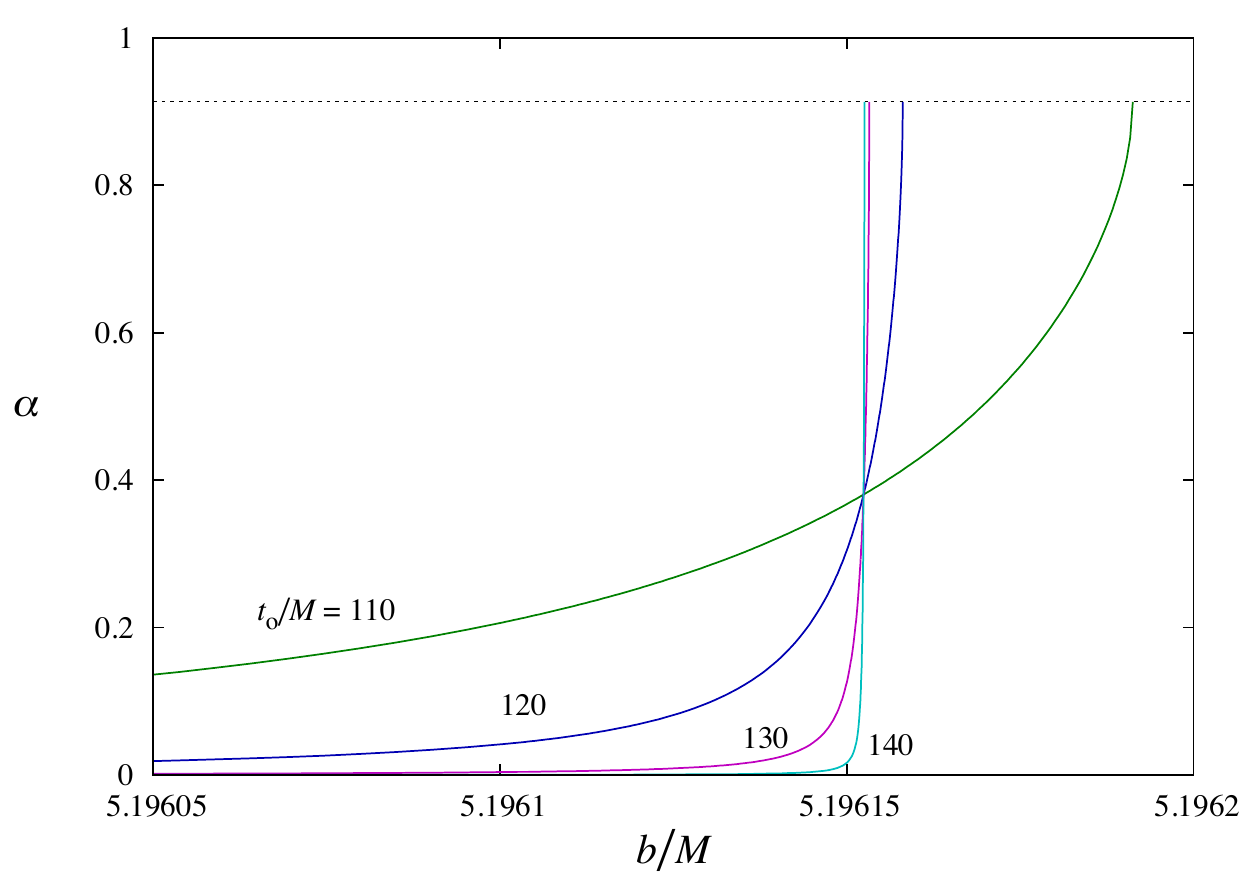}
 \caption{
   Snapshots for the redshift factor $\alpha$
   of observed photons as a function of the impact parameter
   $b$ (in the unit $M=1$).
   Top left panel: The snapshots for $t_{\rm o}/M=2$, $4$, ..., $20$.
   The location of the edge of the inner disk, $b_{\rm id}$, 
   and the location of the limb, $b_{\rm limb}$, are shown for $t_{\rm o}=2M$.
   Top right panel: The snapshots for $t_{\rm o}/M=10$, $20$, ..., $80$.
   The location of the limb, $b_{\rm limb}$, is shown for $t_{\rm o}/M=30$.
   Bottom left panel: The snapshots for $t_{\rm o}/M=80$, $90$, $100$, and $110$.
   Bottom right panel: The snapshots for $t_{\rm o}/M=110$, $120$, $130$, and $140$. 
   }
 \label{Fig:numerical_results_redshift_snapshots}
\end{figure}
%

Figure~\ref{Fig:numerical_results_redshift_snapshots}
shows snapshots of the redshift factor
$\alpha$
as functions of the impact parameter $b/M$.
The top left panel shows the cases $t_{\rm o}/M=2$, $4$, ..., $20$.
Initially, the redshift factors of all photons which make the image are identical to 
$\alpha_{\rm limb}$ defined as Eq.~\eqref{alpha_limb}. 
The edge of the image of the star
is made by photons with $b=b_{\rm limb}(0)$, 
where $b_{\rm limb}(0)$ is given by Eq.~\eqref{blim0}
($5\sqrt{5}M$ in our parameter choice).

After the collapse begins, the redshift factor begins decreasing in a portion of the image with $0\le b \le b_{\rm id}$, 
which corresponds to the ``inner disk'' $0\le \vartheta_{\rm o}\le \vartheta_{\rm o}^{\rm (id)}$ in the image, 
where $b_{\rm id}=0$ for $t_{\rm o}\leq0$ and $0<b_{\rm id}\leq b_{\rm limb}$ for $t_{\rm o}>0$.  
In this portion, the redshift factor $\alpha$ takes values 
$\alpha_{\rm cent}\le \alpha\le \alpha_{\rm limb}$, where
$\alpha_{\rm cent}$ is defined in Eq.~\eqref{alpha_cent}. 
Outside the inner disk, 
there is a region $b_{\rm id}\le b\le b_{\rm limb}(0)$
with a constant redshift factor $\alpha_{\rm limb}$. 
The inner disk grows bigger as time goes on, and 
at $t_{\rm o}=t_{\rm o}^{\rm (full)}$, it fulfills the whole image.

The value of $t_{\rm o}^{\rm (full)}$ is determined as follows.
Consider a null geodesic with the impact parameter $b=b_{\rm limb}(0)$
that passes through the point $(t_{\rm o}, r_{\rm o})$ of the observation event.  
In the period $0\le t_{\rm o}< t_{\rm o}^{\rm (full)}$,
this null geodesic is tangent to the world sheet of the  
static star surface at the point where they intersect, 
since the star is before the beginning of the collapse 
at the emission event of the corresponding photon, 
i.e., $(t_{\rm e}, R)$ with $t_{\rm e}<0$.  
By contrast, in the period $t_{\rm o}^{\rm (full)}<t_{\rm o}$,
the null geodesic with $b=b_{\rm limb}(0)$
and the world sheet of the star surface 
no longer intersect each other.
Therefore, $t_{\rm o}^{\rm (full)}$ is the observer's time
such that this null geodesic 
becomes tangent to the world sheet
of the star surface precisely
at the point $(t_{\rm e}, R)$ with $t_{\rm e}=0$, i.e.,
just when the star begins the gravitational collapse.
By studying this condition numerically, we have found  
$t_{\rm o}^{\rm (full)}/M\approx 11.8887$.

The top right and bottom two panels of
Fig.~\ref{Fig:numerical_results_redshift_snapshots}
shows the behavior of $\alpha(b)$ for $t_{\rm o}/M=10$, $20$, ..., $140$.
In the top right panel, the location of $b_{\rm limb}$ is also
shown for $t_{\rm o}/M=30$.
In the central region, the redshift factor decays rapidly
to zero.
Throughout the collapse, the redshift factor $\alpha_{\rm cent}$ at the
central point $b=0$ takes the minimum value.
The behavior of $\alpha(b)$ around the central point 
$b=0$ can be
perturbatively studied as 
presented in the Appendix.

The value of $b_{\rm limb}$ is time dependent
for $t>t_{\rm o}^{\rm (full)}$.
It decreases in time and asymptotes to $b_{\rm crit}$. 
The redshift factor $\alpha_{\rm limb}$ at the limb 
is unchanged throughout the collapse.
Note, however, that the derivative
$\partial\alpha/\partial b$ diverges there
for $t_{\rm o}>t_{\rm o}^{\rm (full)}$ 
for the following reason.
Let us express worldlines of photons with the impact parameters
$b$ that pass through 
the observation point $(t_{\rm o}, r_{\rm o})$ 
as $t=t^{(t_{\rm o},r_{\rm o})}(r,b)$.
In order to consider the intersection with the world sheet of the star surface,
we substitute $t=t_{\rm e}$ and $r=r_{\rm e}$ into this formula.
Since $t_{\rm e}$ is a 
function of $r_{\rm e}$ through Eqs.~\eqref{timelike-geodesic-r} and
\eqref{timelike-geodesic-t}, we have the equation for $r_{\rm e}$
as 
\begin{equation}
  t_{\rm e}(r_{\rm e}) = t^{(t_{\rm o}, r_{\rm o})}(r_{\rm e},b).
  \label{Equation-for-re}
\end{equation}
By solving this equation with respect to $r_{\rm e}$,
we have the radius $r_{\rm e}$  of the emission event as a function of $b$
for fixed values of $(t_{\rm o}, r_{\rm o})$,
i.e., $r_{\rm e}=r_{\rm e}^{(t_{\rm o}, r_{\rm o})}(b)$. 
Differentiating Eq.~\eqref{Equation-for-re} with respect to $b$, we find
\begin{equation}
  \frac{dr_{\rm e}}{db} =
  \frac{\displaystyle \left.\frac{\partial t^{(t_{\rm o},r_{\rm o})}}{\partial b}\right|_{r=r_{\rm e}}}
       {\displaystyle \frac{dt_{\rm e}}{dr_{\rm e}}-\left.\frac{\partial t^{(t_{\rm o},r_{\rm o})}}{\partial r}\right|_{r=r_{\rm e}}}.
       \label{Eq:dredb}
\end{equation} 
The worldline of a photon with $b_{\rm limb}$
becomes tangential to the world sheet of the star surface at the emission event
$r=r_{\rm e}$ in the $(r,\ t)$-plane
(see Fig.~\ref{Fig:Photon_trajectory_example}), and
this means that the denominator of Eq.~\eqref{Eq:dredb} becomes zero.
Therefore, $dr_{\rm e}/db$
becomes infinity at $b=b_{\rm limb}$.
Since the redshift factor $\alpha$ is given as a function of $r_{\rm e}$
as discussed before and it can be checked that
$d\alpha/dr_{\rm e}$ is nonvanishing, the value of
$\partial\alpha/\partial b$ also diverges.

The function $\alpha(b)$ becomes steeper near $b=b_{\rm crit}$
as $t_{\rm o}$ is increased. 
In the bottom two panels, all curves
approximately intersect at a point $(b,\alpha)=(b_{\rm crit},\alpha_{\infty})$,
where $\alpha_\infty$ is defined as Eq.~\eqref{alpha_infty}.
For late time, $\alpha(b)$ shrinks exponentially in time
in the direction of $b$ with respect to $b_{\rm crit}$.

%
\begin{figure}[tb]
 \centering
 \includegraphics[width=0.45\textwidth,bb=0 0 415 286]{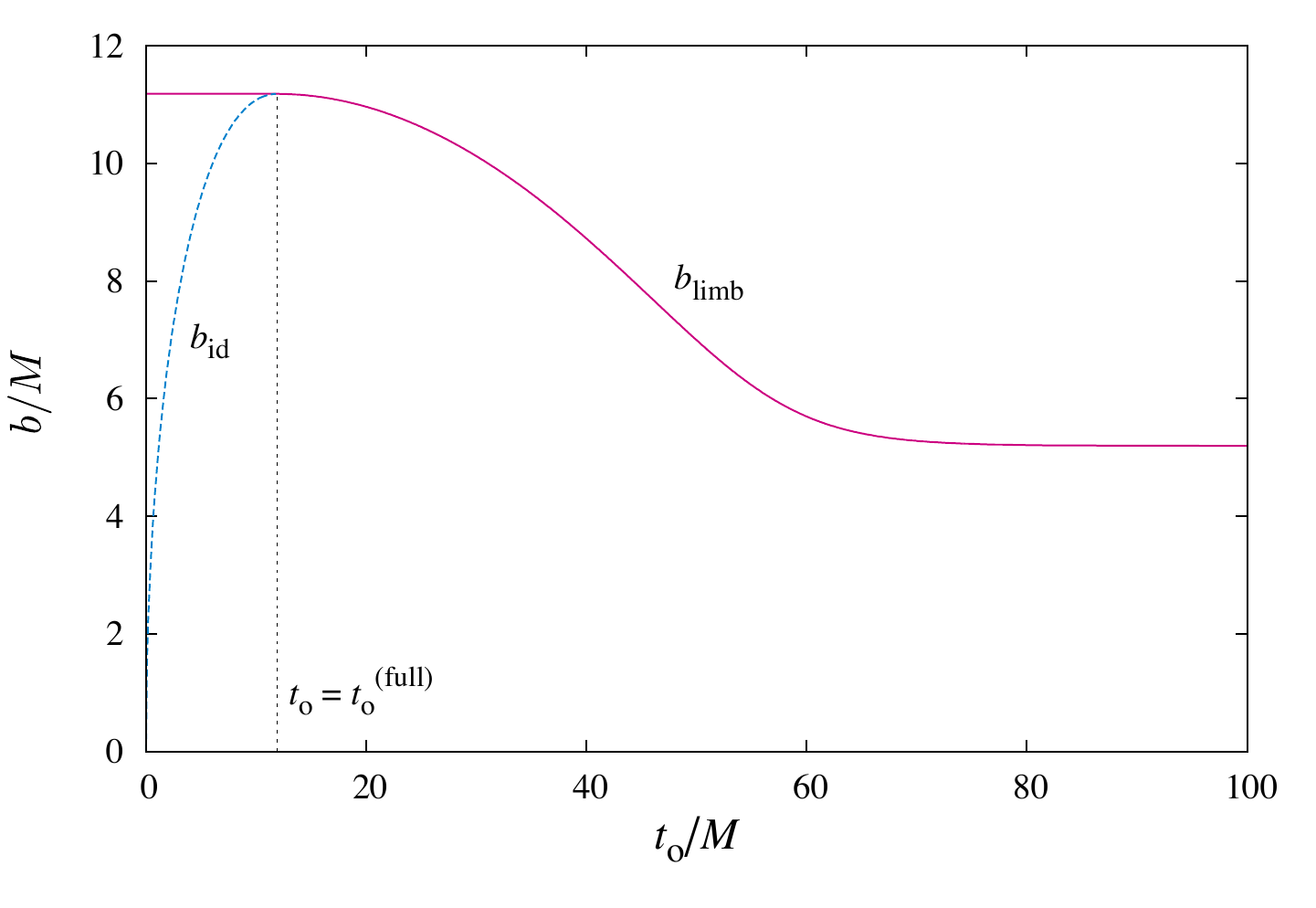}
 \includegraphics[width=0.45\textwidth,bb=0 0 415 286]{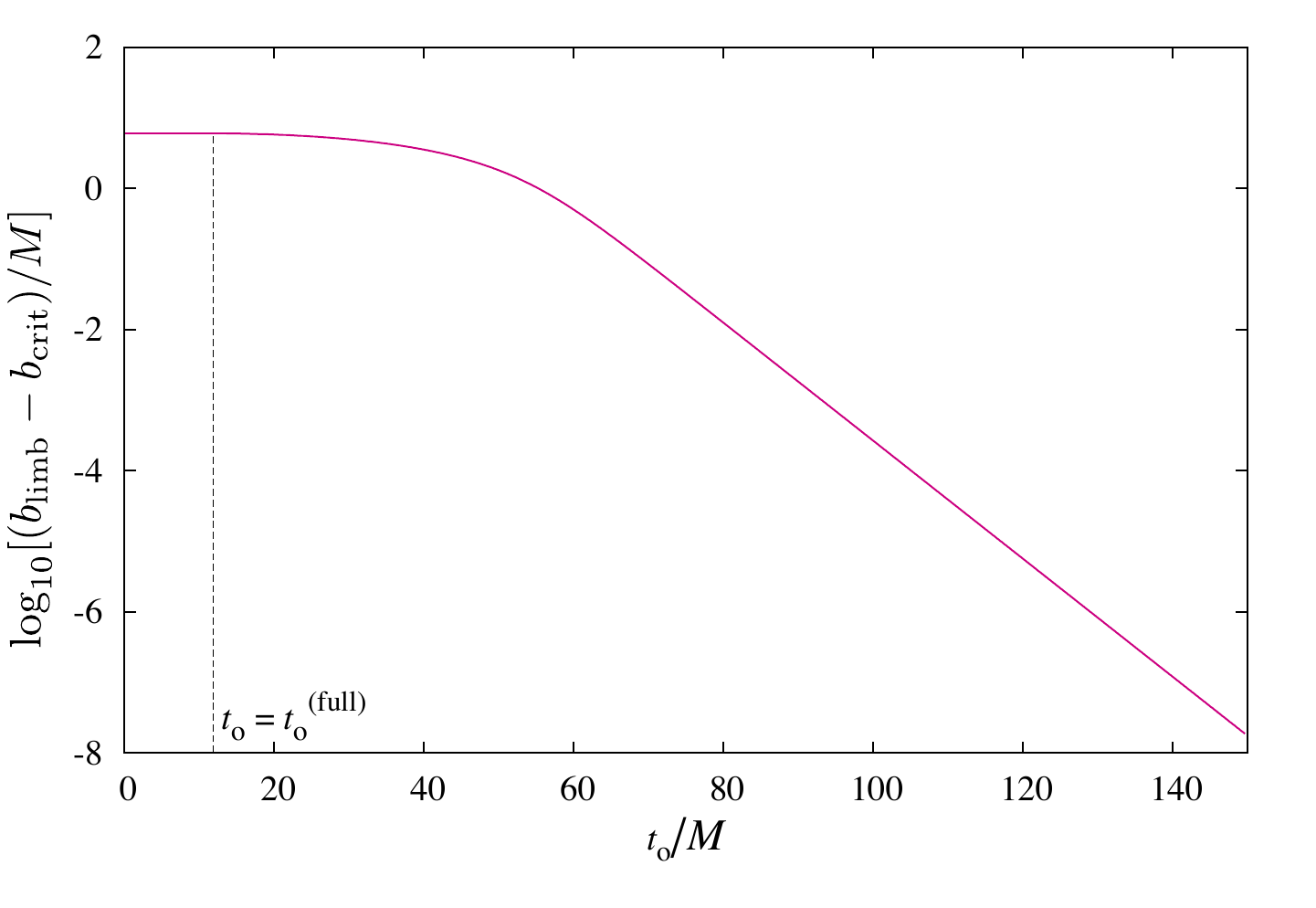}
 \caption{
   Left panel: The impact parameters at 
   the limb and at the edge of the inner disk, $b_{\rm limb}/M$ and
   $b_{\rm id}/M$ respectively, as functions of $t_{\rm o}/M$.
   Right panel: The difference of
   $b_{\rm limb}$ from the critical impact parameter $b_{\rm crit}=3\sqrt{3}M$
   (shown in the logarithmic scale,
   $\log_{10}[(b_{\rm limb}-b_{\rm crit})/M]$)
   as a function of $t_{\rm o}/M$.
}
 \label{Fig:blimb-bid-t}
\end{figure}
%

The left panel of Figure~\ref{Fig:blimb-bid-t} shows the
behavior of $b_{\rm id}$ and $b_{\rm limb}$ as functions
of the observer's time. For $0\le t_{\rm o}\le t_{\rm o}^{\rm (full)}$,
$b_{\rm limb}$ is unchanged.
$b_{\rm id}$ is an increasing function
and merges with the line of $b_{\rm limb}$ at $t_{\rm o}=t_{\rm o}^{\rm (full)}$.
For $t_{\rm o}>t_{\rm o}^{\rm (full)}$,
$b_{\rm limb}$ is dependent on an observer's time $t_{\rm o}$. 
It decreases and asymptotes to $b_{\rm crit}$.
Correspondingly, the limb of the star image 
asymptotes to
\begin{equation}
  \sin\vartheta_{\rm o}^{\rm (limb)}(\infty)=
  \frac{3\sqrt{3f(r_{\rm o})}}{r_{\rm o}}M,
\end{equation}
from Eq.~\eqref{b-thetae-thetao}.
The right panel shows the value of $\log_{10}[(b_{\rm limb}-b_{\rm crit})/M]$
as a function of the observer's time. 
The asymptotic behavior is numerically evaluated as
$b_{\rm limb}-b_{\rm crit}\approx 0.000266\times M\exp[-0.192449\times (t_{\rm o}/M-100)]$. An approximate late time analysis in Sec.~\ref{Sec:VII}
will show $b_{\rm limb}-b_{\rm crit}\propto \exp[-t_{\rm o}/3\sqrt{3}M]$,
and our numerical data agree well with this behavior.

%
\begin{figure}[tb]
 \centering
 \includegraphics[width=0.7\textwidth,bb=0 0 294 286]{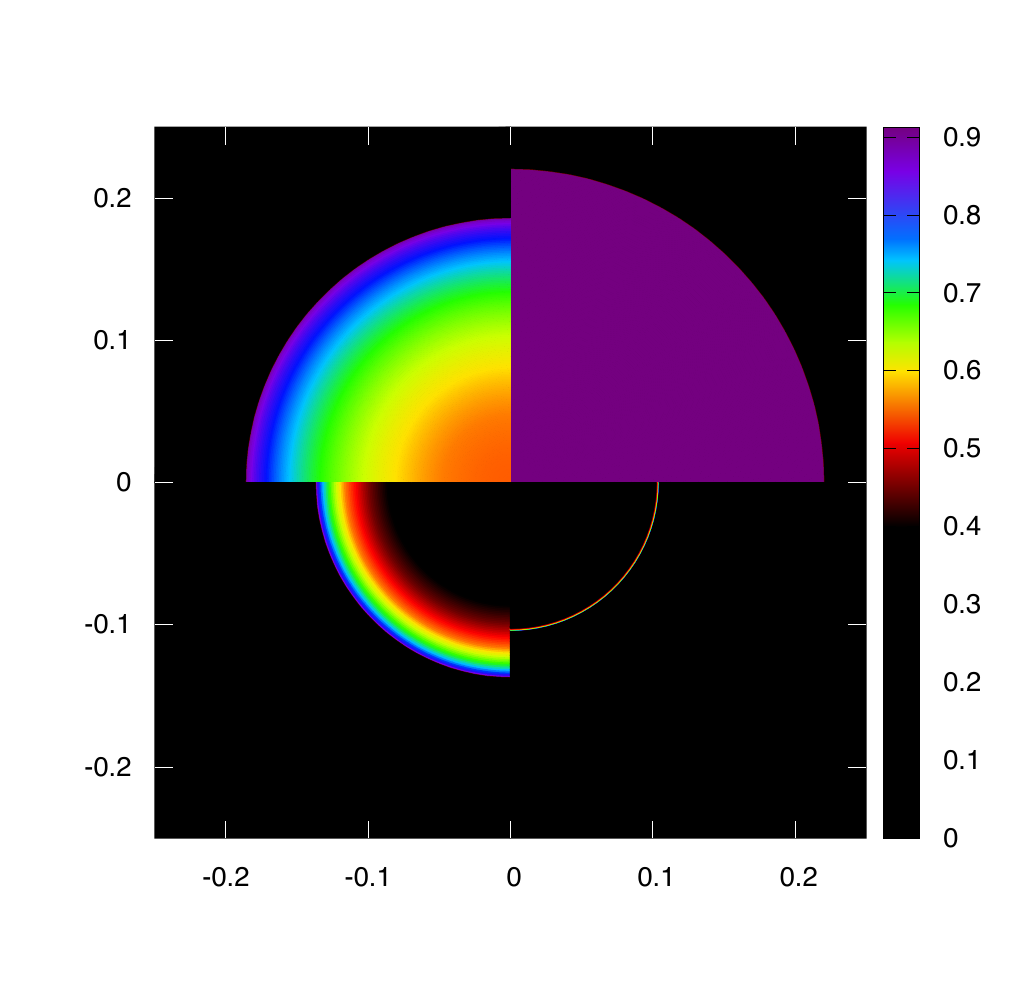}
 \caption{
   Images of the star (with redshift factor indicated by color online)
   for $t_{\rm o}=0$, $35$, $50$, and $69$
   (counterclockwise from top right). These images have been generated by
   KT's code.
}
 \label{Fig:redshift-factor-2D-color}
\end{figure}
%

Figure~\ref{Fig:redshift-factor-2D-color} presents  
a picture that mimics images of the star obtained through  
a telescope at four moments: The first, second, third
and forth quadrants are 
for $t_{\rm o}/M=0$, $30$, $50$, and $69$, respectively.  
In the online version of this paper, 
Fig.~\ref{Fig:redshift-factor-2D-color} is shown by color, and the color
is chosen to be consistent with the case 
where the star surface emits
near ultraviolet monochromatic radiation of the wavelength $\lambda\sim 353~\mathrm{nm}$.
Initially, photons experience only gravitational redshift,
and the whole star image is uniformly purple as shown in
the first quadrant.
After the collapse begin, a red domain appears
near the center as the second quadrant, 
and as time goes on, arriving photons near the center become infrared.
Because infrared photons are invisible to human eyes, 
such a domain is shown by black in the third and fourth quadrants.
After that, the wavelength of arriving photons near the central region
grows unlimitedly large with the observer's time.  
By contrast, in the neighborhood of the limb, photons remain
visible, showing rainbow colors around $t_{\rm o}=50M$
in the third quadrant.
As $t_{\rm o}$ is increased,
the star image becomes very thin as the fourth quadrant 
and arriving photons typically have the wavelength
around that of yellow color
(i.e. $\lambda\sim 600~\mathrm{nm}$).
The wavelength at the limb 
$\vartheta_{\rm o}=\vartheta_{\rm o}^{\rm (limb)}$
is unchanged from purple color throughout the collapse. 
Note that although this figure correctly represents
color, the surface brightness is not very accurate.
The study on the spectral radiant flux 
in Sec.~\ref{Sec:VID1} will show that  
the rainbow colored region is darker
compared to Fig.~\ref{Fig:redshift-factor-2D-color}
in reality.

\subsection{Photon and radiant intensity}
\label{Sec:VIB}

If the radiator's spectral photon intensity 
in the comoving frame does not depend on time as assumed
in Secs.~\ref{Sec:IIB1} and \ref{Sec:IIB2},
the photon intensity~\eqref{photon-intensity-normalized}
and the radiant intensity~\eqref{radiant-intensity} 
do not depend on the angular frequency spectrum of
the radiator because 
$J_{\rm e}^{\rm (N)}(r_{\rm e})=J_{\rm e}^{\rm (N)}(R)$ and $\left.\langle\omega_{\rm e}^\prime\rangle\right|_{r_{\rm e}}
=\left.\langle\omega_{\rm e}^\prime\rangle\right|_{R}$ 
hold throughout the collapse. 
Here, we show the numerical results of the observed intensities in this case.

%
\begin{figure}[tb]
 \centering
 \includegraphics[width=0.45\textwidth,bb=0 0 430 286]{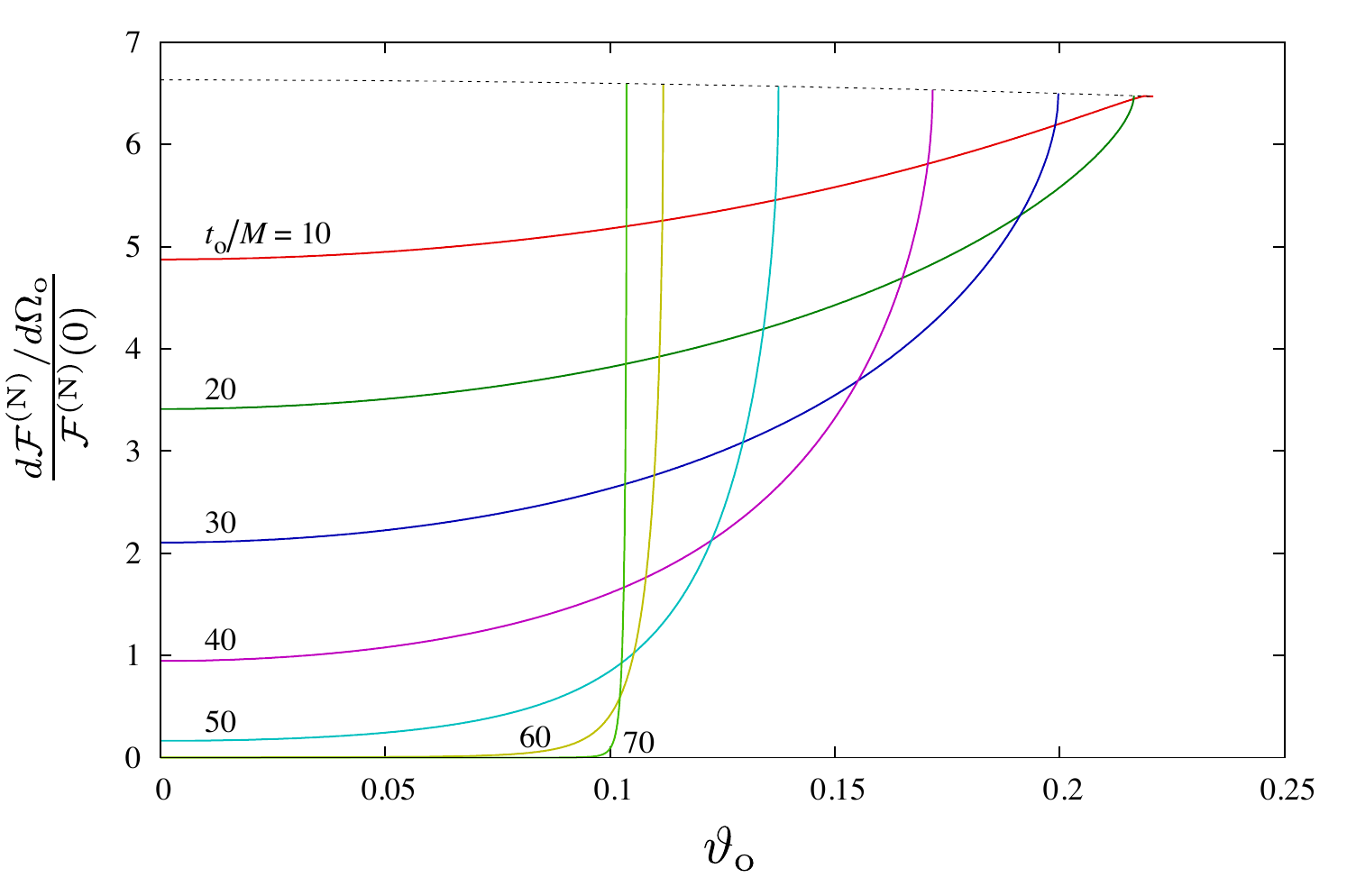}
 \includegraphics[width=0.45\textwidth,bb=0 0 430 286]{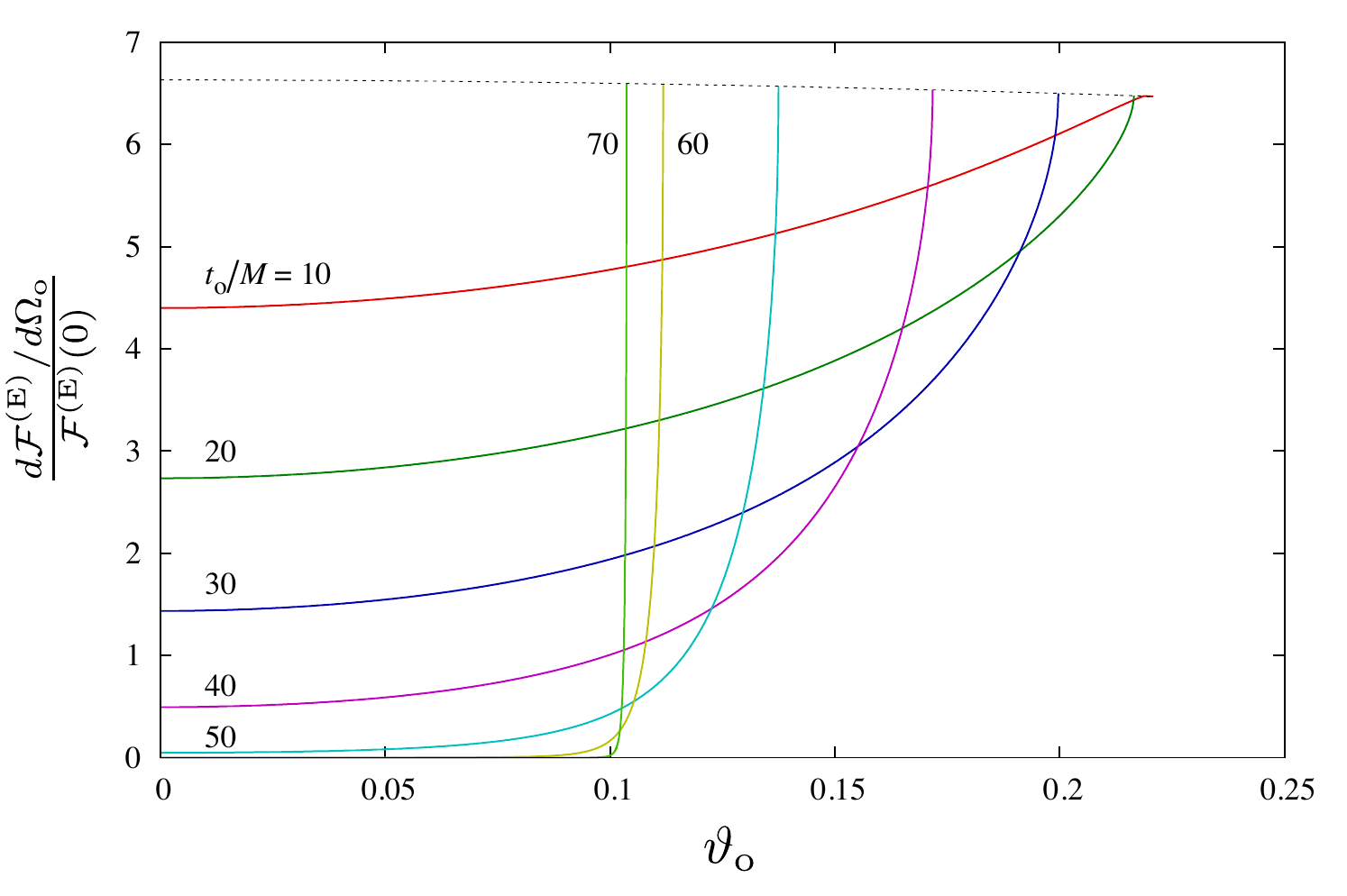}
 \caption{
   Snapshots of the photon intensity (left panel) and the radiant intensity
   (right panel) as functions of the observation direction $\vartheta_{\rm o}$
   for $t_{\rm o}/M=0$ (dotted curves), and $t_{\rm o}/M=10$, $20$, ..., $70$
   (solid curves).}
 \label{Fig:numerical_results_intensity}
\end{figure}
%

The left and right panels
of Fig.~\ref{Fig:numerical_results_intensity} show snapshots
of the photon intensity and the radiant intensity, respectively,
as functions of the angle $\vartheta_{\rm o}$. 
The cases of $t_{\rm o}/M=0$, $10$, ..., $70$
are depicted. From this figure, we see that the values 
of the two kinds of intensity at the limb slightly changes 
during the collapse due to the factor 
of $\cos\vartheta_{\rm o}$ in Eqs.~\eqref{photon-intensity-normalized}
and \eqref{radiant-intensity}.
In the central region, the photon intensity and the radiant intensity
rapidly decrease as $\sim \exp[-(3/4M)t_{\rm o}]$ and
$\sim \exp[-t_{\rm o}/M]$, because they are proportional to
$\alpha^3$ and $\alpha^4$, respectively.
To our eyes, only the neighborhood of the limb of the apparent disk
seems to be shining 
like a ring, and the ring becomes thinner and thinner
as time goes on.

\subsection{Photon and radiant fluxes}
\label{Sec:VIC}

%
\begin{figure}[tb]
 \centering
 \includegraphics[width=0.6\textwidth,bb=0 0 422 286]{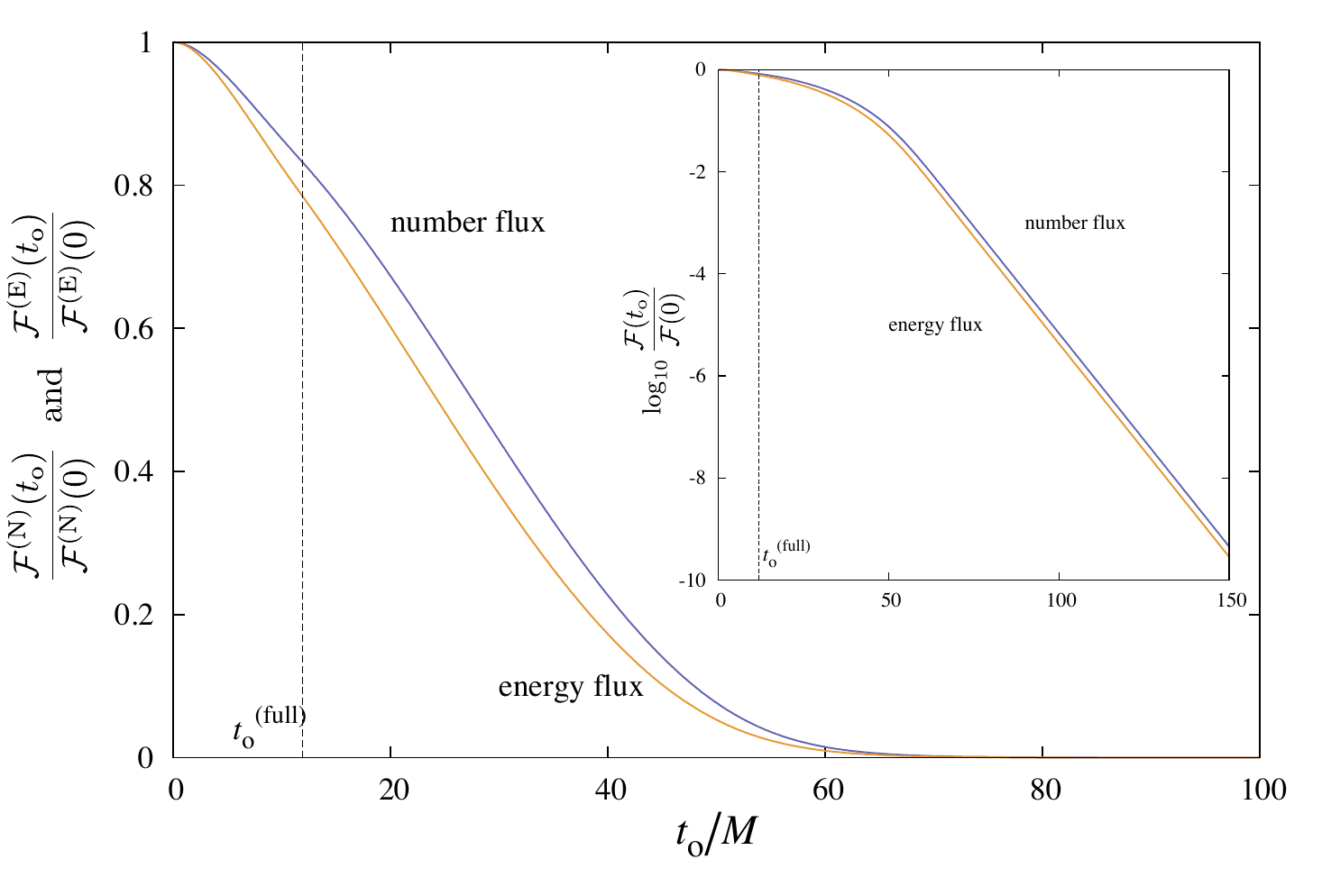}
 \caption{
   Normalized photon flux
   $\mathcal{F}^{\rm (N)}(t_{\rm o})/\mathcal{F}^{\rm (N)}(0)$ and normalized radiant flux
   $\mathcal{F}^{\rm (E)}(t_{\rm o})/\mathcal{F}^{\rm (E)}(0)$
   as functions of the observation time $t_{\rm o}/M$.
}
 \label{Fig:numerical_results_flux}
\end{figure}
%

Figure~\ref{Fig:numerical_results_flux} shows the behavior of
the photon flux~\eqref{photon-flux}
and the radiant flux~\eqref{radiant-flux} as functions of
the observer's time $t_{\rm o}/M$.
The photon flux and the radiant flux are monotonically decreasing functions.
The radiant flux decays faster than the photon
flux due to the 
difference in the power, $\alpha^4$ and $\alpha^3$,
in the integrand of Eqs.~\eqref{photon-flux} and
\eqref{radiant-flux}.
The inset shows the same functions but the vertical axis is
shown in the logarithmic scale.
The photon flux and the radiant flux decay exponentially for large $t_{\rm o}$ as
$\mathcal{F}(t_{\rm o})/\mathcal{F}(0)\approx \hat{\mathcal{F}}_{100}\exp[-\gamma (t_{\rm o}/M-100)]$.
Here, $\hat{\mathcal{F}}_{100}=6.858\times 10^{-6}$ and $\gamma=0.192451$ for
the photon flux, and
$\hat{\mathcal{F}}_{100}=4.308\times 10^{-6}$ and $\gamma=0.192454$ 
for the radiant flux. 
These are consistent with the asymptotic behavior
$\propto \exp[-t_{\rm o}/3\sqrt{3}M]$,
which has already been derived by Ames and Thorne~\cite{Ames:1968}.

\subsection{Spectral photon and radiant fluxes}
\label{Sec:VID}

Here, we show the numerical results of the spectral photon flux
and the spectral radiant flux. Since these quantities
depend on the spectral property of the radiator,
we discuss the cases of monochromatic radiation
and  blackbody radiation separately.

\subsubsection{The case of monochromatic radiation}
\label{Sec:VID1}

In the case of monochromatic radiation
whose spectral radiance is given by the delta-functional form, 
Eq.~\eqref{intensity-monochromatic-wave-emission},
the integration in the formulas for the 
spectral photon flux and the spectral radiant flux,
Eqs.~\eqref{spectrum-photon-flux} and \eqref{spectral-radiant-flux},
cannot be done numerically. 
For this reason, we rewrite these formulas to make them tractable with
numerical calculations. 
If we normalize the observed angular frequency
$\omega_{\rm o}$ with the angular frequency $\bar{\omega}_{\rm e}^\prime$ at the emission event,
it corresponds to the redshift factor as
$\alpha=\omega_{\rm o}/\bar{\omega}_{\rm e}^\prime$.
For this reason, we calculate
${d\mathcal{F}^{\rm (N)}/d\alpha}=\bar{\omega}_{\rm e}^\prime
\ {d\mathcal{F}^{\rm (N)}/d\omega_{\rm o}}$
and ${d\mathcal{F}^{\rm (E)}/d\alpha}=\bar{\omega}_{\rm e}^\prime
\ {d\mathcal{F}^{\rm (E)}/d\omega_{\rm o}}$
as functions of $\alpha$.

For $0\le t_{\rm o}\le t_{\rm o}^{\rm (full)}$, it is convenient
to divide the integration domain 
into two parts,
the first one for $0\le b\le b_{\rm id}$
and the second one for $b_{\rm id}\le b\le b_{\rm limb}$.
The second integral is easy because in this region, the argument
of the delta function $\delta(\omega_{\rm o}/\alpha-\bar{\omega}_{\rm e}^\prime)$
does not depend on $b$. 
In order to proceed with the first integral, we consider 
the inverse relation of $\alpha=\alpha(b, t_{\rm o})$
for a fixed $t_{\rm o}$, i.e., $b=b(\alpha, t_{\rm o})$ and rewrite 
the integral as the one with respect to $\alpha$.
Since $\delta(\omega_{\rm o}/\alpha-\bar{\omega}_{\rm e}^\prime)=(\alpha^2/\omega_{\rm o})\delta(\alpha-\omega_{\rm o}/\bar{\omega}_{\rm e}^\prime)$ holds,  
we obtain
\begin{subequations}
\begin{equation}
  \hat{\mathcal{F}}^{\rm (N)}_{\alpha}(t_{\rm o})
  :=
  \frac{d\mathcal{F}^{\rm (N)}/d\alpha}{\mathcal{F}^{\rm (N)}(0) }
  =\frac{2}{R^2}
  \sqrt{\frac{f(r_{\rm o})^3}{f(R)}}
  \frac{J_{\rm e}^{\rm (N)}(r_{\rm e})}{J_{\rm e}^{\rm (N)}(R)}
  \alpha^3b\frac{db}{d\alpha}
  +
  \left(1-\frac{b_{\rm id}^2}{b_{\rm limb}^2}\right)\delta(\alpha-\alpha_{\rm limb}),
  \label{spectral-photon-flux-monochromatic-case}
\end{equation}
\begin{equation}
  \hat{\mathcal{F}}^{\rm (E)}_{\alpha}(t_{\rm o})
  :=
  \frac{d\mathcal{F}^{\rm (E)}/d\alpha}{\mathcal{F}^{\rm (E)}(0) }
  =
  \frac{2}{R^2}
  {\frac{f(r_{\rm o})^2}{f(R)}}
  \frac{J_{\rm e}^{\rm (N)}(r_{\rm e})}{J_{\rm e}^{\rm (N)}(R)}
  \frac{\left.\langle\omega_{\rm e}^\prime\rangle\right|_{r_{\rm e}}}{\left.\langle\omega_{\rm e}^\prime\rangle\right|_{R}}
  \alpha^4b\frac{db}{d\alpha}
  +
  \left(1-\frac{b_{\rm id}^2}{b^2_{\rm limb}}\right)\delta(\alpha-\alpha_{\rm limb}),
  \label{spectral-radiant-flux-monochromatic-case}
\end{equation}
\end{subequations}
after integration, 
with the constraint $\alpha=\omega_{\rm o}/\bar{\omega}_{\rm e}^\prime$.
Remember that $\alpha_{\rm limb}$ 
in the second term is the initial redshift factor
defined in Eq.~\eqref{alpha_limb}. 
The formula for $t_{\rm o}\ge t_{\rm o}^{\rm (full)}$
is obtained by omitting the second term.
Note that by virtue of Eq.~\eqref{constancy-of-intensity}, 
the formulas~\eqref{spectral-photon-flux-monochromatic-case} and
\eqref{spectral-radiant-flux-monochromatic-case}
are further simplified because
$J_{\rm e}^{\rm (N)}(r_{\rm e})=J_{\rm e}^{\rm (N)}(R)$ and 
$\left.\langle\omega_{\rm e}^\prime\rangle\right|_{r_{\rm e}}
=\left.\langle\omega_{\rm e}^\prime\rangle\right|_{R}$
hold.
We also remark that the first terms of
Eqs.~\eqref{spectral-photon-flux-monochromatic-case}
and \eqref{spectral-radiant-flux-monochromatic-case}
can be derived from the formulas for
the photon intensity \eqref{photon-intensity-normalized}
and the radiant intensity \eqref{radiant-intensity}
just by rewriting $\vartheta_{\rm o}$ with $\alpha$,
because there is one to one
correspondence between $\vartheta_{\rm o}$ and $\alpha$
for a fixed $t_{\rm o}$. 

%
\begin{figure}[tb]
 \centering
 \includegraphics[width=0.45\textwidth,bb=0 0 429 295]{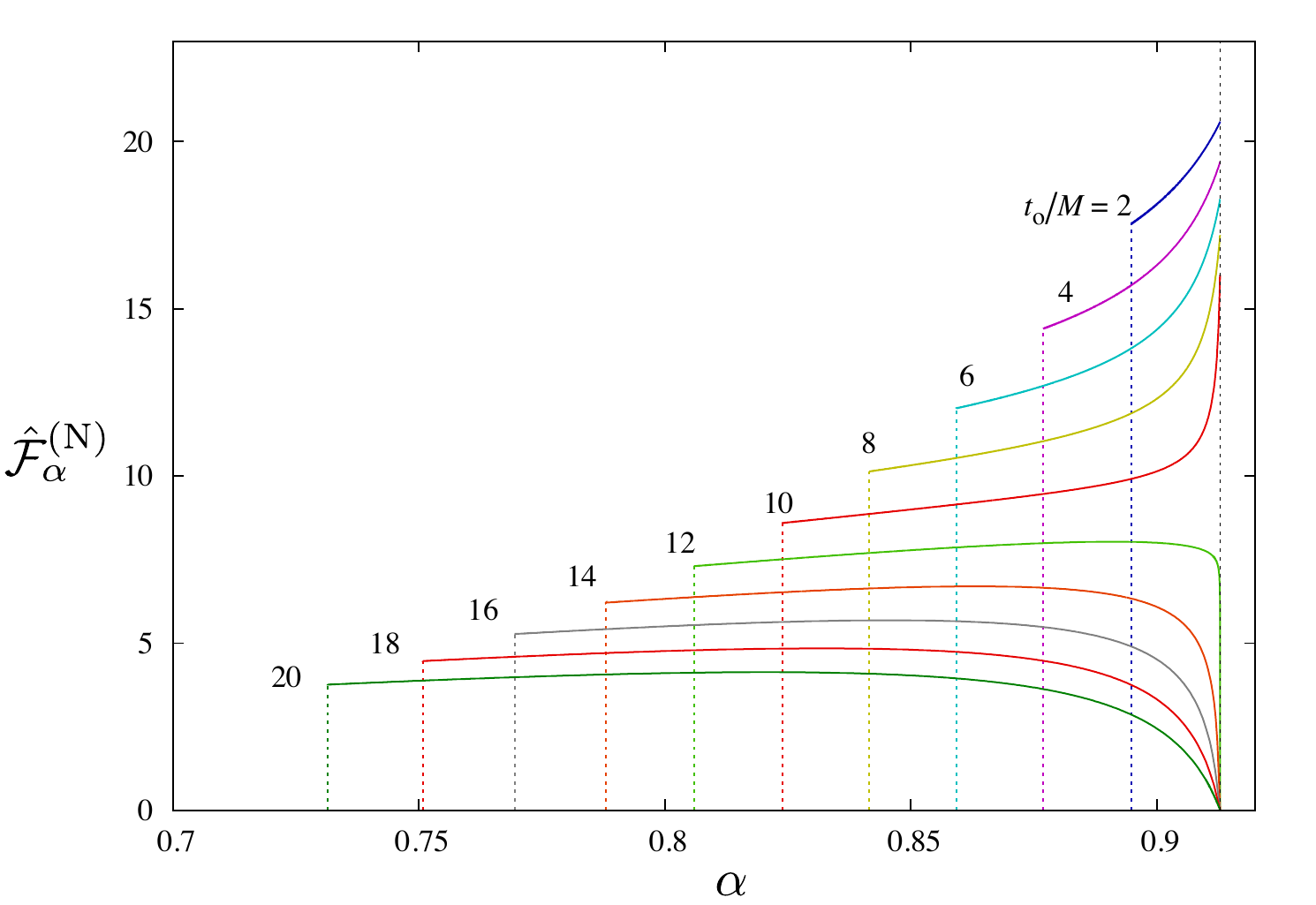}
 \includegraphics[width=0.45\textwidth,bb=0 0 429 295]{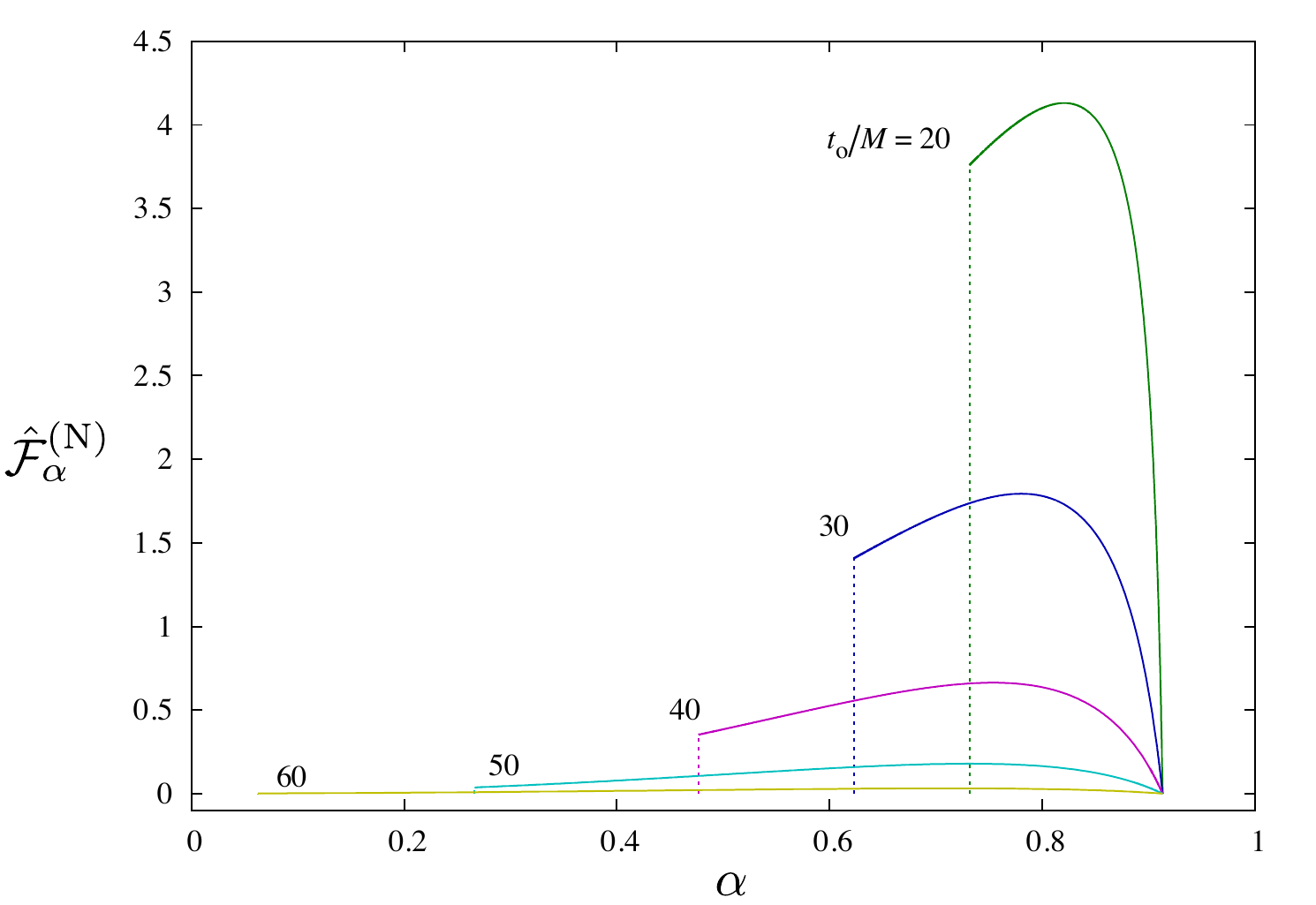}
 \includegraphics[width=0.50\textwidth,bb=0 0 314 302]{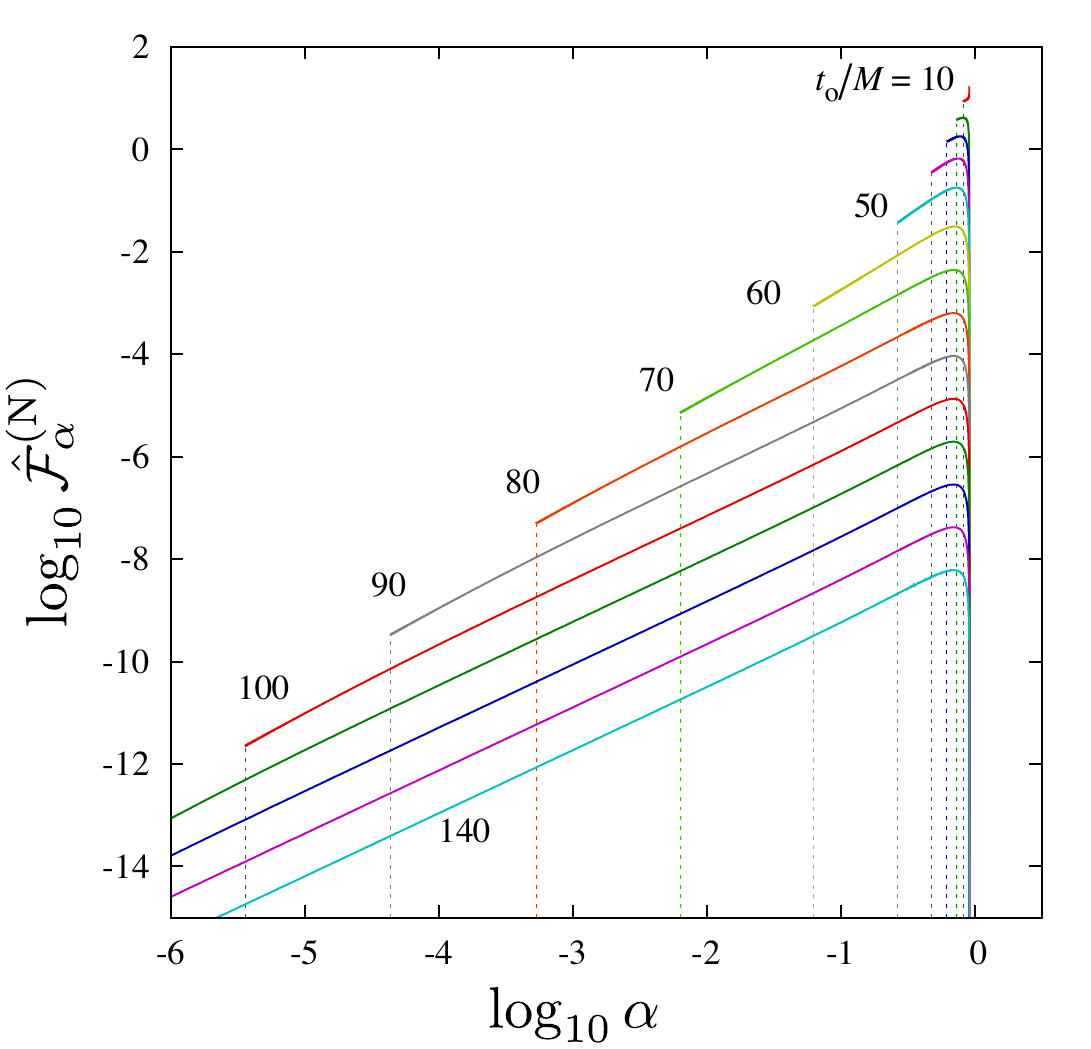}
 \caption{
   Snapshots of the normalized
   spectral photon flux $\hat{\mathcal{F}}^{\rm (N)}_{\alpha}$ as
   a function of
   $\alpha:=\omega_{\rm o}/\bar{\omega}_{\rm e}^\prime$
   for the case of monochromatic radiator. 
   Here, only the first terms of
   Eq.~\eqref{spectral-photon-flux-monochromatic-case}
   is shown.
   Top left panel: Snapshots for $t_{\rm o}/M=2$, $4$, ..., $20$.
   Top right panel: Snapshots for $t_{\rm o}/M=20$, $30$, ..., $60$.
   Bottom panel: Snapshots for $t_{\rm o}/M=10$, $20$, ..., $140$, where
   horizontal and vertical axes are shown in the logarithmic scale.
}
 \label{Fig:numerical_results_spectral_numberflux_monochromatic}
\end{figure}
%

Top left and top right panels of 
Fig.~\ref{Fig:numerical_results_spectral_numberflux_monochromatic} 
show snapshots of $\hat{\mathcal{F}}^{\rm (N)}_{\alpha}$
as a function of $\alpha$
for the moments $t_{\rm o}/M=0$, $2$, ..., $20$ and $t_{\rm o}/M=20$,
$30$, ..., $70$, respectively. 
Because we cannot plot the delta-functional profile,
only the first term of Eqs.~\eqref{spectral-photon-flux-monochromatic-case}
is shown. The range where the values of $\hat{\mathcal{F}}^{\rm (N)}_{\alpha}$
is nonzero 
is limited within $\alpha_{\rm cent}\le \alpha\le \alpha_{\rm limb}$,
where $\alpha_{\rm cent}$ is defined as Eq.~\eqref{alpha_cent} and is regarded as 
a function of only the observer's time $t_{\rm o}$. 
When $t_{\rm o}$ is small, $\hat{\mathcal{F}}^{\rm (N)}_{\alpha}$
is a monotonically
increasing function of $\alpha$.
But the value of $\hat{\mathcal{F}}^{\rm (N)}_{\alpha}$
at the limb $\alpha=\alpha_{\rm limb}$
decreases because at the edge $b=b_{\rm id}$ of the inner disk $0\leq b \leq b_{\rm id}$,
the value of $d\alpha/db$ grows larger
as time goes on 
(see Fig.~\ref{Fig:numerical_results_redshift_snapshots})
and becomes infinity 
after the edge arrives at the limb of the image, i.e.,
$t_{\rm o}>t_{\rm o}^{\rm (full)}$. 
As a result, 
$\hat{\mathcal{F}}^{\rm (N)}_{\alpha}$ is always zero
at $\alpha=\alpha_{\rm limb}$ for $t_{\rm o}>t_{\rm o}^{\rm (full)}$.
Since $\hat{\mathcal{F}}^{\rm (N)}_{\alpha}$
is an increasing function 
in the neighborhood of $\alpha=\alpha_{\rm cent}$ 
(see the Appendix for an analytic explanation), 
a peak appears for $t_{\rm o}>t_{\rm o}^{\rm (full)}$.
The bottom panel of 
Fig.~\ref{Fig:numerical_results_spectral_numberflux_monochromatic}
shows the relation between $\hat{\mathcal{F}}^{\rm (N)}_{\alpha}$
and $\alpha$, but both axes are shown in the logarithmic scale.
It can be seen that for late time, 
the curve approximately keeps its shape and size  
(except for the region of very small $\alpha\sim \alpha_{\rm cent}$)
and shifts down in a parallel manner as $t_{\rm o}$ is increased.
This means that the spectral photon flux decays exponentially in time.
$\hat{\mathcal{F}}^{\rm (N)}_{\alpha}$
is approximately proportional to $\alpha$ in small region of $\alpha$,
and there remains a peak in the spectrum.  
The asymptotic position of the peak is $\alpha\approx 0.69638$,
and it is consistent with $\alpha \approx 0.69625$
that will be derived in the asymptotic analysis in the next section.
Our results indicate that 
the redshift factor of most of the
arriving photons from the star remains finite,
and a star becomes invisible because
the rate of photon detection becomes slower.

%
\begin{figure}[tb]
 \centering
 \includegraphics[width=0.45\textwidth,bb=0 0 429 295]{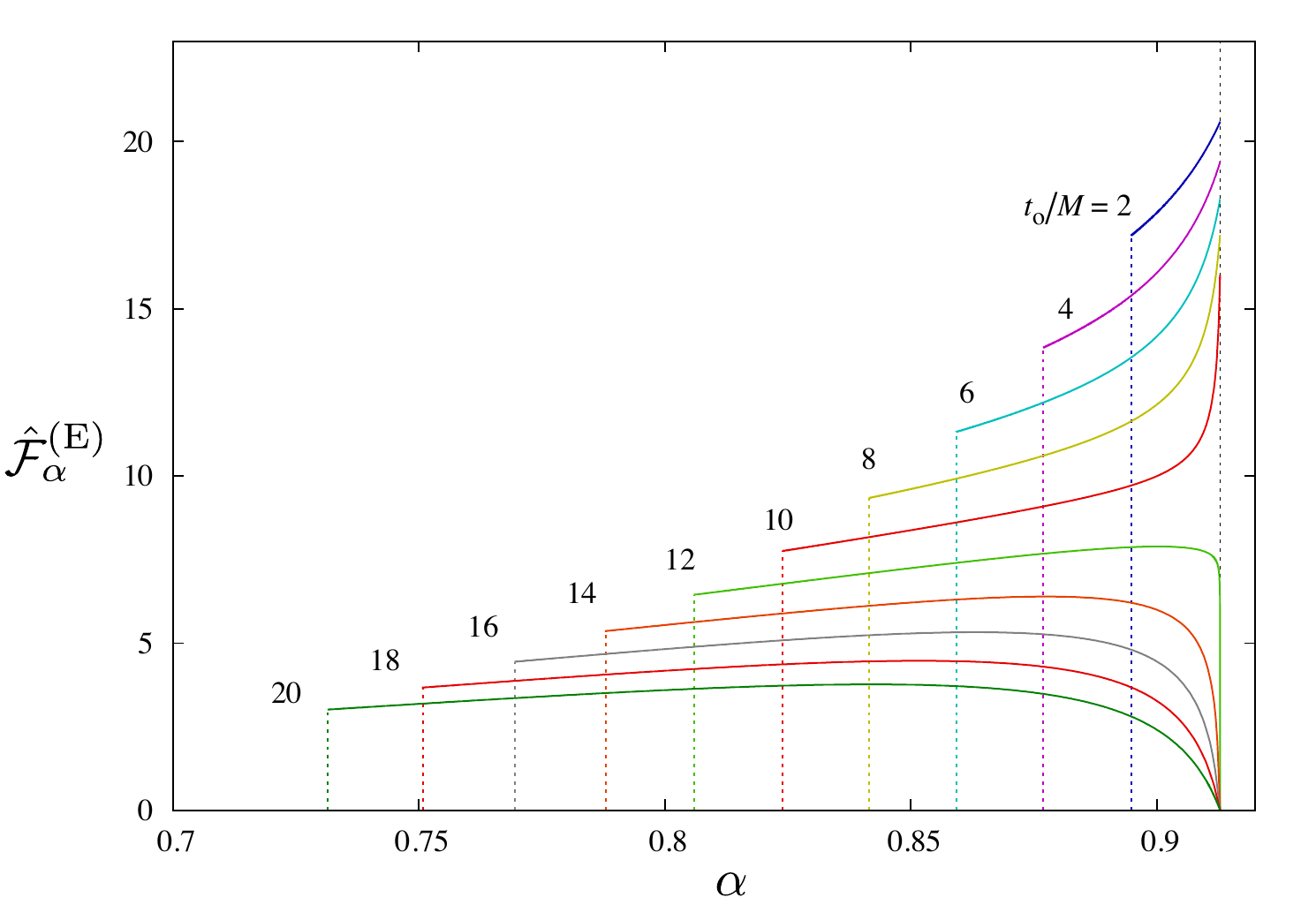}
 \includegraphics[width=0.45\textwidth,bb=0 0 429 295]{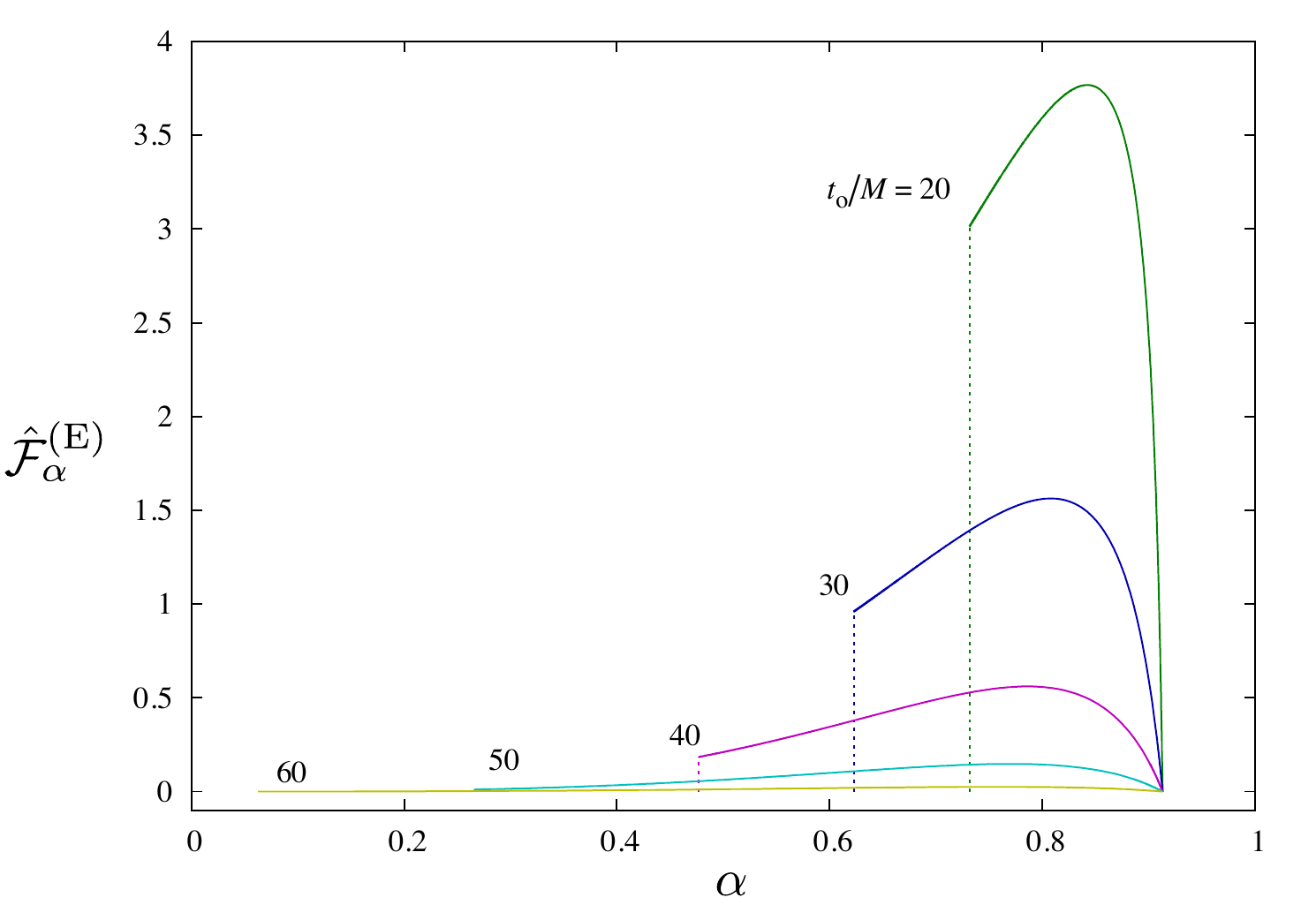}
 \includegraphics[width=0.50\textwidth,bb=0 0 315 301]{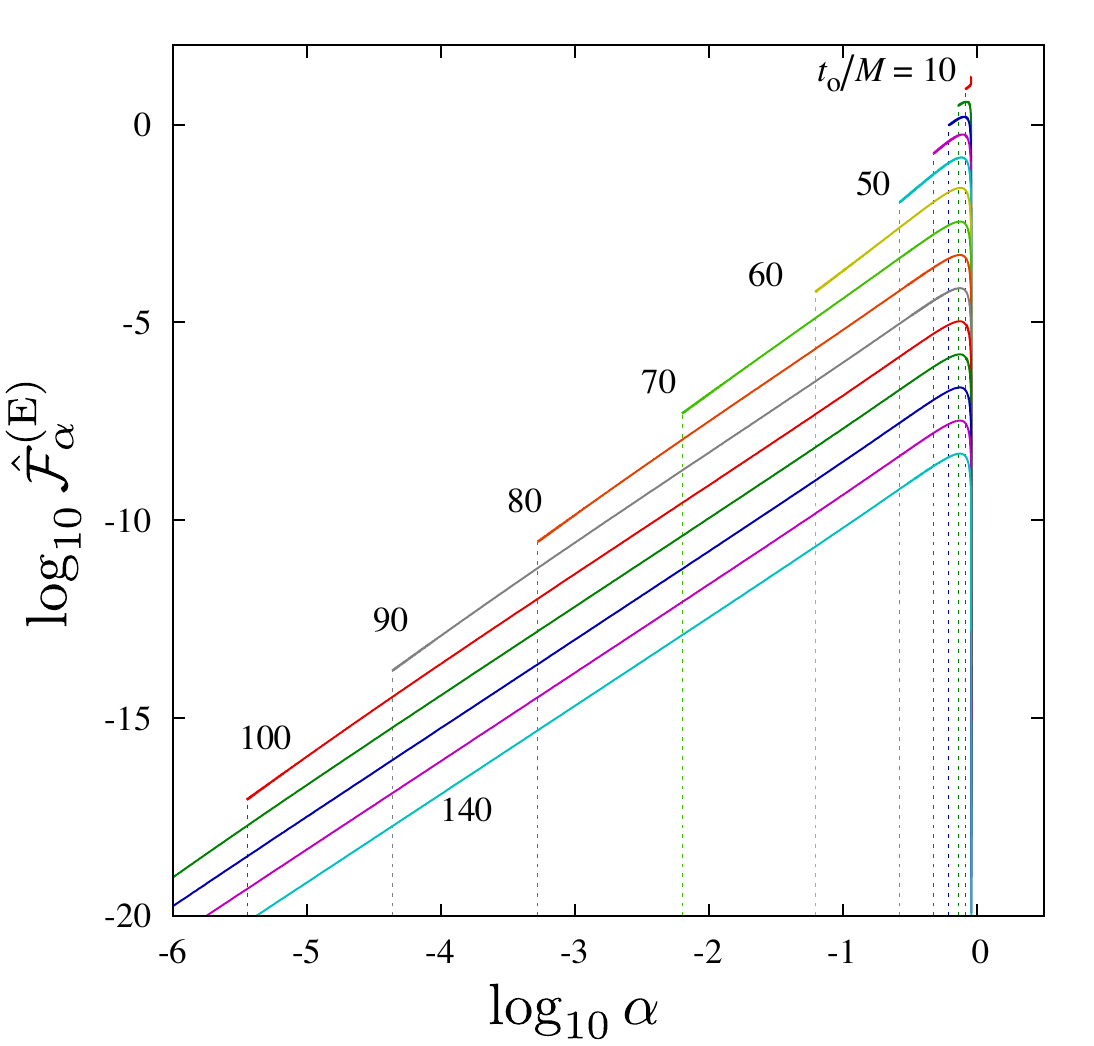}
 \caption{
   Same as Fig.~\ref{Fig:numerical_results_spectral_numberflux_monochromatic}
   but for the (normalized)
   spectral radiant flux $\hat{\mathcal{F}}^{\rm (E)}_{\alpha}$.
   Here, only the first term of
   Eq.~\eqref{spectral-radiant-flux-monochromatic-case} is shown. 
}
 \label{Fig:numerical_results_spectral_energyflux_monochromatic}
\end{figure}
%

Figure~\ref{Fig:numerical_results_spectral_energyflux_monochromatic}
is the same as
Fig.~\ref{Fig:numerical_results_spectral_numberflux_monochromatic}
but for the spectral radiant flux, $\hat{\mathcal{F}}^{\rm (E)}_{\alpha}$.
Qualitatively similar behavior to the spectral photon flux
$\hat{\mathcal{F}}^{\rm (N)}_{\alpha}$ is observed. 
For late time, 
$\hat{\mathcal{F}}^{\rm (E)}_{\alpha}$
is approximately proportional to $\alpha^2$ in small region of $\alpha$,
and a peak remains in the spectrum.
The asymptotic position of the peak is $\alpha\approx 0.74948$,
and it is consistent with $\alpha \approx 0.74944$
that will be derived in the asymptotic analysis in the next section.
The bottom panel
plots the same relation as Fig.~3 by Ames and Thorne \cite{Ames:1968}.
Here, we have presented the ``modern version''
generated by huge power of a recent computer.

\subsubsection{The case of blackbody radiation}
\label{Sec:VID2}

Now, we turn our attention to the case of blackbody radiation
with constant temperature given by
Eqs.~\eqref{intensity-Planck-radiation}--\eqref{constancy-of-temperature}.
Since the star surface emits photons with typical angular frequency
$\omega_{\rm e}^\prime\sim T_{\rm e}^\prime$, it is convenient
to normalize the observed angular frequency as
$\hat{\omega}_{\rm o}:=\omega_{\rm o}/T_{\rm e}^\prime$. In terms
of $\hat{\omega}_{\rm o}$, the spectral formulas
\eqref{spectrum-photon-flux} and \eqref{spectral-radiant-flux} become
\begin{subequations}
\begin{eqnarray}
  \hat{\mathcal{F}}^{\rm (N)}_{\hat{\omega}_{\rm o}}(t_{\rm o})
  =\frac{d\mathcal{F}^{\rm (N)}/d\hat{\omega}_{\rm o}}{\mathcal{F}^{\rm (N)}(0)}
&=&
\frac{\zeta(3)^{-1}}{R^2} \sqrt{\frac{f(r_{\rm o})^{3}}{f(R)}}
  \ \hat{\omega}_{\rm o}^2\ I(\hat{\omega}_{\rm o}),
  \label{spectrum-numberflux-Planck}
  \\
\hat{\mathcal{F}}^{\rm (E)}_{\hat{\omega}_{\rm o}}(t_{\rm o})
  =\frac{d\mathcal{F}^{\rm (N)}/d\hat{\omega}_{\rm o}}{\mathcal{F}^{\rm (N)}(0)}
    &=&
    \frac{30/\pi^4}{R^2}
  \frac{f(r_{\rm o})^2}{f(R)}\ 
  \ \hat{\omega}_{\rm o}^3\ I(\hat{\omega}_{\rm o}),
  \label{spectrum-energyflux-Planck}
\end{eqnarray}
\end{subequations}
where
\begin{equation}
  I(\hat{\omega}_{\rm o}):=
  \int_0^{b_{\rm limb}}\frac{bdb}{\exp(\hat{\omega}_{\rm o}/\alpha)-1}
  \label{spectrum-fluxes-Planck-integral}
\end{equation}

%
\begin{figure}[tb]
 \centering
 \includegraphics[width=0.60\textwidth,bb=0 0 427 291]{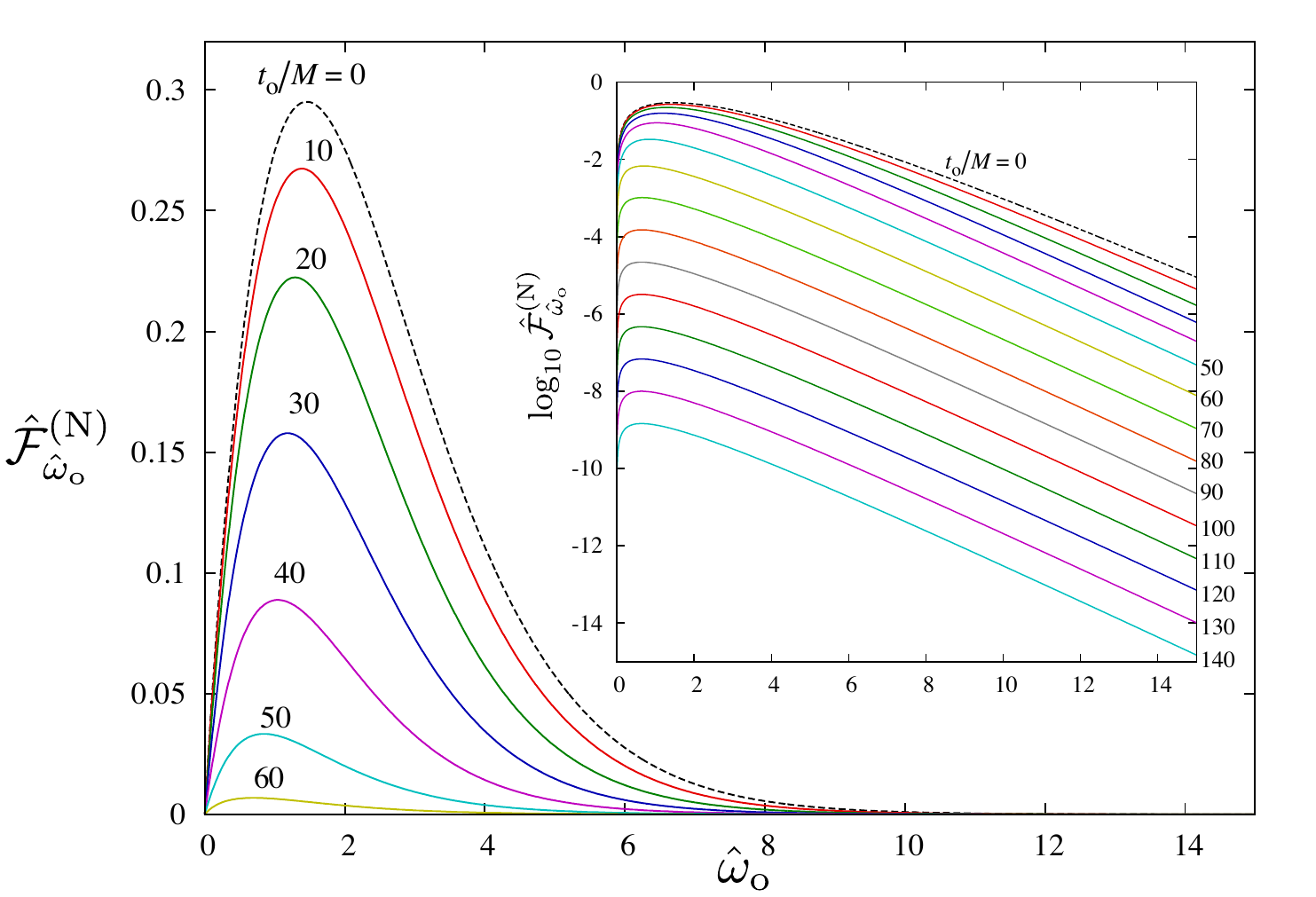}
 \caption{Snapshots of the normalized
   spectral photon flux $\hat{\mathcal{F}}^{\rm (N)}_{\hat{\omega}_{\rm o}}$ as
   a function of
   $\hat{\omega}_{\rm o}:=\omega_{\rm o}/T_{\rm e}^\prime$
   for the case of blackbody radiator
   for $t_{\rm o}/M=10$, $20$, ..., $60$.
   The inset shows snapshots for $t_{\rm o}/M=10$, $20$, ..., $140$, where
   the vertical axis is shown in the logarithmic scale.
}
 \label{Fig:numerical_results_spectral_numberflux_Planck}
\end{figure}
%

%
\begin{figure}[tb]
 \centering
 \includegraphics[width=0.60\textwidth,bb=0 0 427 291]{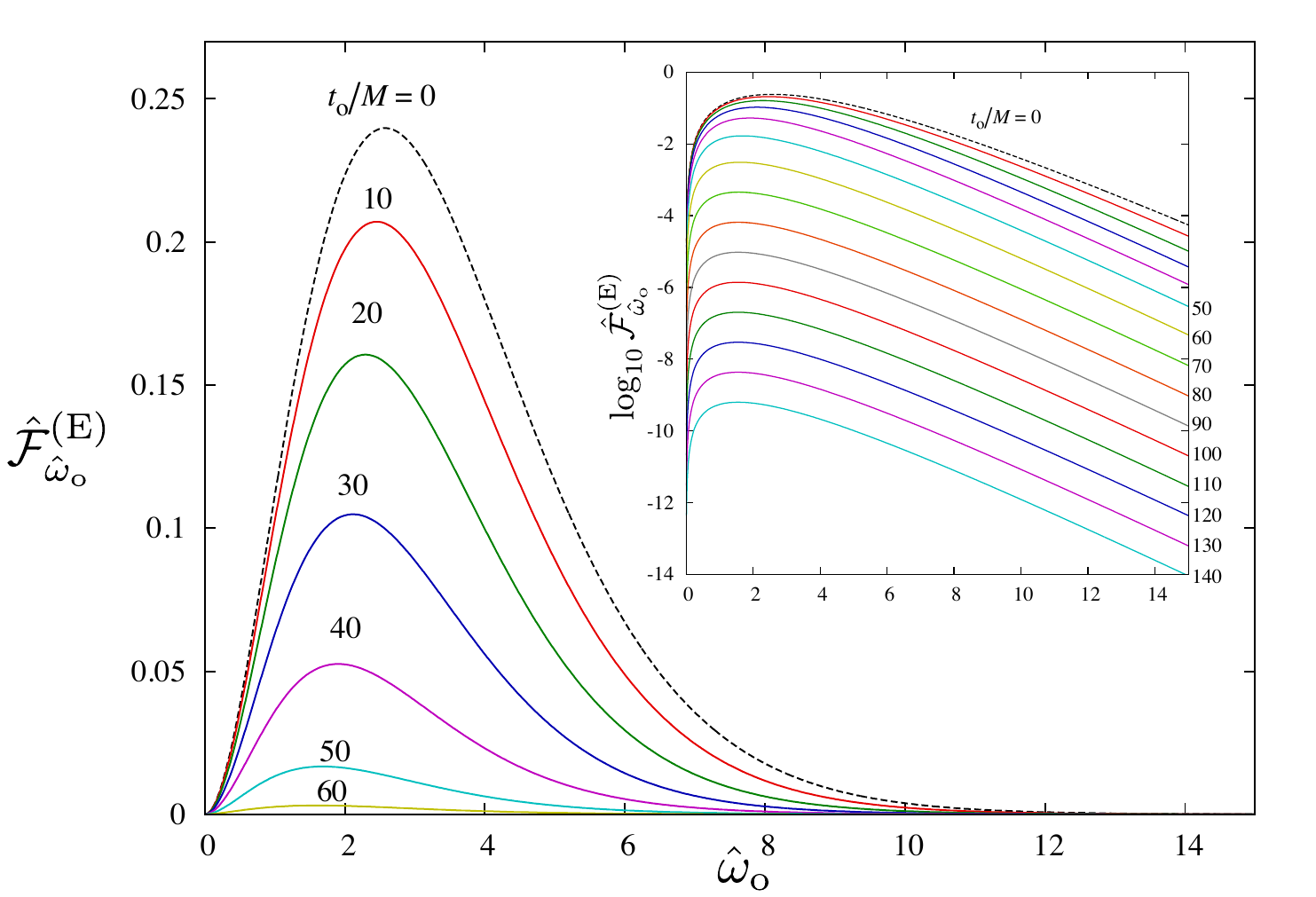}
 \caption{
   Same as Fig.~\ref{Fig:numerical_results_spectral_numberflux_Planck}
   but for the normalized
   spectral radiant flux $\hat{\mathcal{F}}^{\rm (E)}_{\hat{\omega}_{\rm o}}$. 
}
 \label{Fig:numerical_results_spectral_energyflux_Planck}
\end{figure}
%

Figures~\ref{Fig:numerical_results_spectral_numberflux_Planck}
and \ref{Fig:numerical_results_spectral_energyflux_Planck}
show snapshots of $\hat{\mathcal{F}}^{\rm (N)}_{\hat{\omega}_{\rm o}}$
and $\hat{\mathcal{F}}^{\rm (E)}_{\hat{\omega}_{\rm o}}$, respectively,
as functions of $\hat{\omega}_{\rm o}$
for the moments $t_{\rm o}/M=0$, $10$, ..., $60$.
The inset of each figure shows the moments $t_{\rm o}/M=0$, $10$,
..., $140$ but the vertical axis is shown in the logarithmic scale.
Throughout the evolution,
each of the two curves, 
$\hat{\mathcal{F}}^{\rm (N)}_{\alpha}(\hat{\omega}_{\rm o})$ and
$\hat{\mathcal{F}}^{\rm (E)}_{\alpha}(\hat{\omega}_{\rm o})$,
has a peak throughout the evolution. 
Initially, the peak locations are $\hat{\omega}_{\rm o}\approx 1.455$
and $\approx 2.576$, respectively. 
The location of each peak
shifts toward a smaller value of $\hat{\omega}_{\rm o}$ a bit,
but remains a nonzero finite value.
The asymptotic peak location is
$\hat{\omega}_{\rm o}\approx 0.63973$ and $1.55367$ for 
$\hat{\mathcal{F}}^{\rm (N)}_{\hat{\omega}_{\rm o}}$ and
$\hat{\mathcal{F}}^{\rm (E)}_{\hat{\omega}_{\rm o}}$, respectively.
The shapes of $\log_{10}\hat{\mathcal{F}}^{\rm (N)}$ and
$\log_{10}\hat{\mathcal{F}}^{\rm (E)}$ 
scarcely change, and the curves
shift to the lower direction. This means that 
$\hat{\mathcal{F}}^{\rm (N)}_{\hat{\omega}_{\rm o}}$ and
$\hat{\mathcal{F}}^{\rm (E)}_{\hat{\omega}_{\rm o}}$
decay exponentially in time.
Similarly to the the case of monochromatic radiation, 
a star becomes invisible because
the rate of photon detection becomes lower.

%
\begin{figure}[tb]
 \centering
 \includegraphics[width=0.60\textwidth,bb=0 0 420 293]{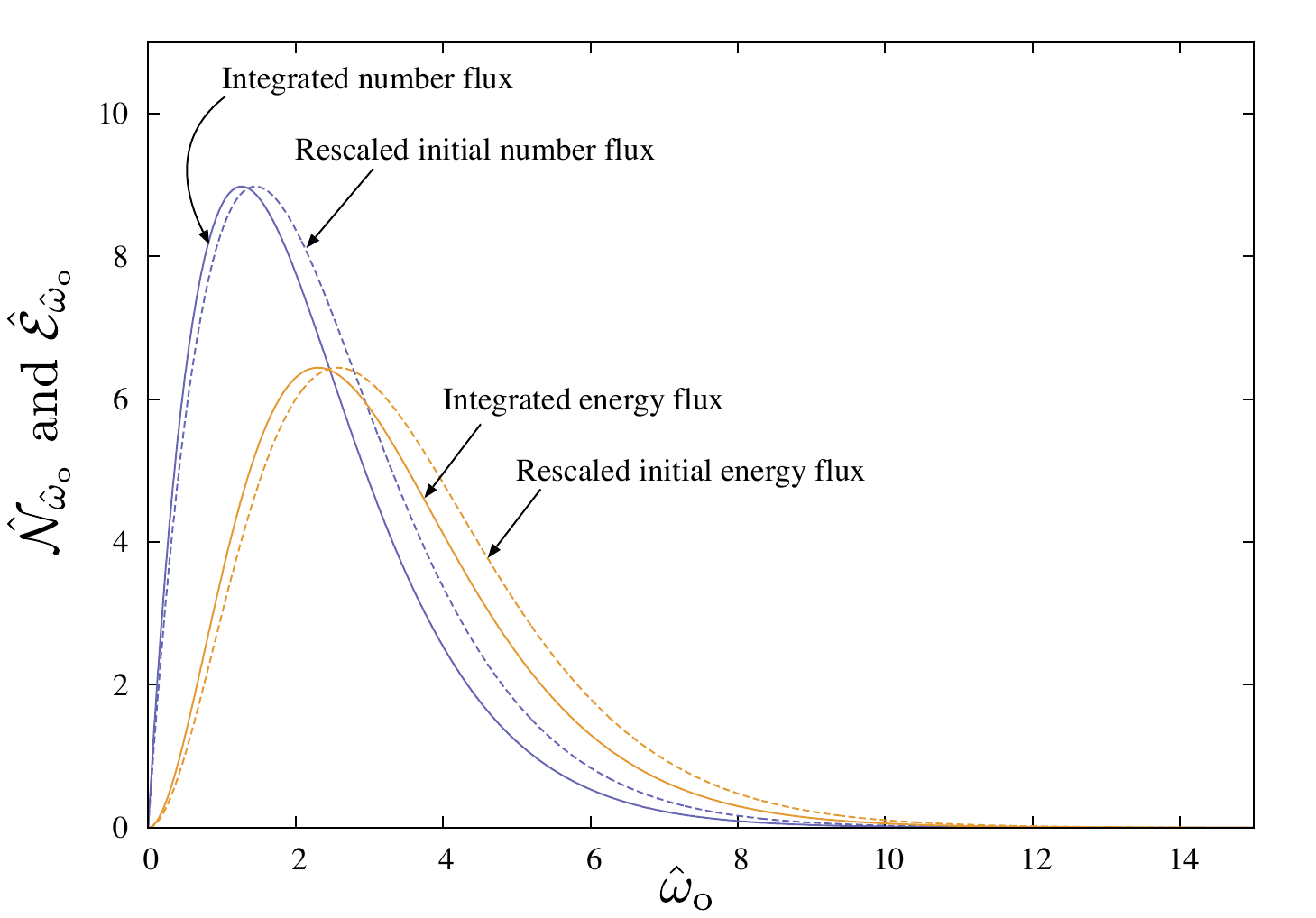}
 \caption{
   Photon number spectrum $\hat{\mathcal{N}}_{\hat{\omega}_{\rm o}}$
   and
   energy spectrum $\hat{\mathcal{E}}_{\hat{\omega}_{\rm o}}$
   as functions of
   $\hat{\omega}_{\rm o}:=\omega_{\rm o}/T_{\rm e}^\prime$
   integrated in the period $0\le t_{\rm o}/M\le 150$.
}
 \label{Fig:numerical_results_spectra_Planck}
 \end{figure}
%

The photon number spectrum
$dN/d\omega_{\rm o}$
and energy spectrum $dE/d\omega_{\rm o}$ (in the period $t_{\rm o}\ge 0$)
are obtained by 
integrating the spectral flux $d\mathcal{F}^{\rm (N)}/d\omega_{\rm o}$ and
$d\mathcal{F}^{\rm (E)}/d\omega_{\rm o}$ with
respect to the observer's proper time $\tau_{\rm o}=\sqrt{f(r_{\rm o})}t_{\rm o}$.
As nondimensional quantities, we calculate
\begin{eqnarray}
  \hat{\mathcal{N}}_{\hat{\omega}_{\rm o}}&:=&\frac{\sqrt{f(r_{\rm o})}}{M}
  \int_0^{t_{\rm o}^{\rm (cut)}} \hat{\mathcal{F}}^{\rm (N)}_{\hat{\omega}_{\rm o}}
  dt_{\rm o},
  \label{normalized-total-photon-number-spectrum}
  \\
  \hat{\mathcal{E}}_{\hat{\omega}_{\rm o}}&:=&\frac{\sqrt{f(r_{\rm o})}}{M}
  \int_0^{t_{\rm o}^{\rm (cut)}} \hat{\mathcal{F}}^{\rm (E)}_{\hat{\omega}_{\rm o}}
  dt_{\rm o},
  \label{normalized-total-energy-spectrum}
\end{eqnarray}
where $t_{\rm o}^{\rm (cut)}$ indicates an upper
cutoff value of the integration domain, which
we choose $t_{\rm o}^{\rm (cut)}=150M$.
Solid curves of Fig.~\ref{Fig:numerical_results_spectra_Planck} show
these quantities as functions of $\hat{\omega}_{\rm o}$.
For comparison, the rescaled initial values of the two kinds of flux,
$30.43\times\hat{\mathcal{F}}^{\rm (N)}_{\hat{\omega}_{\rm o}}$ and 
$26.87\times\hat{\mathcal{F}}^{\rm (E)}_{\hat{\omega}_{\rm o}}$, are
shown by dashed curves. 
The photon number spectrum and the radiant energy spectrum
have peaks at $\hat{\omega}_{\rm o}\approx 1.265$ and
$\hat{\omega}_{\rm o}\approx 2.303$, respectively.
Compared to 
the peak positions of two kinds of initial flux,
$\hat{\omega}_{\rm o}\approx 1.455$
and $\hat{\omega}_{\rm o}\approx 2.576$, 
the peak positions shift toward smaller values of $\hat{\omega}_{\rm o}$.

%
%
\section{Approximate analysis on asymptotic behavior}
\label{Sec:VII}

In our numerical results, as $t_{\rm o}$ is increased,
each of the two kinds of spectral flux 
asymptotes to a configuration with a peak at some fixed value
of angular frequency, and its magnitude
decays exponentially with time $t_{\rm o}$.
In this section, we develop a new approximate method to
give analytic formulas for late time behavior. 

\subsection{Method}

The primary contribution to the spectral photon and radiant flux
for the late time behavior 
comes from photons with the impact parameter $b$
close to the critical value $b_{\rm crit}$.
Hence, we study the approximate behavior
for worldlines of photons for small 
$\delta b$ compared to $M$, where $\delta b$
is defined as Eq.~\eqref{Delta-b}. 
Our first trial was to treat the deviation
from the worldline of a photon with $b=b_{\rm crit}$, which
is given analytically as
Eqs.~\eqref{orbit-critical-impact-parameter-2}--\eqref{bcrit_geodesic_outer_region}, 
as a perturbation. 
However, although we can obtain an analytic formula for the perturbative quantity
as well, it includes a strongly divergent
term proportional to $(r-3M)^{-2}$. 
Therefore, the perturbative scheme breaks down 
in the vicinity of the photon sphere, $\left|r-3M\right|\lesssim\sqrt{M\delta b}$.

%
\begin{figure}[tb]
 \centering
 \includegraphics[width=0.4\textwidth,bb=0 0 164 286]{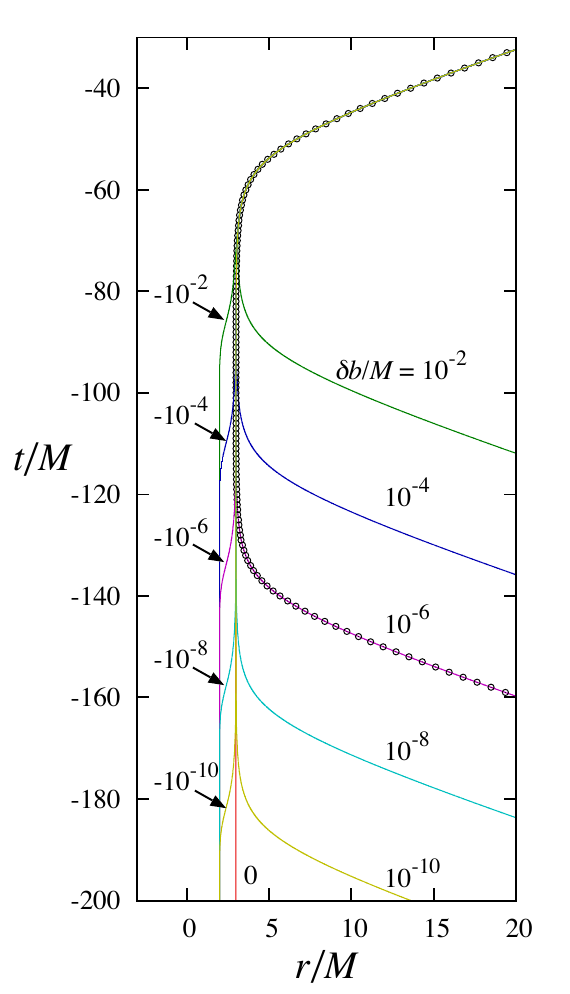}
 \caption{Worldlines of photons with impact parameters
   $\delta b/M=0$, $\pm 10^{-2}$, $\pm 10^{-4}$, $\pm 10^{-6}$, $\pm 10^{-8}$, and $\pm 10^{-10}$
   that pass through the observation point $t_{\rm o}=0$
   and $r_{\rm o}=50M$. Here, $\delta b$ is defined by 
   $\delta b:=b-b_{\rm crit}$. Except for the case $\delta b=0$, 
   numerically generated data are shown. For the case of $\delta b/M=10^{-6}$,
   the worldline obtained by the approximate formula
   \eqref{approximate-solution-positive-deltab} 
   is shown by circles ($\odot$). The approximate worldline
   coincides with the numerical one very well.
}
 \label{Fig:near_critical_impact}
\end{figure}
%

Instead of perturbation, we develop a method of nonperturbative
approximation here. Before doing this, let us look at worldlines of
photons with $|\delta b/M|\ll 1$. 
Since a worldline with $\delta b>0$ has a turning point in
general (a pericenter in the case of a worldline in the region $r>3M$),
it is convenient to regard $r$ as a function of $t$ in order to describe it,
because $r(t)$ becomes a single-valued function. 
The equation for $r(t)$ is given by Eq.~\eqref{nullgeodesic-equation-r-and-t}.
Figure~\ref{Fig:near_critical_impact} shows the
worldlines of photons with $\delta b/M = 0$, $\pm 10^{-2}$, $\pm 10^{-4}$,
$\pm 10^{-6}$, $\pm 10^{-8}$, and $\pm 10^{-10}$
that pass through the point $(t_{\rm o}, r_{\rm o})=(0,\ 50M)$. 
The photon motion has three phases: (i) {\it Approaching phase}
where a photon with $\delta b>0$ (respectively, $\delta b<0$) 
becomes closer to the photon sphere $r=3M$
from the outside (respectively, inside) region;
(ii) {\it Orbiting phase} where the photon orbits approximately
on the photon sphere; and (iii) {\it Escaping phase} where the
photon leaves the photon sphere and propagates to the observer.
As $|\delta b/M|$ becomes smaller, duration of the orbiting
phase becomes longer. 
The worldline in the escaping phase
is approximately the same as that of $b=b_{\rm crit}$,
with the error of $O(\delta b/M)$. 
Since the worldline of the approaching phase for $\delta b>0$
is just the time reversal of the escaping phase, it also 
can be approximated in the same manner.
For $\delta b<0$, it is possible to
approximate its worldline with an analytic formula 
of the photon worldline with $b=b_{\rm crit}$
that approaches the photon sphere from the inside of the photon sphere 
(see below).

These observations lead to the following approximate method. 
First, we consider the case $\delta b>0$.
We introduce a function $r-3M=\tilde{F}(t)$
that is equivalent to the analytic worldline
$t=F^{\rm (out)}(r)$ of an outwardly propagating photon with the critical impact parameter $b_{\rm crit}$,
where $F^{\rm (out)}(r)$ is
given by Eq.~\eqref{bcrit_geodesic_outer_region} with
Eqs.~\eqref{orbit-critical-impact-parameter-2} and
\eqref{orbit-critical-impact-parameter-3}.
For a later convenience, we introduce $G(r):=-F^{\rm (out)}(r)$. 
As mentioned in Sec.~\ref{Sec:IIIB}, $t=G(r)$
describes a worldline of
a photon with the critical impact parameter $b_{\rm crit}$
that asymptotes 
to the photon sphere $r=3M$ from the outside region.
We introduce a function $r-3M=\tilde{G}(t)$
that is equivalent to $t=G(r)$.
Then, these two solutions are combined as
\begin{equation}
  r-3M=\tilde{F}(t-t_{\rm o}+F^{\rm (out)}(r_{\rm o})) +\tilde{G}(t+D).
  \label{approximate-solution-positive-deltab}
\end{equation}
to approximate the worldline with $\delta b>0$.
Here, $D$ is a parameter related to the length
of the orbiting phase that will be 
related to the value of the impact parameter $\delta b$.

The formula \eqref{approximate-solution-positive-deltab}
actually gives an approximate worldline
in the following sense. In the escaping (respectively, approaching) phase,
the second term $\tilde{G}(t+D)$
[respectively, the first term $\tilde{F}(t-t_{\rm o}+F^{\rm (out)}(r_{\rm o}))$]
is exponentially small
and approximately zero. Thus, the formula approximately
satisfies 
Eq.~\eqref{nullgeodesic-equation-r-and-t}. 
In the orbiting phase, the
formula \eqref{approximate-solution-positive-deltab}
gives a correct perturbative behavior from $r=3M$.
Introducing $\delta r:=r-3M$, Eq.~\eqref{nullgeodesic-equation-r-and-t}
is linearized as
\begin{equation}
  \frac{d^2}{dt^2}\delta r = \frac{1}{27M^2} \delta r.
  \label{nullgeodesic-perturbative-equation}
\end{equation}
The formula \eqref{approximate-solution-positive-deltab}
approximately behaves as
\begin{eqnarray}
  \frac{r}{M}-3 &\approx&\exp\left[\frac{t-t_{\rm o}-C+F^{\rm (out)}(r_{\rm o})}{3\sqrt{3}M}\right]
  +\sigma\ \exp\left[-\frac{t+C+D}{3\sqrt{3}M}\right],
  \label{perturbative-behavior-positive-deltab}
\end{eqnarray}
with $\sigma=+1$ in the orbiting phase, 
which satisfies the perturbative
equation \eqref{nullgeodesic-perturbative-equation}.
Here, $C$ is defined by Eq.~\eqref{definition-C}.

Now, we relate $D$ with $\delta b$. From Eq.~\eqref{geodesic-drdt},
the radial coordinate $r=r_{\rm p}$ of the pericenter of the photon worldline
is determined by the equation $r^2=b^2f$,
and it is perturbatively solved as
$r_{\rm p}/M-3=(4/3)^{1/4}(\delta b/M)^{1/2}+O(\delta b/M)$.
By contrast, the minimum value of the right-hand side of
Eq.~\eqref{perturbative-behavior-positive-deltab} is
$\exp\left[\frac{(-t_{\rm o}+F^{\rm (out)}(r_{\rm o})-D)/2-C}{3\sqrt{3}M}\right]$.
Equating these two, we have
\begin{equation}
  D=-t_{\rm o}+F^{\rm (out)}(r_{\rm o})-2C-3\sqrt{3}M\log\left|\frac{\delta b}{M}\right|+\frac{3\sqrt{3}}{2}
  (\log12) M.
  \label{relation-D-deltab}
\end{equation}
Since this perturbative quantity
is $O(\sqrt{\delta b/M})$ in the orbiting phase, the error
in this approximation is $O(\delta b/M)$ in all phases.
In Fig.~\ref{Fig:near_critical_impact}, the behavior of the approximate
formula \eqref{approximate-solution-positive-deltab}
with \eqref{relation-D-deltab} 
is shown by circles for $\delta b/M=10^{-6}$
and is compared with the numerical solution. 
The approximate formula remarkably coincides with
the numerical solution. 
If a photon is emitted from the point  $(t, r)=(t_{\rm e}, r_{\rm e})$
in the approaching phase,
we have $r_{\rm e}-3M\approx \tilde{G}(t_{\rm e}+D)$
since the first term of the right-hand side of
Eq.~\eqref{approximate-solution-positive-deltab}
is approximately zero.
This is rewritten as
\begin{equation}
  \frac{\delta b}{M}
  =2\sqrt{3}
  \left(\frac{\sqrt{3r_{\rm e}}-\sqrt{r_{\rm e}+6M}}
       {\sqrt{3r_{\rm e}}+\sqrt{r_{\rm e}+6M}}\right)
  \exp\left[\frac{t_{\rm e}+\bar{F}(r_{\rm e})-2C-t_{\rm o}+F^{\rm (out)}(r_{\rm o})}{3\sqrt{3}M}\right],
  \label{asymptotic-deltab}
\end{equation}
where $\bar{F}$ is defined in Eq.~\eqref{orbit-critical-impact-parameter-3}.
This relation means that 
a photon traveling from 
the emission event at $(t_{\rm e}, r_{\rm e})$
to the observation event at  $(t_{\rm o}, r_{\rm o})$ 
must have the impact parameter $b=b_{\rm crit}+\delta b$
given by this formula.
The coordinate values of the observation point $(t_{\rm o}, r_{\rm o})$ appear
in Eq.~\eqref{asymptotic-deltab} 
only through the factor $\exp[(-t_{\rm o}+F^{\rm (out)}(r_{\rm o}))/3\sqrt{3}M]$.
As the observer's time $t_{\rm o}$ is increased, the required deviation $\delta b$
from the critical impact parameter becomes exponentially small.

Next, let us consider the case $\delta b<0$. For this case,
we define $G(r)$ as $G(r):=-F^{\rm (in)}(r)$, where $F^{\rm (in)}$
is defined in Eq.~\eqref{bcrit_geodesic_inner_region} with
Eqs.~\eqref{orbit-critical-impact-parameter-2} and \eqref{orbit-critical-impact-parameter-3}.
As mentioned in Sec.~\ref{Sec:IIIB}, $t=G(r)$ 
represents the worldline of a photon with $b=b_{\rm crit}$ 
that asymptotes to the photon sphere $r=3M$
from the inside region. 
This analytic solution approximates the worldline
with $\delta b<0$ in the approaching phase.
Requiring $r-3M=\tilde{G}(t)$ to be equivalent to $t=G(r)$,
we obtain the approximate worldline with $\delta b<0$ by the
same formula as Eq.~\eqref{approximate-solution-positive-deltab}.
The behavior of this function in the orbiting phase
is given by Eq.~\eqref{perturbative-behavior-positive-deltab}
with $\sigma=-1$. In order to relate $D$ with $\delta b$,
we evaluate the value of $dr/dt$
from Eq.~\eqref{perturbative-behavior-positive-deltab}
at the point where the curve intersects $r=3M$ and require it to be
equal to
$dr/dt=(2/27\sqrt{3})^{1/2}(-\delta b/M)^{1/2}+O(\delta b/M)$
evaluated from Eq.~\eqref{geodesic-drdt} up to the order of $\delta b^{1/2}$.
This gives the relation \eqref{relation-D-deltab}.
Rewriting Eq.~\eqref{relation-D-deltab},
we are led to exactly the same formula
for $\delta b/M$ as Eq.~\eqref{asymptotic-deltab}.

We examine implication of Eq.~\eqref{asymptotic-deltab}.
Since the emission event at $(t_{\rm e},r_{\rm e})$ is on 
the world sheet of the star surface given 
by Eqs.~\eqref{timelike-geodesic-r} and \eqref{timelike-geodesic-t}
in the parametrized form with respect to $\xi$,
the value of the impact parameter $b$ is determined only
by the radius $r_{\rm e}$ of the star surface
at the emission event, or equivalently, the value of $\xi$
for a fixed observation point $(t_{\rm o}, r_{\rm o})$.
In the region $2M<r_{\rm e}\le 3\sqrt{3f(R)}M$, $\delta b$
is a monotonically increasing function of $r_{\rm e}$ and
takes the maximum value $b_{\rm limb}$
at $r_{\rm e}=3\sqrt{3f(R)}M$ that corresponds to the limb of the observed
image. 
In the region $3\sqrt{3f(R)}M\le r_{\rm e}\le R$, $\delta b$
is a monotonically decreasing function of $r_{\rm e}$.
We do not consider the region $3\sqrt{3f(R)}M\le r_{\rm e}\le R$,
because Eqs.~\eqref{b-thetae-thetao} and \eqref{relation_thetae_thetaeprime}
imply that photons with $b\approx b_{\rm crit}$
emitted from this region have the angle
$\vartheta_{\rm e}^\prime$ 
greater than $\pi/2$ in the comoving frame,
which is unrealistic in our setup. 
Restricting our attention
to the region $2M<r_{\rm e}\le 3\sqrt{3f(R)}M$, 
Eq.~\eqref{asymptotic-deltab} gives a one-to-one relation
between $b$ and $r_{\rm e}$.

In order to describe  the redshift factor
$\alpha$ in this approximation, 
we use the formula obtained 
by eliminating $\cos\vartheta_{\rm e}$ in 
Eq.~\eqref{redshift-in-terms-of-thetae} through Eq.~\eqref{b-thetae-thetao}. 
Here, we set $b=b_{\rm crit}$ in this relation 
since our approximation allows the error of $O(\delta b/M)$.
This leads to
\begin{equation}
 \alpha =\frac{f(r_{\rm e})}{\sqrt{f(r_{\rm o})f(R)}}
 \left[1
   -\left(1-\frac{3M}{r_{\rm e}}\right)
   \sqrt{\left(1+\frac{6M}{r_{\rm e}}\right)
     \left(1-\frac{f(r_{\rm e})}{f(R)}\right)}\right]^{-1} +O\left(\delta b/M\right).
\label{asymptotic-alpha}
\end{equation}
Now, for a fixed value of $R$ and the observation point  
$(t_{\rm o}, r_{\rm o})$, the redshift factor $\alpha$ is also expressed
as a function of $r_{\rm e}$.
Therefore, we have found the asymptotic relation between 
$\delta b$ and $\alpha$ that is parametrically given by
$r_{\rm e}$.

\subsection{Results}

%
\begin{figure}[tb]
 \centering
 \includegraphics[width=0.6\textwidth,bb=0 0 312 229]{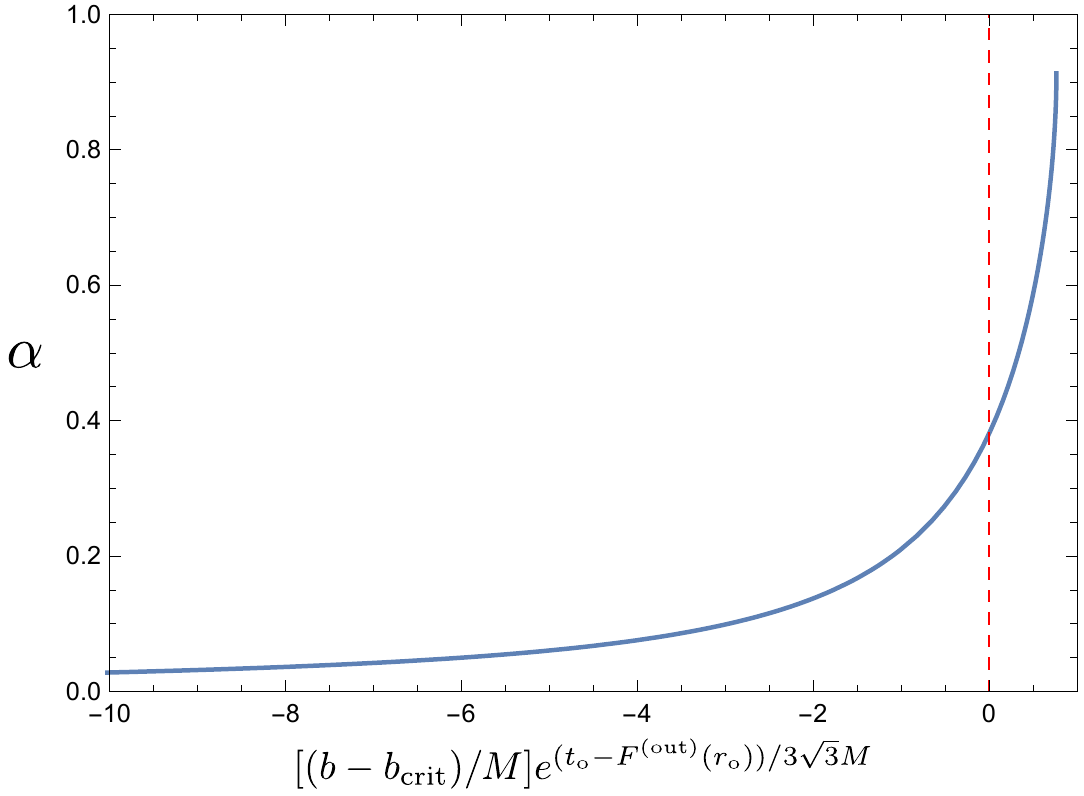}
 \caption{Asymptotic relation between
   the (shifted and scaled) impact parameter
   $[(b-b_{\rm crit})/M]e^{(t_{\rm o}-F^{\rm (out)}(r_{\rm o}))/3\sqrt{3}M}$
   and the redshift factor $\alpha$ for $t_{\rm o}- F^{\rm (out)}(r_{\rm o})\gg M$.
}
 \label{Fig:asymptotic_behavior}
\end{figure}
%

Now, we show the asymptotic behaviors of observable quantities
(as already mentioned, we focus on the case $R=10M$ and $r_{\rm o}=50M$).
Figure~\ref{Fig:asymptotic_behavior} shows the relation between
the redshift factor $\alpha$ and the scaled impact parameter
$(\delta b/M)e^{(t_{\rm o}-F^{\rm (out)}(r_{\rm o}))/3\sqrt{3}M}$. 
The redshift factor $\alpha(b)$ is a monotonically
increasing function.  
At $b=b_{\rm limb}$, $\alpha$ takes
the value $\alpha_{\rm limb}$ defined in Eq.~\eqref{alpha_limb}
while the derivative $d\alpha/db$ diverges. 
If we plot the function 
$\alpha(b)$ in the $(b,\alpha)$-plane for large $t_{\rm o}$,
the curve always passes through the point
$(b_{\rm crit}, \alpha_{\infty})$, where $\alpha_{\infty}$ is defined
by Eq.~\eqref{alpha_infty}, 
and continues to shrink in the horizontal direction
as $t_{\rm o}$ is increased. This explains the late time behavior
of our numerical results of $\alpha(b)$
in Fig.~\ref{Fig:numerical_results_redshift_snapshots}.

%
\begin{figure}[tb]
 \centering
 \includegraphics[width=0.6\textwidth,bb=0 0 313 218]{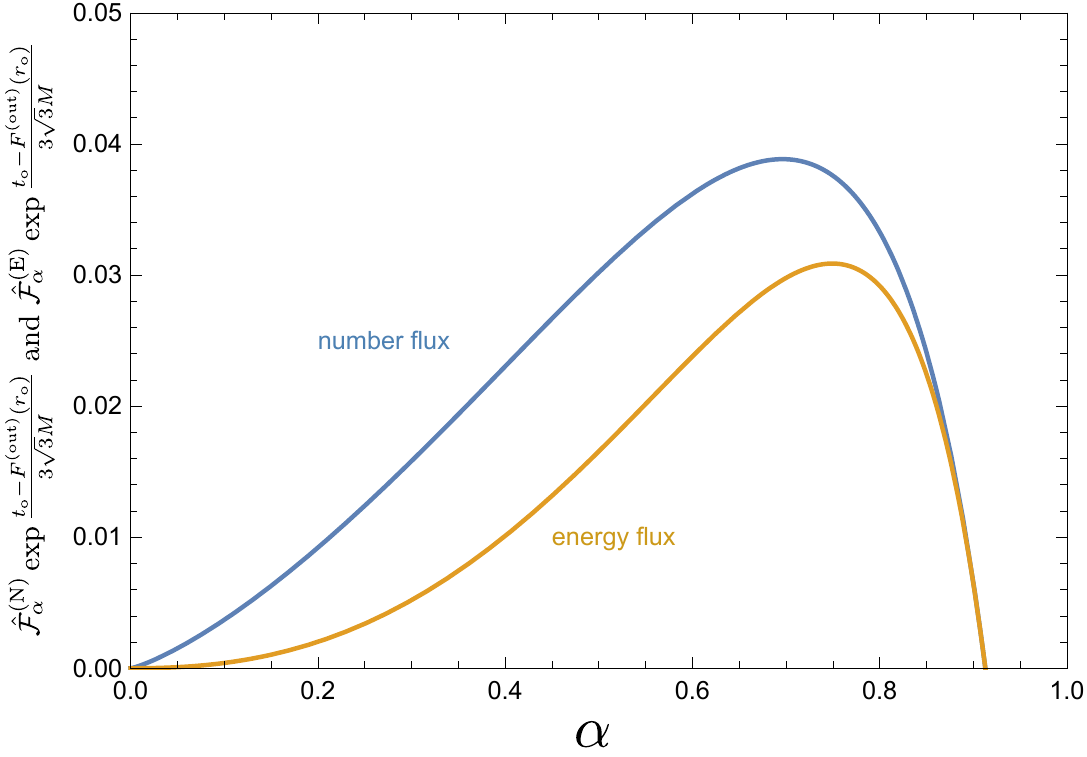}
 \caption{Asymptotic behavior
   of the (rescaled) spectral photon/radiant flux
   for the case of a monochromatic radiator 
   for $t_{\rm o}-F^{\rm (out)}(r_{\rm o})\gg M$. Here, 
   $\hat{\mathcal{F}}_{\alpha}^{\rm (N)}e^{(t_{\rm o}-F^{\rm (out)}(r_{\rm o}))/3\sqrt{3}M}$
   and $\hat{\mathcal{F}}_{\alpha}^{\rm (E)}e^{(t_{\rm o}-F^{\rm (out)}(r_{\rm o}))/3\sqrt{3}M}$
   are shown as functions of the normalized angular frequency
   $\alpha:=\omega_{\rm o}/\omega_{\rm e}^\prime$
   (indicated by ``number flux'' and ``energy flux'', respectively).
}
 \label{Fig:asymptotic_behavior_flux}
\end{figure}
%

Now, we show the spectral photon flux and the spectral radiant flux  
in the monochromatic case 
given in Eqs.~\eqref{spectral-photon-flux-monochromatic-case}
and 
\eqref{spectral-radiant-flux-monochromatic-case}.
We evaluate only the first terms of these formulas
because the second terms vanish for sufficiently large $t_{\rm o}$.
Since the emission from the star surface is assumed to be
time independent as assumed in Secs.~\ref{Sec:IIB1} and \ref{Sec:IIB2},
we set 
$J_{\rm e}^{\rm (N)}(r_{\rm e})=J_{\rm e}^{\rm (N)}(R)$
and 
$\left.\langle\omega_{\rm e}^\prime\rangle\right|_{r_{\rm e}}=\left.\langle\omega_{\rm e}^\prime\rangle\right|_{R}$.
We approximate these formulas by substituting $b=b_{\rm crit}$ and
calculating
$db/d\alpha = [d(\delta b)/dr_{\rm e}]/(d\alpha/dr_{\rm e})$.
The quantity $dt_{\rm e}/dr_{\rm e}$, which appears in the calculation
of $d(\delta b)/dr_{\rm e}$, is evaluated using Eqs.~\eqref{Eq:geodesic_dottre}
and \eqref{Eq:geodesic_dortre}.
Here, the factor $\exp[(-t_{\rm o}+F^{\rm (out)}(r_{\rm o}))/3\sqrt{3}M]$ appears
from $d(\delta b)/dr_{\rm e}$, and thus, the spectral photon/radiant
flux decays as the observer's time $t_{\rm o}$ is increased.
Unfortunately, we could not find simple formulas for these quantities,
but it is possible to calculate them and plot a figure
on {\it Mathematica}. 
Figure~\ref{Fig:asymptotic_behavior_flux} shows the behavior
of the two kinds of rescaled spectral flux, 
$\hat{\mathcal{F}}_{\alpha}^{\rm (N)}e^{(t_{\rm o}-F^{\rm (out)}(r_{\rm o}))/3\sqrt{3}M}$
and $\hat{\mathcal{F}}_{\alpha}^{\rm (E)}e^{(t_{\rm o}-F^{\rm (out)}(r_{\rm o}))/3\sqrt{3}M}$.
They are zero at $\alpha=0$ and $\alpha_{\rm limb}$ 
and have peaks at $\alpha\approx 0.696$ and $0.749$,
respectively.
Integrating $\hat{\mathcal{F}}^{\rm (N)}_{\alpha}$
with respect to $\alpha$, it is found that  
approximately 82\% of photons have the redshift factor
$\alpha_{\infty}\le \alpha\le \alpha_{\rm limb}$,
and thus, they are emitted while the star radius
is within the range $3M\le r_{\rm e}\le 3\sqrt{3f(R)}M$.

%
\begin{figure}[tb]
 \centering
 \includegraphics[width=0.6\textwidth,bb=0 0 314 194]{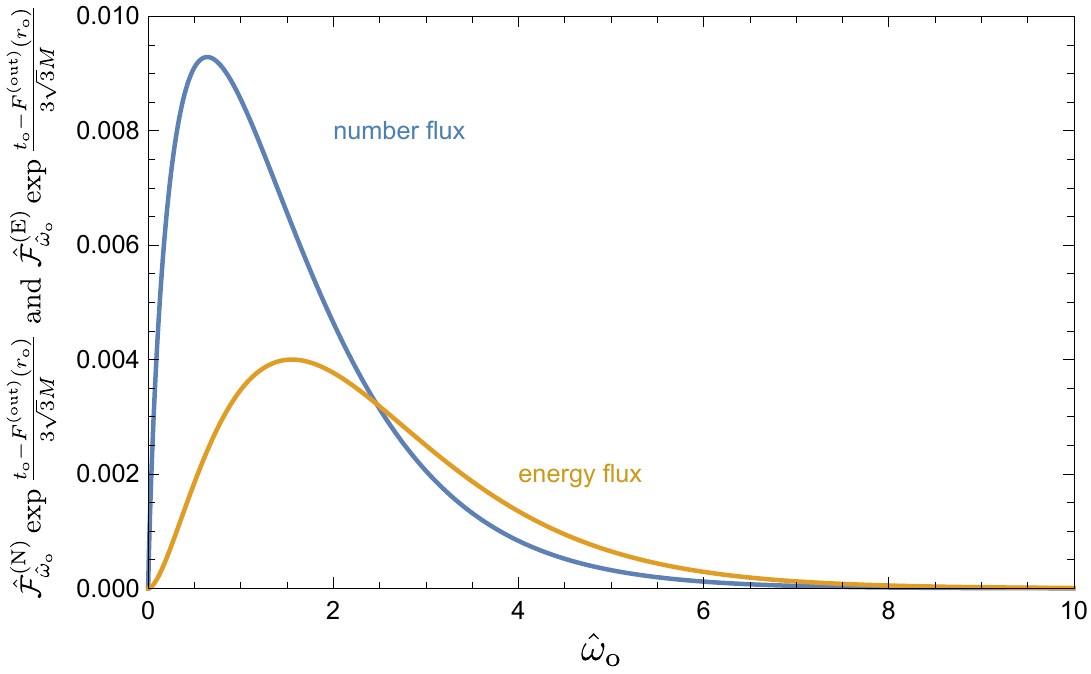}
 \caption{Same as Fig.~\ref{Fig:asymptotic_behavior_flux} but
   for the case of a blackbody radiator.
   The horizontal axis is angular frequency normalized
   by the temperature of the star surface,
   $\hat{\omega}_{\rm o}:=\omega_{\rm o}/T_{\rm e}^\prime$.
}
 \label{Fig:asymptotic_behavior_Planck}
\end{figure}
%

Let us discuss the two kinds of spectral flux in the case
of a blackbody radiator. We use the
formulas~\eqref{spectrum-numberflux-Planck} and
\eqref{spectrum-energyflux-Planck}
with \eqref{spectrum-fluxes-Planck-integral},
and rewrite $I(\hat{\omega}_{\rm o})$ as
\begin{equation}
  I(\hat{\omega}_{\rm o})
  \approx
  \int_{\xi_{\rm h}}^{\xi_{\rm limb}}\frac{3\sqrt{3}M}{\exp(\hat{\omega}_{\rm o}/\alpha)-1}
  \left[\frac{d(\delta b)}{dr_{\rm e}}\right]
  \left(\frac{dr_{\rm e}}{d\xi}\right)
  d\xi,
\end{equation}
where $\xi_{\rm h}$ is defined in Eq.~\eqref{horizon_crossing_xih}
and $\xi_{\rm limb}:=2\arccos\left(27M^2f(R)/R^2\right)^{1/4}$
is the value of $\xi$ that
corresponds to $r_{\rm e}=3\sqrt{3f(R)}M$. 
Here, ${d(\delta b)}/{dr_{\rm e}}$
gives the factor $\exp[(-t_{\rm o}+F^{\rm (out)}(r_{\rm o}))/3\sqrt{3}M]$,
and thus, $\hat{\mathcal{F}}^{\rm (N)}_{\hat{\omega}_{\rm o}}$
and $\hat{\mathcal{F}}^{\rm (E)}_{\hat{\omega}_{\rm o}}$
decay with respect to an observer's time $t_{\rm o}$. 
Although the integral $I(\hat{\omega}_{\rm o})$ cannot be proceeded
analytically, it is easily integrated numerically
by {\it Mathematica} because the integrand can be given
by an analytic formula.
The results
for $\hat{\mathcal{F}}^{\rm (N)}_{\hat{\omega}_{\rm o}}e^{(t_{\rm o}-F^{\rm (out)}(r_{\rm o}))/3\sqrt{3}M}$ and
$\hat{\mathcal{F}}^{\rm (E)}_{\hat{\omega}_{\rm o}}e^{(t_{\rm o}-F^{\rm (out)}(r_{\rm o}))/3\sqrt{3}M}$
are shown in Fig.~\ref{Fig:asymptotic_behavior_Planck}.
The two kinds of spectral flux 
have peaks at $\hat{\omega}_{\rm o}=0.639$ and $1.553$, respectively,
resembling Planck's law with a temperature that
is about half of that of the star surface.

\subsection{Interpretation}

The above derivation of the asymptotic behavior
gives a supportive evidence to our numerical results.
From the above analysis, the mechanism that causes the asymptotic behavior
has become clear.
For large $t_{\rm o}$, almost all of the observed photons
have the impact parameter
$b\approx b_{\rm crit}$, and they arrive at the observer
after orbiting in the neighborhood of the photon sphere $r=3M$.
Since most of them are emitted when the radius of the
star is within the range $3M\lesssim r_{\rm e}\le 3\sqrt{3f(R)}M$,
their redshift factor $\alpha$ remains $O(1)$
according to Eq.~\eqref{asymptotic-alpha}.
Therefore, the observed photon intensity
(i.e., the observed photon number flux per unit solid angle)
is also $O(1)$ around the limb of the star image.
But, as an observer's time $t_{\rm o}$ is increased,
the difference of the impact parameter $\delta b$ of
most of the observable photons 
from the critical one becomes smaller 
because photons must orbit around $r=3M$ for a longer time.
Due to the relation between $b$ and 
$\vartheta_{\rm o}$ given by Eq.~\eqref{b-thetae-thetao},
the bright ring-shaped portion of the star image becomes thinner and thinner, 
leading to the decay of the total photon number flux.
Since the typical scale of $\delta b$ of arriving photons 
decays exponentially as
$\sim \exp[(-t_{\rm o}+F^{\rm (out)}(r_{\rm o}))/3\sqrt{3}M]$ 
reflecting the instability time scale of the circular orbit,
the photon number also decays in the same way.

The discussion here was partly given in Ref.~\cite{Ames:1968}. 
In that paper, focusing attention to 
photons that are emitted from the inside region of the photon sphere
$r=3M$ and orbit around the black hole for a long time
(class I and II in their definitions), 
gross features of time evolution of the spectral radiant flux  
were derived. 
Our analysis completes their discussion by
adding the contribution of backward emitted photons
and giving the analytic formulas~\eqref{asymptotic-deltab}
and \eqref{asymptotic-alpha} for asymptotic behavior
for the first time.

%
%
\section{Summary and discussion}
\label{Sec:VIII}

In this paper,
we have studied the propagation of photons
from a star under the gravitational collapse and 
developed a formalism
for calculating observable quantities related to photon counting
and radiometry.
Although this is an old problem,
we have revisited this problem 
taking primary attention to generation of accurate numerical data
of the observable quantities. 
The formulas \eqref{spectrum-photon-flux} and \eqref{spectral-radiant-flux}
for the spectral photon flux and
the spectral radiant flux have been derived 
in the case that the star surface
emits radiation that obeys Lambert's cosine law
with arbitrary spectral intensity 
(Sec.~\ref{Sec:IV}). After explaining our numerical method
in Sec.~\ref{Sec:V}, 
we have applied our formalism to two types of spectra,
monochromatic and blackbody radiation, and
calculated observable quantities numerically 
(Sec.~\ref{Sec:VI}).
There, it has been reconfirmed that at late stage of the collapse, 
each of the two kinds of spectral flux has a peak at some finite
value of angular frequency $\omega_{\rm o}$: 
The star becomes invisible not by infinite redshift
but by decay of the photon flux.
This late time behavior has also been confirmed
by an analytic approximate method developed in Sec.~\ref{Sec:VII}.

Here, we discuss to what extent our formalism is general.
In the setup in Sec.~\ref{Sec:II}, we modeled the collapsing star
by the Oppenheimer-Snyder solution. 
But the results of this paper can be applied
to arbitrary gravitational collapse where the world sheet of the
star surface coincides with a collection of radial timelike geodesics
in a Schwarzschild spacetime. Such a situation is realized
when pressure of the star is negligible, that is, the
Tolman-Bondi collapse \cite{Tolman:1934,Bondi:1947}. It is known that depending
on the initial condition, the Tolman-Bondi collapse leads to
the formation of a naked singularity at the center.
Our results can be applied to a star under the Tolman-Bondi collapse, 
as long as nothing is emitted from a naked singularity.

How about the collapse of a star having nonzero pressure?
In this case, some of the formulas
that uses the speciality of the motion of the star surface,
e.g., the third equality in Eq.~\eqref{definition-beta}, 
must be modified.
In particular, the constancy of the redshift factor
on the limb of the image does not hold for a star surface with nonzero
four-acceleration. 
It would be interesting to extend our formalism to 
collapses with nonzero pressure,
and this issue will be discussed elsewhere. 
Of course, our method highly depends on the spherical symmetry
of the Schwarzschild spacetime. 
It is an important future problem to extend our formalism
to more general cases like axisymmetric collapse of a rotating star.

Does our analysis have a practical meaning for 
realistic gravitational collapses of stars?
As commented in Sec.~\ref{Sec:I}, we expect that 
our analysis would have meaning 
in the context of neutrino observations 
rather than electromagnetic observations. 
Although many simulations have been performed up to now, 
the mechanism of supernova explosions is
still under debate. A very naive expected scenario 
is as follows. During the collapse,  
due to the increase in the density of the star,
a neutrino-sphere is formed in which
neutrinos are trapped. The core bounce happens
due to the formation of a proto-neutron star,
and a shock propagates toward outward direction.
In the case that the supernova explosion happens, 
the shock stalls during the propagation,
but it revives by some mechanisms (including reheating
by neutrinos) and blows the star surface off.
The proto-neutron star at the center gradually cools down 
and becomes a neutron star. By contrast, if the star fails to explode, 
which is called a failed supernova, 
matter of the star continues to be accreted onto 
the proto-neutron star. As a result, the proto-neutron star
is expected to begin gravitational collapse 
and become a black hole. In this case,  
the neutrino sphere would fall into the black hole as well.
In Refs.~\cite{Baumgarte:1996,Beacom:2000,Sasaqui:2005}, it was argued that
if the neutrino flux is measurably high
in this stage, the neutrino flux would be terminated abruptly.
Our study in this paper is a necessary step toward
predicting the detailed process of this
neutrino flux truncation including general relativistic
effects on neutrino propagation.

Another interesting direction is the multimessenger observations.
In Refs.~\cite{Sotani:2007,Sotani:2009}, gravitational waves
generated by contracting magnetic fields during 
the Oppenheimer-Snyder collapse were studied 
by treating both electromagnetic fields and gravitational waves
as perturbations. 
Simultaneous observations of gravitational waves and neutrinos
would give more information on a collapsing star
under strong gravity environment.

The most fascinating possibility is the 
black hole formation in the gravitational collapse of Betelgeuse,
which is a red giant
nearing the end of its life.
Its distance from the Earth is estimated
as $197\pm 45~\mathrm{pc}$ \cite{Harper:2008}, and 
if a failed supernova happens at such a close position, 
observations will give us various valuable information
including general relativistic
effects near the black hole.
The study of this paper is a necessary step
to prepare for such a possibility.

%
%

\acknowledgments

We thank Yudai Suwa, Takaaki Kajita, 
Tomohiro Harada, Hirotada Okawa and Hajime Sotani for helpful comments.
We also thank the International Molecule-type Workshop
``Dynamics in Strong Gravity Universe'' (YITP-T-18-05)
held at Yukawa Institute 
for Theoretical Physics at Kyoto University 
 for hospitality.
The work of H.Y. was supported by the Grant-in-Aid for
Scientific Research (C) (Grant No. JP18K03654) from Japan Society for
the Promotion of Science (JSPS).
This work was partly supported by
Osaka City University Advanced Mathematical Institute
(MEXT Joint Usage/Research Center on Mathematics and Theoretical Physics).

\appendix

%
%

\section{Approximate analysis near the center of the image}
\label{Appendix_A}

In this appendix, we study perturbative approximation
around the center of the disk image.
To be specific, the redshift factor 
$\alpha(b)$ is studied
for small $b$ values (i.e., small $\vartheta_{\rm o}$ values)
by a perturbative method, and  
the two kinds of spectral flux, $\hat{\mathcal{F}}_{\alpha}^{\rm (N)}$
  and $\hat{\mathcal{F}}_{\alpha}^{\rm (E)}$, 
defined in Eqs.~\eqref{spectral-photon-flux-monochromatic-case} and 
\eqref{spectral-radiant-flux-monochromatic-case}
in the neighborhood of the central value 
of the redshift factor are studied 
for a monochromatic radiator.
The late time behavior of these quantities is also presented.

\subsection{The redshift factor}

From Eq.~\eqref{redshift-in-terms-of-thetae},
varying $b$ from zero causes the change in the
redshift factor through the two effects:
The change in the emitted angle $\vartheta_{\rm e}$
and the change in the radius $r_{\rm e}$
at the emission event.
Because the first effect can be absorbed into the second effect
by the relation \eqref{b-thetae-thetao},
we begin with examining
the dependence of the radius $r_{\rm e}$ on $b$.
By perturbative expansion 
with respect to $b^2$, the geodesic equation for photons,
Eq.~\eqref{geodesic-drdt}, with the plus sign becomes
\begin{eqnarray}
  \frac{dt}{dr} 
  = f^{-1}+\frac{b^2}{2r^2} +\frac{3b^4}{8r^4}f + O\left(b^6\right).
\end{eqnarray}
Integrating this equation and substituting 
$(t,\ r)=(t_{\rm e}(r_{\rm e}), r_{\rm e})$, we have
\begin{multline}
  t_{\rm e}(r_{\rm e})-t_{\rm o}
 =
  r_{\rm e}-r_{\rm o} + 2M\log\left(\frac{r_{\rm e}-2M}{r_{\rm o}-2M}\right)
  -\frac{b^2}{2}\left(\frac{1}{r_{\rm e}}-\frac{1}{r_{\rm o}}\right)
  \\
  +b^4\left[
    -\frac18\left(\frac{1}{r_{\rm e}^3}-\frac{1}{r_{\rm o}^3}\right)
    +\frac{3M}{16}\left(\frac{1}{r_{\rm e}^4}-\frac{1}{r_{\rm o}^4}\right)
    \right]
  +O\left(b^6\right)
  \label{relation-to-re-nonzerob-secondorder}
\end{multline}
We give the expansion form of $r_{\rm e}$ as
\begin{equation}
  r_{\rm e}(b)=r_{\rm e}^{(0)} + r_{\rm e}^{(1)}b^2 + r_{\rm e}^{(2)}b^4
  +O(b^6),
  \label{expand_re_smallb}
\end{equation}
where $r_{\rm e}^{(n)}$ denotes the $n$-th order quantity
with respect to the small parameter $b^2$.
Here, the zeroth-order quantity 
${r}_{\rm e}^{(0)}$ satisfies 
\begin{equation}
  t_{\rm o} = t_{\rm e}({r}_{\rm e}^{(0)})-{r}_{\rm e}^{(0)}+r_{\rm o} - 2M\log\left(\frac{{r}_{\rm e}^{(0)}-2M}{r_{\rm o}-2M}\right).
  \label{radial-geodesic-relation}
\end{equation}
From this formula, together with Eqs.~\eqref{timelike-geodesic-r}
and \eqref{timelike-geodesic-t},
$t_{\rm o}$ is expressed as a function of $r_{\rm e}^{(0)}$. 
We substitute Eq.~\eqref{expand_re_smallb}
into Eq.~\eqref{relation-to-re-nonzerob-secondorder}
and collect terms with the same order.
In expanding $t_{\rm e}(r_{\rm e})$, the quantities
$\left.dt_{\rm e}/dr_{\rm e}\right|_{r_{\rm e}=r_{\rm e}^{(0)}}$ and
$\left.d^2t_{\rm e}/dr_{\rm e}^2\right|_{r_{\rm e}=r_{\rm e}^{(0)}}$ appear,
and we use Eqs.~\eqref{Eq:geodesic_dottre} and \eqref{Eq:geodesic_dortre}
in order to evaluate these quantities.
As a result, we find
\begin{equation}
  r_{\rm e}^{(1)} = \frac{(1/2)f}
  {1+\left(1-{f}/{f(R)}\right)^{-1/2}}
  \left(\frac{1}{r_{\rm e}^{(0)}}-\frac{1}{r_{\rm o}}\right),
\end{equation}
\begin{multline}
  r_{\rm e}^{(2)} =\frac{f}
  {1+\left(1-{f}/{f(R)}\right)^{-1/2}}
  \left\{
  \frac{f^{\prime}}{4f^{2}}
  \left[2+\frac{2-3f/f(R)}
    {\left(1-{f}/{f(R)}\right)^{3/2}}\right]r_{\rm e}^{(1)2}
  \right.
  \\
  \left.
  -\frac{r_{\rm e}^{(1)}}{2r_{\rm e}^{(0)2}}
  +\frac18\left(\frac{1}{r_{\rm e}^{(0)3}}-\frac{1}{r_{\rm o}^3}\right)
    -\frac{3M}{16}\left(\frac{1}{r_{\rm e}^{(0)4}}-\frac{1}{r_{\rm o}^4}\right)
    \right\},
\end{multline}
where $f$ and $f^{\prime}$ are evaluated at $r_{\rm e}^{(0)}$.
Since $r_{\rm e}(b)$ is also expressed in terms of $r_{\rm e}^{(0)}$,
we have obtained the relation between $r_{\rm e}(b)$ and $t_{\rm o}$.

%
\begin{figure}[tb]
 \centering
 \includegraphics[width=0.5\textwidth,bb=0 0 360 252]{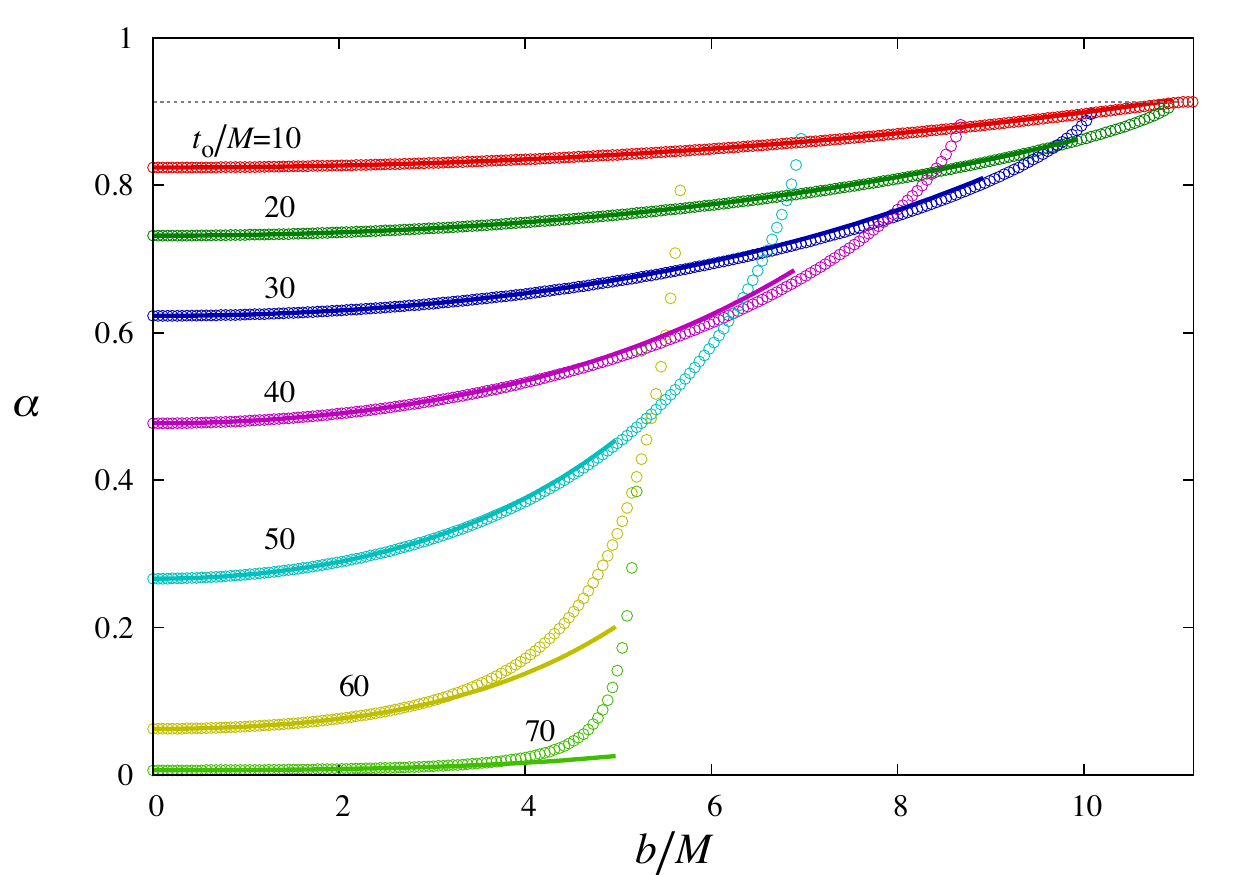}
 \caption{The behavior of the
   redshift factor $\alpha$ as a function of $b$
   near the center $b=0$ obtained by the approximate
   formula~\eqref{near_center_redshift} (solid curves) 
   and the numerical data (circles,
   $\circ$). The data are shown 
   for the observer's time $t_{\rm o}/M=10$, $20$, ..., $70$
   in the case $R=10M$ and $r_{\rm o}=50M$.
}
 \label{Fig:nearcenter_alpha}
\end{figure}
%

Now, we examine the expansion formula of $\alpha$.
For small values of $b^2$, 
the redshift factor $\alpha$ is given by
\begin{equation}
\alpha = \frac{f(r_{\rm e})}{\sqrt{f(r_{\rm o})f(R)}}
  \left[1+\sqrt{\left(1-\frac{f(r_{\rm e})}{f(R)}\right)
      \left(1-\frac{f(r_{\rm e})}{r_{\rm e}^2}b^2\right)}
    \right]^{-1}
  \label{redshift_factor_smallb}
\end{equation}
from Eqs.~\eqref{b-thetae-thetao} and \eqref{redshift-in-terms-of-thetae}.
Substituting Eq.~\eqref{expand_re_smallb} and expanding
with respect to $b$, we find the expansion form,
\begin{equation}
  \alpha =
         {\alpha}^{(0)} + {\alpha}^{(1)}b^2 + {\alpha}^{(2)}b^4 +O(b^6),
 \label{near_center_redshift}
\end{equation}
where $\alpha^{(n)}$ denotes the $n$-th order term
with respect to the expansion parameter $b^2$.
Here, $\alpha^{(0)}$ coincides with the value at the center of the image,
$\alpha_{\rm cent}$ in the main article,
and it is trivially obtained by
substituting $r_{\rm e}=r_{\rm e}^{(0)}$ and $b=0$
into Eq.~\eqref{redshift_factor_smallb}. $\alpha^{(1)}$
is calculated as
\begin{multline}
  \alpha^{(1)}=\frac{f}{2\sqrt{f(r_{\rm o})f(R)}\ r_{\rm e}^{(0)2}}
\left[1+\sqrt{1-\frac{f}{f(R)}}\right]^{-2}
  \\
  \times
\left[\frac{M}{r_{\rm e}^{(0)}}-\frac{M}{r_{\rm o}}
  +\sqrt{1-\frac{f}{f(R)}}
  \left(1-\frac{M}{r_{\rm e}^{(0)}}-\frac{M}{r_{\rm o}}\right)\right].
\end{multline}
Although we do not present the complicated form of $\alpha^{(2)}$ here,
it is easily calculated  by, e.g., {\it Mathematica}.
Figure~\ref{Fig:nearcenter_alpha} shows the
behavior of the approximate formula \eqref{near_center_redshift}
of $\alpha(b)$ 
for $t_{\rm o}/M=10$, ..., $70$ in the case $R=10M$ and $r_{\rm o}=50M$.
For $t_{\rm o}\lesssim 10M$, 
this formula approximately coincides with the numerical data
of $\alpha(b)$ in a fairly large domain of $b$ (up to $b\lesssim 10M$).
Although such domain becomes
smaller as the observer's time is increased, the numerical
data agree well with the approximate formula for $b\lesssim 2M$.

\subsection{Spectral flux for a monochromatic radiator}

%
\begin{figure}[tb]
 \centering
 \includegraphics[width=0.45\textwidth,bb=0 0 396 256]{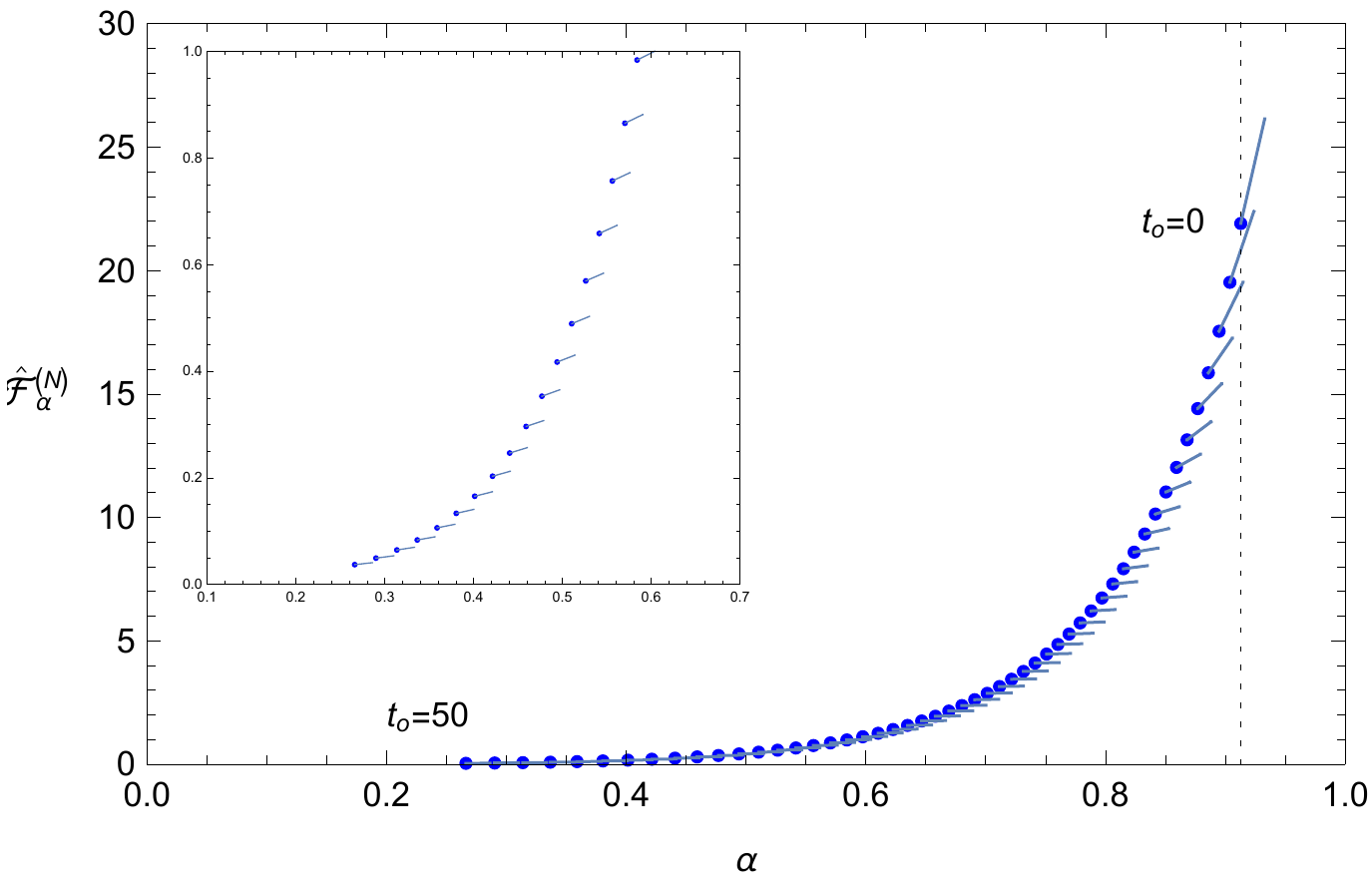}
 \includegraphics[width=0.45\textwidth,bb=0 0 396 256]{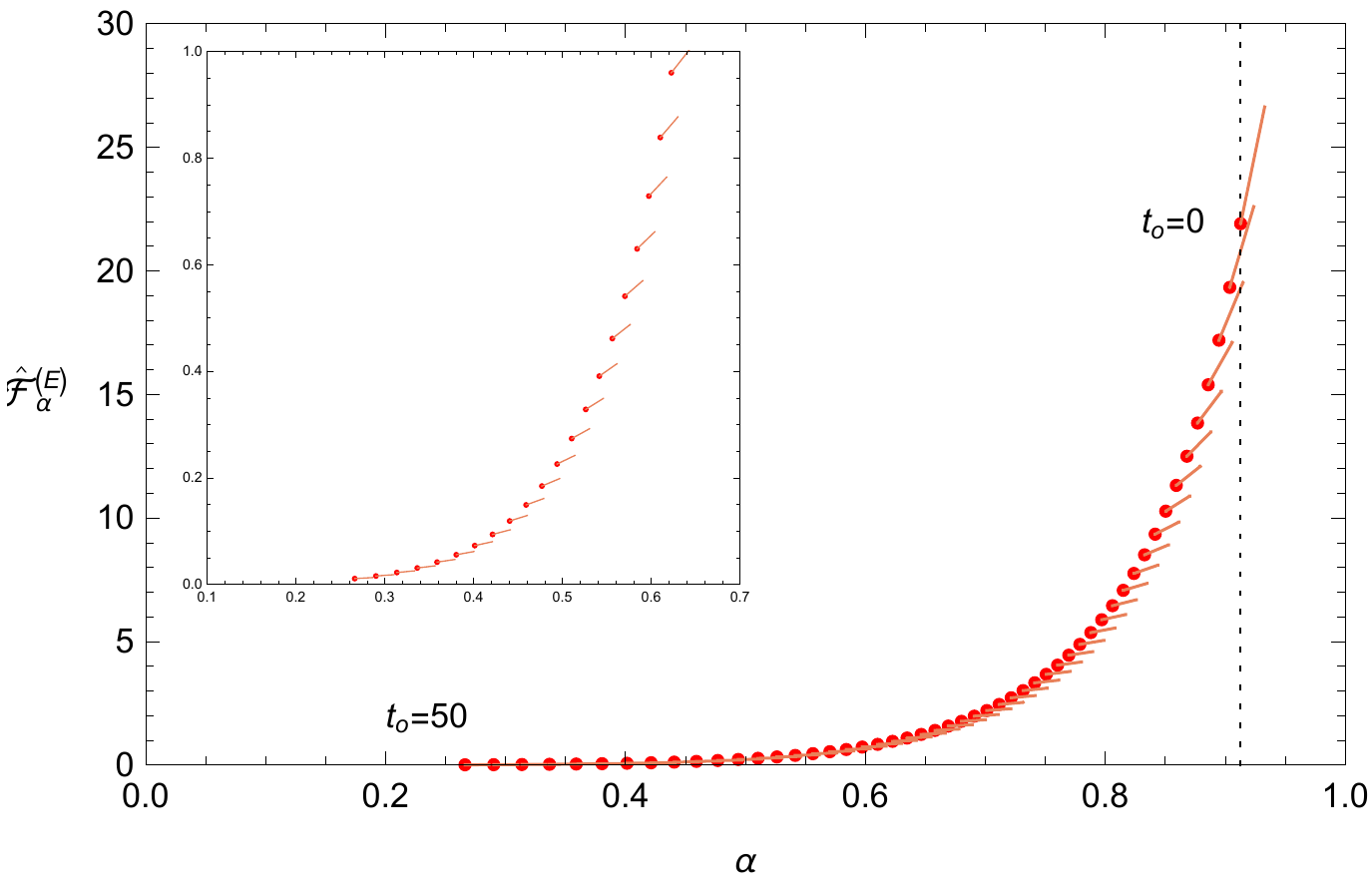}
 \caption{The behavior of $\hat{\mathcal{F}}_{\alpha}^{\rm (N)}$ (left panel)
and $\hat{\mathcal{F}}_{\alpha}^{\rm (E)}$ (right panel)
in the neighborhood of $\alpha=\alpha^{(0)}$ obtained by
the approximate formulas \eqref{spectral_flux_number_alpha_expand_b}
and \eqref{spectral_flux_energy_alpha_expand_b}
for $t_{\rm o}=0$, $1$, ..., $50$ (in the unit $M=1$).
Inset of each panel enlarges the late time data. 
}
 \label{Fig:Falpha-alpha-linear3}
\end{figure}
%

In the case of a monochromatic radiator,
the spectral photon/radiant flux is zero in the range
$0\le \alpha< \alpha^{(0)}$ and has nonzero
values for $\alpha^{(0)}\le \alpha\le\alpha_{\rm limb}$.
In the nonzero region,
the two kinds of spectral flux are  
given by the first terms of the 
formulas \eqref{spectral-photon-flux-monochromatic-case}
and \eqref{spectral-radiant-flux-monochromatic-case}
with $J_{\rm e}^{\rm (N)}(r_{\rm e})=J_{\rm e}^{\rm (N)}(R)$ and
$\left.\langle\omega_{\rm e}^\prime\rangle\right|_{r_{\rm e}}
=\left.\langle\omega_{\rm e}^\prime\rangle\right|_{R}$.
Substituting the expansion formula for $\alpha$,
they are rewritten as
\begin{subequations}
\begin{eqnarray}
\hat{\mathcal{F}}_{\alpha}^{\rm (N)} &\approx &
  \frac{f(r_{\rm o})^{3/2}}{R^2f(R)^{1/2}}\ 
  \frac{\alpha^{(0)3}}{\alpha^{(1)}}
  \left[1+
    \frac{1}{\alpha^{(1)}}\left(\frac{3\alpha^{(1)}}{\alpha^{(0)}}-\frac{2\alpha^{(2)}}{\alpha^{(1)}}\right)(\alpha-\alpha^{(0)})\right],
  \label{spectral_flux_number_alpha_expand_b}
  \\
\hat{\mathcal{F}}_{\alpha}^{\rm (E)} &\approx & 
  \frac{f(r_{\rm o})^2}{R^2f(R)}\
  \frac{\alpha^{(0)4}}{\alpha^{(1)}}
  \left[1+
    \frac{2}{\alpha^{(1)}}\left(\frac{2\alpha^{(1)}}{\alpha^{(0)}}-\frac{\alpha^{(2)}}{\alpha^{(1)}}\right)(\alpha-\alpha^{(0)})\right],
  \label{spectral_flux_energy_alpha_expand_b}
\end{eqnarray}
\end{subequations}
in the neighborhood of $\alpha=\alpha^{(0)}$.
Figure~\ref{Fig:Falpha-alpha-linear3} shows the behavior
of $\hat{\mathcal{F}}_{\alpha}^{\rm (N)}$ (left panel)
and $\hat{\mathcal{F}}_{\alpha}^{\rm (E)}$ (right panel)
in the neighborhood of $\alpha=\alpha^{(0)}$ obtained by these formulas
for $t_{\rm o}=0$, $1$, ..., $50$ (in the unit $M=1$).

\subsection{Late time behavior of observable quantities}

We study late time behavior of the redshift factor
$\alpha^{(0)}$ at the central point
and the two kinds of normalized spectral flux, 
$\hat{\mathcal{F}}_{\alpha}^{\rm (N)}$ 
and 
$\hat{\mathcal{F}}_{\alpha}^{\rm (E)}$, at $\alpha=\alpha^{(0)}$.
Since photons that come to  the center of the disk image
are emitted from the near horizon region $r\approx 2M$,
we keep only leading order terms of $(r_{\rm e}^{(0)}-2M)$
in this approximation. 
The world sheet \eqref{timelike-geodesic-r}
and \eqref{timelike-geodesic-t} of the star surface is approximately given by
\begin{equation}
  t_{\rm e}+\mathcal{T}
  \approx 
  2M\log\left[\frac{4f(R)}{f(r_{\rm e})}\right]
  +4MC^\prime,
\end{equation}
with
\begin{equation}
C^\prime :=\left(\frac{R}{2M}-1\right)+
\frac{\xi_{\rm h}}{2}\sqrt{\frac{R}{2M}-1}\left(1+\frac{R}{4M}\right),
\end{equation}
where $\mathcal{T}$ and $\xi_{\rm h}$
are defined in Eqs.~\eqref{te0} and \eqref{horizon_crossing_xih},
respectively. 
Substituting $r_{\rm e}=r_{\rm e}^{(0)}$ into this formula and 
combining with the relation \eqref{radial-geodesic-relation},
we have
\begin{equation}
  f(r_{\rm e}^{(0)})\approx
  \sqrt{\frac{2R}{M}}f(R)e^{C^\prime}e^{-t_{\rm o}/4M}.
\end{equation}
Since $\alpha^{(0)}$
and $\alpha^{(1)}$
are approximated as
\begin{equation}
  \alpha^{(0)}\approx \frac{f(r_{\rm e}^{(0)})}{2\sqrt{f(r_{\rm o})f(R)}}
\quad \textrm{and} \quad
\alpha^{(1)}\approx \sqrt{\frac{f(r_{\rm o})}{f(R)}}\ \frac{f(r_{\rm e}^{(0)})}{32M^2},
\end{equation}
respectively, we obtain the late time behavior for the redshift factor,
\begin{equation}
  \alpha^{(0)}\approx
  \sqrt{\frac{Rf(R)}{2Mf(r_{\rm o})}}e^{C^\prime}e^{-t_{\rm o}/4M},
\end{equation}
and the two kinds of spectral flux,
\begin{subequations}
\begin{eqnarray}
  \left.\hat{\mathcal{F}}_{\alpha}^{\rm (N)}\right|_{\alpha^{(0)}}
  &\approx &
  \frac{8M}{R}\sqrt{\frac{f(R)}{f(r_{\rm o})}}
  e^{2{C^\prime}}e^{-t_{\rm o}/2M},
  \\
  \left.\hat{\mathcal{F}}_{\alpha}^{\rm (E)}\right|_{\alpha^{(0)}}
  &\approx &
  4\sqrt{\frac{2Mf(R)}{Rf(r_{\rm o})}}
  e^{3{C^\prime}}e^{-3t_{\rm o}/4M}.
\end{eqnarray}
\end{subequations}
at the central point $b=0$ (i.e., $\vartheta_{\rm o}=0$).



\end{document}